\documentclass[longauth]{aa}
\usepackage{graphicx}
\usepackage{natbib}
\usepackage{scalerel}
\usepackage{svg}
\usepackage[table]{xcolor}
\usepackage{xspace}
\usepackage{txfonts}
\usepackage[pdfencoding=auto,psdextra]{hyperref}
\hypersetup{
    colorlinks=true,
    linkcolor=blue,
    filecolor=magenta,      
    urlcolor=blue,
    citecolor=blue
}
\usepackage{textgreek}
\usepackage[dvipsnames]{xcolor}

\newcommand{\Nii}{[\ion{N}{ii}]\xspace}
\newcommand{\Revolver}{\textsc{Revolver}\xspace}
\newcommand{\Disperse}{\textsc{DisPerSE}\xspace}
\newcommand{\Sparkling}{\textsc{Sparkling}\xspace}
\newcommand{\twodvf}{\textsc{2D void-finder}\xspace}
\newcommand{\Ei}{\ensuremath{\mathcal{E}_1}\xspace}

\newcommand{\Eiii}{\ensuremath{\mathcal{E}_3}\xspace}

\newcommand{\Ev}{\ensuremath{\mathcal{E}_5}\xspace}
\newcommand{\Evi}{\ensuremath{\mathcal{E}_6}\xspace}
\newcommand{\Evii}{\ensuremath{\mathcal{E}_7}\xspace}

\makeatletter
\renewcommand*\aa@pageof{, page \thepage{} of \pageref*{LastPage}}
\makeatother

\usepackage[utf8]{inputenc}
\usepackage{enumitem} 

\usepackage[switch, modulo]{lineno}

\renewcommand{\linenumbers}[0]{}
\usepackage{euclid}

\usepackage{collcell}
\usepackage{siunitx} 

\definecolor{low_p}{HTML}{57bb8a}   
\definecolor{mid_p}{HTML}{ffd966}   
\definecolor{high_p}{HTML}{e67c73}  

\definecolor{voids}{HTML}{3A86FF}
\definecolor{walls}{HTML}{FF006E}
\definecolor{highd}{HTML}{000000}

\begin{document}

\vspace{-0.09cm}
\title{\Euclid preparation.}
\subtitle{Probing galaxy evolution within cosmic voids in \Euclid-like simulations} 

\newcommand{\orcid}[1]{}   
\author{Euclid Collaboration: G.~Papini\orcid{0000-0002-5810-318X}\thanks{\email{giulia.papini6@unibo.it}}\inst{\ref{aff1},\ref{aff2}}
\and O.~Cucciati\orcid{0000-0002-9336-7551}\inst{\ref{aff1}}
\and M.~Bolzonella\orcid{0000-0003-3278-4607}\inst{\ref{aff1}}
\and S.~Contarini\orcid{0000-0002-9843-723X}\inst{\ref{aff3}}
\and S.~Sartori\orcid{0009-0000-5585-8336}\inst{\ref{aff4}}
\and K.~Kraljic\orcid{0000-0001-6180-0245}\inst{\ref{aff5}}
\and C.~M.~Correa\orcid{0000-0003-1596-5254}\inst{\ref{aff3}}
\and P.~Vielzeuf\orcid{0000-0003-2035-9339}\inst{\ref{aff4}}
\and G.~De~Lucia\orcid{0000-0002-6220-9104}\inst{\ref{aff6}}
\and A.~Pisani\orcid{0000-0002-6146-4437}\inst{\ref{aff4}}
\and J.~G.~Sorce\orcid{0000-0002-2307-2432}\inst{\ref{aff7},\ref{aff8}}
\and M.~Magliocchetti\orcid{0000-0001-9158-4838}\inst{\ref{aff9}}
\and C.~Schimd\orcid{0000-0002-0003-1873}\inst{\ref{aff10},\ref{aff6}}
\and F.~Fontanot\orcid{0000-0003-4744-0188}\inst{\ref{aff6},\ref{aff11}}
\and E.~Sarpa\orcid{0000-0002-1256-655X}\inst{\ref{aff12},\ref{aff13},\ref{aff6}}
\and L.~Pozzetti\orcid{0000-0001-7085-0412}\inst{\ref{aff1}}
\and A.~Enia\orcid{0000-0002-0200-2857}\inst{\ref{aff1}}
\and E.~Pouyer\inst{\ref{aff10}}
\and M.~Hirschmann\orcid{0000-0002-3301-3321}\inst{\ref{aff14}}
\and M.~Spinelli\orcid{0000-0003-0148-3254}\inst{\ref{aff15},\ref{aff6},\ref{aff16}}
\and L.~Xie\orcid{0000-0003-3864-068X}\inst{\ref{aff17}}
\and G.~Zamorani\orcid{0000-0002-2318-301X}\inst{\ref{aff1}}
\and M.~Fumagalli\orcid{0000-0001-6676-3842}\inst{\ref{aff6},\ref{aff18}}
\and M.~Fossati\orcid{0000-0002-9043-8764}\inst{\ref{aff18},\ref{aff19}}
\and M.~Aubert\orcid{0009-0002-7667-8814}\inst{\ref{aff20},\ref{aff21}}
\and P.~Awad\orcid{0000-0002-0428-849X}\inst{\ref{aff22}}
\and A.~W.~S.~Man\orcid{0000-0003-2475-124X}\inst{\ref{aff23}}
\and A.~Kov\'acs\orcid{0000-0002-5825-579X}\inst{\ref{aff24},\ref{aff25}}
\and K.~V.~Nedkova\orcid{0000-0001-5294-8002}\inst{\ref{aff26}}
\and B.~Altieri\orcid{0000-0003-3936-0284}\inst{\ref{aff27}}
\and A.~Amara\inst{\ref{aff28}}
\and S.~Andreon\orcid{0000-0002-2041-8784}\inst{\ref{aff19}}
\and N.~Auricchio\orcid{0000-0003-4444-8651}\inst{\ref{aff1}}
\and C.~Baccigalupi\orcid{0000-0002-8211-1630}\inst{\ref{aff11},\ref{aff6},\ref{aff29},\ref{aff12}}
\and M.~Baldi\orcid{0000-0003-4145-1943}\inst{\ref{aff30},\ref{aff1},\ref{aff31}}
\and S.~Bardelli\orcid{0000-0002-8900-0298}\inst{\ref{aff1}}
\and P.~Battaglia\orcid{0000-0002-7337-5909}\inst{\ref{aff1}}
\and A.~Biviano\orcid{0000-0002-0857-0732}\inst{\ref{aff6},\ref{aff11}}
\and E.~Branchini\orcid{0000-0002-0808-6908}\inst{\ref{aff32},\ref{aff33},\ref{aff19}}
\and M.~Brescia\orcid{0000-0001-9506-5680}\inst{\ref{aff34},\ref{aff35}}
\and S.~Camera\orcid{0000-0003-3399-3574}\inst{\ref{aff36},\ref{aff37},\ref{aff38}}
\and V.~Capobianco\orcid{0000-0002-3309-7692}\inst{\ref{aff38}}
\and C.~Carbone\orcid{0000-0003-0125-3563}\inst{\ref{aff39}}
\and V.~F.~Cardone\inst{\ref{aff40},\ref{aff41}}
\and J.~Carretero\orcid{0000-0002-3130-0204}\inst{\ref{aff42},\ref{aff43}}
\and S.~Casas\orcid{0000-0002-4751-5138}\inst{\ref{aff44},\ref{aff45}}
\and M.~Castellano\orcid{0000-0001-9875-8263}\inst{\ref{aff40}}
\and G.~Castignani\orcid{0000-0001-6831-0687}\inst{\ref{aff1}}
\and S.~Cavuoti\orcid{0000-0002-3787-4196}\inst{\ref{aff35},\ref{aff46}}
\and K.~C.~Chambers\orcid{0000-0001-6965-7789}\inst{\ref{aff47}}
\and A.~Cimatti\inst{\ref{aff48}}
\and C.~Colodro-Conde\inst{\ref{aff49}}
\and G.~Congedo\orcid{0000-0003-2508-0046}\inst{\ref{aff50}}
\and L.~Conversi\orcid{0000-0002-6710-8476}\inst{\ref{aff51},\ref{aff27}}
\and Y.~Copin\orcid{0000-0002-5317-7518}\inst{\ref{aff21}}
\and A.~Costille\inst{\ref{aff10}}
\and F.~Courbin\orcid{0000-0003-0758-6510}\inst{\ref{aff52},\ref{aff53},\ref{aff54}}
\and H.~M.~Courtois\orcid{0000-0003-0509-1776}\inst{\ref{aff55}}
\and M.~Cropper\orcid{0000-0003-4571-9468}\inst{\ref{aff56}}
\and A.~Da~Silva\orcid{0000-0002-6385-1609}\inst{\ref{aff57},\ref{aff58}}
\and H.~Degaudenzi\orcid{0000-0002-5887-6799}\inst{\ref{aff59}}
\and S.~de~la~Torre\inst{\ref{aff10}}
\and H.~Dole\orcid{0000-0002-9767-3839}\inst{\ref{aff8}}
\and F.~Dubath\orcid{0000-0002-6533-2810}\inst{\ref{aff59}}
\and X.~Dupac\inst{\ref{aff27}}
\and S.~Dusini\orcid{0000-0002-1128-0664}\inst{\ref{aff60}}
\and S.~Escoffier\orcid{0000-0002-2847-7498}\inst{\ref{aff4}}
\and M.~Farina\orcid{0000-0002-3089-7846}\inst{\ref{aff9}}
\and S.~Farrens\orcid{0000-0002-9594-9387}\inst{\ref{aff61}}
\and S.~Ferriol\inst{\ref{aff21}}
\and F.~Finelli\orcid{0000-0002-6694-3269}\inst{\ref{aff1},\ref{aff62}}
\and P.~Fosalba\orcid{0000-0002-1510-5214}\inst{\ref{aff63},\ref{aff64}}
\and S.~Fotopoulou\orcid{0000-0002-9686-254X}\inst{\ref{aff65}}
\and N.~Fourmanoit\orcid{0009-0005-6816-6925}\inst{\ref{aff4}}
\and M.~Frailis\orcid{0000-0002-7400-2135}\inst{\ref{aff6}}
\and E.~Franceschi\orcid{0000-0002-0585-6591}\inst{\ref{aff1}}
\and M.~Fumana\orcid{0000-0001-6787-5950}\inst{\ref{aff39}}
\and S.~Galeotta\orcid{0000-0002-3748-5115}\inst{\ref{aff6}}
\and K.~George\orcid{0000-0002-1734-8455}\inst{\ref{aff66}}
\and B.~Gillis\orcid{0000-0002-4478-1270}\inst{\ref{aff50}}
\and C.~Giocoli\orcid{0000-0002-9590-7961}\inst{\ref{aff1},\ref{aff31}}
\and J.~Gracia-Carpio\orcid{0000-0003-4689-3134}\inst{\ref{aff3}}
\and A.~Grazian\orcid{0000-0002-5688-0663}\inst{\ref{aff67}}
\and F.~Grupp\inst{\ref{aff3},\ref{aff68}}
\and S.~V.~H.~Haugan\orcid{0000-0001-9648-7260}\inst{\ref{aff69}}
\and W.~Holmes\inst{\ref{aff70}}
\and F.~Hormuth\inst{\ref{aff71}}
\and A.~Hornstrup\orcid{0000-0002-3363-0936}\inst{\ref{aff72},\ref{aff73}}
\and K.~Jahnke\orcid{0000-0003-3804-2137}\inst{\ref{aff74}}
\and M.~Jhabvala\inst{\ref{aff75}}
\and B.~Joachimi\orcid{0000-0001-7494-1303}\inst{\ref{aff76}}
\and S.~Kermiche\orcid{0000-0002-0302-5735}\inst{\ref{aff4}}
\and A.~Kiessling\orcid{0000-0002-2590-1273}\inst{\ref{aff70}}
\and B.~Kubik\orcid{0009-0006-5823-4880}\inst{\ref{aff21}}
\and M.~K\"ummel\orcid{0000-0003-2791-2117}\inst{\ref{aff68}}
\and M.~Kunz\orcid{0000-0002-3052-7394}\inst{\ref{aff77}}
\and H.~Kurki-Suonio\orcid{0000-0002-4618-3063}\inst{\ref{aff78},\ref{aff79}}
\and A.~M.~C.~Le~Brun\orcid{0000-0002-0936-4594}\inst{\ref{aff80}}
\and S.~Ligori\orcid{0000-0003-4172-4606}\inst{\ref{aff38}}
\and P.~B.~Lilje\orcid{0000-0003-4324-7794}\inst{\ref{aff69}}
\and V.~Lindholm\orcid{0000-0003-2317-5471}\inst{\ref{aff78},\ref{aff79}}
\and I.~Lloro\orcid{0000-0001-5966-1434}\inst{\ref{aff81}}
\and G.~Mainetti\orcid{0000-0003-2384-2377}\inst{\ref{aff82}}
\and E.~Maiorano\orcid{0000-0003-2593-4355}\inst{\ref{aff1}}
\and O.~Mansutti\orcid{0000-0001-5758-4658}\inst{\ref{aff6}}
\and S.~Marcin\inst{\ref{aff83}}
\and O.~Marggraf\orcid{0000-0001-7242-3852}\inst{\ref{aff84}}
\and M.~Martinelli\orcid{0000-0002-6943-7732}\inst{\ref{aff40},\ref{aff41}}
\and N.~Martinet\orcid{0000-0003-2786-7790}\inst{\ref{aff10}}
\and F.~Marulli\orcid{0000-0002-8850-0303}\inst{\ref{aff2},\ref{aff1},\ref{aff31}}
\and R.~J.~Massey\orcid{0000-0002-6085-3780}\inst{\ref{aff85}}
\and S.~Maurogordato\inst{\ref{aff15}}
\and E.~Medinaceli\orcid{0000-0002-4040-7783}\inst{\ref{aff1}}
\and S.~Mei\orcid{0000-0002-2849-559X}\inst{\ref{aff86},\ref{aff87}}
\and M.~Meneghetti\orcid{0000-0003-1225-7084}\inst{\ref{aff1},\ref{aff31}}
\and E.~Merlin\orcid{0000-0001-6870-8900}\inst{\ref{aff40}}
\and G.~Meylan\inst{\ref{aff88}}
\and A.~Mora\orcid{0000-0002-1922-8529}\inst{\ref{aff89}}
\and M.~Moresco\orcid{0000-0002-7616-7136}\inst{\ref{aff2},\ref{aff1}}
\and L.~Moscardini\orcid{0000-0002-3473-6716}\inst{\ref{aff2},\ref{aff1},\ref{aff31}}
\and R.~Nakajima\orcid{0009-0009-1213-7040}\inst{\ref{aff84}}
\and C.~Neissner\orcid{0000-0001-8524-4968}\inst{\ref{aff90},\ref{aff43}}
\and S.-M.~Niemi\orcid{0009-0005-0247-0086}\inst{\ref{aff91}}
\and C.~Padilla\orcid{0000-0001-7951-0166}\inst{\ref{aff90}}
\and S.~Paltani\orcid{0000-0002-8108-9179}\inst{\ref{aff59}}
\and F.~Pasian\orcid{0000-0002-4869-3227}\inst{\ref{aff6}}
\and K.~Pedersen\inst{\ref{aff92}}
\and V.~Pettorino\orcid{0000-0002-4203-9320}\inst{\ref{aff91}}
\and S.~Pires\orcid{0000-0002-0249-2104}\inst{\ref{aff61}}
\and G.~Polenta\orcid{0000-0003-4067-9196}\inst{\ref{aff93}}
\and M.~Poncet\inst{\ref{aff94}}
\and L.~A.~Popa\inst{\ref{aff95}}
\and F.~Raison\orcid{0000-0002-7819-6918}\inst{\ref{aff3}}
\and A.~Renzi\orcid{0000-0001-9856-1970}\inst{\ref{aff96},\ref{aff60}}
\and J.~Rhodes\orcid{0000-0002-4485-8549}\inst{\ref{aff70}}
\and G.~Riccio\inst{\ref{aff35}}
\and E.~Romelli\orcid{0000-0003-3069-9222}\inst{\ref{aff6}}
\and M.~Roncarelli\orcid{0000-0001-9587-7822}\inst{\ref{aff1}}
\and C.~Rosset\orcid{0000-0003-0286-2192}\inst{\ref{aff86}}
\and R.~Saglia\orcid{0000-0003-0378-7032}\inst{\ref{aff68},\ref{aff3}}
\and Z.~Sakr\orcid{0000-0002-4823-3757}\inst{\ref{aff97},\ref{aff98},\ref{aff99}}
\and D.~Sapone\orcid{0000-0001-7089-4503}\inst{\ref{aff100}}
\and B.~Sartoris\orcid{0000-0003-1337-5269}\inst{\ref{aff68},\ref{aff6}}
\and P.~Schneider\orcid{0000-0001-8561-2679}\inst{\ref{aff84}}
\and T.~Schrabback\orcid{0000-0002-6987-7834}\inst{\ref{aff101}}
\and M.~Scodeggio\inst{\ref{aff39}}
\and A.~Secroun\orcid{0000-0003-0505-3710}\inst{\ref{aff4}}
\and G.~Seidel\orcid{0000-0003-2907-353X}\inst{\ref{aff74}}
\and E.~Sihvola\orcid{0000-0003-1804-7715}\inst{\ref{aff102}}
\and P.~Simon\inst{\ref{aff84}}
\and C.~Sirignano\orcid{0000-0002-0995-7146}\inst{\ref{aff96},\ref{aff60}}
\and G.~Sirri\orcid{0000-0003-2626-2853}\inst{\ref{aff31}}
\and L.~Stanco\orcid{0000-0002-9706-5104}\inst{\ref{aff60}}
\and P.~Tallada-Cresp\'{i}\orcid{0000-0002-1336-8328}\inst{\ref{aff42},\ref{aff43}}
\and A.~N.~Taylor\inst{\ref{aff50}}
\and I.~Tereno\orcid{0000-0002-4537-6218}\inst{\ref{aff57},\ref{aff103}}
\and N.~Tessore\orcid{0000-0002-9696-7931}\inst{\ref{aff56}}
\and S.~Toft\orcid{0000-0003-3631-7176}\inst{\ref{aff104},\ref{aff105}}
\and R.~Toledo-Moreo\orcid{0000-0002-2997-4859}\inst{\ref{aff106}}
\and F.~Torradeflot\orcid{0000-0003-1160-1517}\inst{\ref{aff43},\ref{aff42}}
\and I.~Tutusaus\orcid{0000-0002-3199-0399}\inst{\ref{aff64},\ref{aff63},\ref{aff98}}
\and J.~Valiviita\orcid{0000-0001-6225-3693}\inst{\ref{aff78},\ref{aff79}}
\and T.~Vassallo\orcid{0000-0001-6512-6358}\inst{\ref{aff6},\ref{aff66}}
\and G.~Verdoes~Kleijn\orcid{0000-0001-5803-2580}\inst{\ref{aff107}}
\and Y.~Wang\orcid{0000-0002-4749-2984}\inst{\ref{aff26}}
\and J.~Weller\orcid{0000-0002-8282-2010}\inst{\ref{aff68},\ref{aff3}}
\and F.~M.~Zerbi\orcid{0000-0002-9996-973X}\inst{\ref{aff19}}
\and E.~Zucca\orcid{0000-0002-5845-8132}\inst{\ref{aff1}}
\and V.~Allevato\orcid{0000-0001-7232-5152}\inst{\ref{aff35}}
\and M.~Ballardini\orcid{0000-0003-4481-3559}\inst{\ref{aff108},\ref{aff109},\ref{aff1}}
\and E.~Bozzo\orcid{0000-0002-8201-1525}\inst{\ref{aff59}}
\and C.~Burigana\orcid{0000-0002-3005-5796}\inst{\ref{aff110},\ref{aff62}}
\and R.~Cabanac\orcid{0000-0001-6679-2600}\inst{\ref{aff98}}
\and M.~Calabrese\orcid{0000-0002-2637-2422}\inst{\ref{aff111},\ref{aff39}}
\and A.~Cappi\inst{\ref{aff15},\ref{aff1}}
\and T.~Castro\orcid{0000-0002-6292-3228}\inst{\ref{aff6},\ref{aff29},\ref{aff11},\ref{aff13}}
\and J.~A.~Escartin~Vigo\inst{\ref{aff3}}
\and L.~Gabarra\orcid{0000-0002-8486-8856}\inst{\ref{aff112}}
\and J.~Macias-Perez\orcid{0000-0002-5385-2763}\inst{\ref{aff113}}
\and R.~Maoli\orcid{0000-0002-6065-3025}\inst{\ref{aff114},\ref{aff40}}
\and J.~Mart\'{i}n-Fleitas\orcid{0000-0002-8594-569X}\inst{\ref{aff115}}
\and N.~Mauri\orcid{0000-0001-8196-1548}\inst{\ref{aff48},\ref{aff31}}
\and R.~B.~Metcalf\orcid{0000-0003-3167-2574}\inst{\ref{aff2},\ref{aff1}}
\and P.~Monaco\orcid{0000-0003-2083-7564}\inst{\ref{aff116},\ref{aff6},\ref{aff29},\ref{aff11}}
\and A.~A.~Nucita\inst{\ref{aff117},\ref{aff118},\ref{aff119}}
\and A.~Pezzotta\orcid{0000-0003-0726-2268}\inst{\ref{aff19}}
\and M.~P\"ontinen\orcid{0000-0001-5442-2530}\inst{\ref{aff78}}
\and I.~Risso\orcid{0000-0003-2525-7761}\inst{\ref{aff19},\ref{aff33}}
\and V.~Scottez\orcid{0009-0008-3864-940X}\inst{\ref{aff120},\ref{aff121}}
\and M.~Sereno\orcid{0000-0003-0302-0325}\inst{\ref{aff1},\ref{aff31}}
\and M.~Tenti\orcid{0000-0002-4254-5901}\inst{\ref{aff31}}
\and M.~Tucci\inst{\ref{aff59}}
\and M.~Viel\orcid{0000-0002-2642-5707}\inst{\ref{aff11},\ref{aff6},\ref{aff12},\ref{aff29},\ref{aff13}}
\and M.~Wiesmann\orcid{0009-0000-8199-5860}\inst{\ref{aff69}}
\and Y.~Akrami\orcid{0000-0002-2407-7956}\inst{\ref{aff97},\ref{aff122}}
\and I.~T.~Andika\orcid{0000-0001-6102-9526}\inst{\ref{aff66}}
\and G.~Angora\orcid{0000-0002-0316-6562}\inst{\ref{aff35},\ref{aff108}}
\and S.~Anselmi\orcid{0000-0002-3579-9583}\inst{\ref{aff60},\ref{aff96},\ref{aff123}}
\and M.~Archidiacono\orcid{0000-0003-4952-9012}\inst{\ref{aff124},\ref{aff125}}
\and F.~Atrio-Barandela\orcid{0000-0002-2130-2513}\inst{\ref{aff126}}
\and L.~Bazzanini\orcid{0000-0003-0727-0137}\inst{\ref{aff108},\ref{aff1}}
\and D.~Bertacca\orcid{0000-0002-2490-7139}\inst{\ref{aff96},\ref{aff67},\ref{aff60}}
\and M.~Bethermin\orcid{0000-0002-3915-2015}\inst{\ref{aff5}}
\and F.~Beutler\orcid{0000-0003-0467-5438}\inst{\ref{aff50}}
\and A.~Blanchard\orcid{0000-0001-8555-9003}\inst{\ref{aff98}}
\and L.~Blot\orcid{0000-0002-9622-7167}\inst{\ref{aff127},\ref{aff80}}
\and S.~Borgani\orcid{0000-0001-6151-6439}\inst{\ref{aff116},\ref{aff11},\ref{aff6},\ref{aff29},\ref{aff13}}
\and M.~L.~Brown\orcid{0000-0002-0370-8077}\inst{\ref{aff128}}
\and S.~Bruton\orcid{0000-0002-6503-5218}\inst{\ref{aff129}}
\and A.~Calabro\orcid{0000-0003-2536-1614}\inst{\ref{aff40}}
\and B.~Camacho~Quevedo\orcid{0000-0002-8789-4232}\inst{\ref{aff11},\ref{aff12},\ref{aff6}}
\and F.~Caro\inst{\ref{aff40}}
\and C.~S.~Carvalho\inst{\ref{aff103}}
\and F.~Cogato\orcid{0000-0003-4632-6113}\inst{\ref{aff2},\ref{aff1}}
\and A.~R.~Cooray\orcid{0000-0002-3892-0190}\inst{\ref{aff130}}
\and S.~Davini\orcid{0000-0003-3269-1718}\inst{\ref{aff33}}
\and F.~De~Paolis\orcid{0000-0001-6460-7563}\inst{\ref{aff117},\ref{aff118},\ref{aff119}}
\and G.~Desprez\orcid{0000-0001-8325-1742}\inst{\ref{aff107}}
\and A.~D\'iaz-S\'anchez\orcid{0000-0003-0748-4768}\inst{\ref{aff131}}
\and S.~Di~Domizio\orcid{0000-0003-2863-5895}\inst{\ref{aff32},\ref{aff33}}
\and J.~M.~Diego\orcid{0000-0001-9065-3926}\inst{\ref{aff132}}
\and P.-A.~Duc\orcid{0000-0003-3343-6284}\inst{\ref{aff5}}
\and V.~Duret\orcid{0009-0009-0383-4960}\inst{\ref{aff4}}
\and M.~Y.~Elkhashab\orcid{0000-0001-9306-2603}\inst{\ref{aff6},\ref{aff29},\ref{aff116},\ref{aff11}}
\and Y.~Fang\orcid{0000-0002-0334-6950}\inst{\ref{aff68}}
\and A.~Finoguenov\orcid{0000-0002-4606-5403}\inst{\ref{aff78}}
\and A.~Franco\orcid{0000-0002-4761-366X}\inst{\ref{aff118},\ref{aff117},\ref{aff119}}
\and K.~Ganga\orcid{0000-0001-8159-8208}\inst{\ref{aff86}}
\and T.~Gasparetto\orcid{0000-0002-7913-4866}\inst{\ref{aff40}}
\and E.~Gaztanaga\orcid{0000-0001-9632-0815}\inst{\ref{aff64},\ref{aff63},\ref{aff133}}
\and F.~Giacomini\orcid{0000-0002-3129-2814}\inst{\ref{aff31}}
\and F.~Gianotti\orcid{0000-0003-4666-119X}\inst{\ref{aff1}}
\and G.~Gozaliasl\orcid{0000-0002-0236-919X}\inst{\ref{aff134},\ref{aff78}}
\and A.~Gruppuso\orcid{0000-0001-9272-5292}\inst{\ref{aff1},\ref{aff31}}
\and M.~Guidi\orcid{0000-0001-9408-1101}\inst{\ref{aff30},\ref{aff1}}
\and C.~M.~Gutierrez\orcid{0000-0001-7854-783X}\inst{\ref{aff49},\ref{aff135}}
\and A.~Hall\orcid{0000-0002-3139-8651}\inst{\ref{aff50}}
\and C.~Hern\'andez-Monteagudo\orcid{0000-0001-5471-9166}\inst{\ref{aff135},\ref{aff49}}
\and H.~Hildebrandt\orcid{0000-0002-9814-3338}\inst{\ref{aff136}}
\and J.~Hjorth\orcid{0000-0002-4571-2306}\inst{\ref{aff92}}
\and J.~J.~E.~Kajava\orcid{0000-0002-3010-8333}\inst{\ref{aff137},\ref{aff138},\ref{aff139}}
\and Y.~Kang\orcid{0009-0000-8588-7250}\inst{\ref{aff59}}
\and V.~Kansal\orcid{0000-0002-4008-6078}\inst{\ref{aff140},\ref{aff141}}
\and D.~Karagiannis\orcid{0000-0002-4927-0816}\inst{\ref{aff108},\ref{aff16}}
\and K.~Kiiveri\inst{\ref{aff102}}
\and J.~Kim\orcid{0000-0003-2776-2761}\inst{\ref{aff112}}
\and C.~C.~Kirkpatrick\inst{\ref{aff102}}
\and S.~Kruk\orcid{0000-0001-8010-8879}\inst{\ref{aff27}}
\and M.~Lembo\orcid{0000-0002-5271-5070}\inst{\ref{aff142}}
\and F.~Lepori\orcid{0009-0000-5061-7138}\inst{\ref{aff143}}
\and G.~F.~Lesci\orcid{0000-0002-4607-2830}\inst{\ref{aff2},\ref{aff1}}
\and J.~Lesgourgues\orcid{0000-0001-7627-353X}\inst{\ref{aff44}}
\and T.~I.~Liaudat\orcid{0000-0002-9104-314X}\inst{\ref{aff144}}
\and X.~Lopez~Lopez\orcid{0009-0008-5194-5908}\inst{\ref{aff1}}
\and F.~Mannucci\orcid{0000-0002-4803-2381}\inst{\ref{aff145}}
\and C.~J.~A.~P.~Martins\orcid{0000-0002-4886-9261}\inst{\ref{aff146},\ref{aff147}}
\and L.~Maurin\orcid{0000-0002-8406-0857}\inst{\ref{aff8}}
\and M.~Miluzio\inst{\ref{aff27},\ref{aff148}}
\and A.~Montoro\orcid{0000-0003-4730-8590}\inst{\ref{aff64},\ref{aff63}}
\and C.~Moretti\orcid{0000-0003-3314-8936}\inst{\ref{aff6},\ref{aff11},\ref{aff29}}
\and G.~Morgante\inst{\ref{aff1}}
\and S.~Nadathur\orcid{0000-0001-9070-3102}\inst{\ref{aff133}}
\and A.~Navarro-Alsina\orcid{0000-0002-3173-2592}\inst{\ref{aff84}}
\and S.~Nesseris\orcid{0000-0002-0567-0324}\inst{\ref{aff97}}
\and D.~Paoletti\orcid{0000-0003-4761-6147}\inst{\ref{aff1},\ref{aff62}}
\and F.~Passalacqua\orcid{0000-0002-8606-4093}\inst{\ref{aff96},\ref{aff60}}
\and K.~Paterson\orcid{0000-0001-8340-3486}\inst{\ref{aff74}}
\and L.~Patrizii\inst{\ref{aff31}}
\and D.~Potter\orcid{0000-0002-0757-5195}\inst{\ref{aff149}}
\and G.~W.~Pratt\inst{\ref{aff61}}
\and S.~Quai\orcid{0000-0002-0449-8163}\inst{\ref{aff2},\ref{aff1}}
\and M.~Radovich\orcid{0000-0002-3585-866X}\inst{\ref{aff67}}
\and G.~Rodighiero\orcid{0000-0002-9415-2296}\inst{\ref{aff96},\ref{aff67}}
\and W.~Roster\orcid{0000-0002-9149-6528}\inst{\ref{aff3}}
\and S.~Sacquegna\orcid{0000-0002-8433-6630}\inst{\ref{aff150}}
\and M.~Sahl\'en\orcid{0000-0003-0973-4804}\inst{\ref{aff151}}
\and D.~B.~Sanders\orcid{0000-0002-1233-9998}\inst{\ref{aff47}}
\and A.~Schneider\orcid{0000-0001-7055-8104}\inst{\ref{aff149}}
\and M.~Schultheis\inst{\ref{aff15}}
\and D.~Sciotti\orcid{0009-0008-4519-2620}\inst{\ref{aff40},\ref{aff41}}
\and E.~Sellentin\inst{\ref{aff152},\ref{aff22}}
\and F.~Shankar\orcid{0000-0001-8973-5051}\inst{\ref{aff153}}
\and L.~C.~Smith\orcid{0000-0002-3259-2771}\inst{\ref{aff154}}
\and K.~Tanidis\orcid{0000-0001-9843-5130}\inst{\ref{aff155}}
\and C.~Tao\orcid{0000-0001-7961-8177}\inst{\ref{aff4}}
\and F.~Tarsitano\orcid{0000-0002-5919-0238}\inst{\ref{aff156},\ref{aff59}}
\and G.~Testera\inst{\ref{aff33}}
\and R.~Teyssier\orcid{0000-0001-7689-0933}\inst{\ref{aff157}}
\and S.~Tosi\orcid{0000-0002-7275-9193}\inst{\ref{aff32},\ref{aff19},\ref{aff33}}
\and A.~Troja\orcid{0000-0003-0239-4595}\inst{\ref{aff96},\ref{aff60}}
\and A.~Venhola\orcid{0000-0001-6071-4564}\inst{\ref{aff158}}
\and D.~Vergani\orcid{0000-0003-0898-2216}\inst{\ref{aff1}}
\and G.~Verza\orcid{0000-0002-1886-8348}\inst{\ref{aff159},\ref{aff160}}
\and S.~Vinciguerra\orcid{0009-0005-4018-3184}\inst{\ref{aff10}}
\and N.~A.~Walton\orcid{0000-0003-3983-8778}\inst{\ref{aff154}}
\and A.~H.~Wright\orcid{0000-0001-7363-7932}\inst{\ref{aff136}}}
										   
\institute{INAF-Osservatorio di Astrofisica e Scienza dello Spazio di Bologna, Via Piero Gobetti 93/3, 40129 Bologna, Italy\label{aff1}
\and
Dipartimento di Fisica e Astronomia "Augusto Righi" - Alma Mater Studiorum Universit\`a di Bologna, via Piero Gobetti 93/2, 40129 Bologna, Italy\label{aff2}
\and
Max Planck Institute for Extraterrestrial Physics, Giessenbachstr. 1, 85748 Garching, Germany\label{aff3}
\and
Aix-Marseille Universit\'e, CNRS/IN2P3, CPPM, Marseille, France\label{aff4}
\and
Universit\'e de Strasbourg, CNRS, Observatoire astronomique de Strasbourg, UMR 7550, 67000 Strasbourg, France\label{aff5}
\and
INAF-Osservatorio Astronomico di Trieste, Via G. B. Tiepolo 11, 34143 Trieste, Italy\label{aff6}
\and
Univ. Lille, CNRS, Centrale Lille, UMR 9189 CRIStAL, 59000 Lille, France\label{aff7}
\and
Universit\'e Paris-Saclay, CNRS, Institut d'astrophysique spatiale, 91405, Orsay, France\label{aff8}
\and
INAF-Istituto di Astrofisica e Planetologia Spaziali, via del Fosso del Cavaliere, 100, 00100 Roma, Italy\label{aff9}
\and
Aix-Marseille Universit\'e, CNRS, CNES, LAM, Marseille, France\label{aff10}
\and
IFPU, Institute for Fundamental Physics of the Universe, via Beirut 2, 34151 Trieste, Italy\label{aff11}
\and
SISSA, International School for Advanced Studies, Via Bonomea 265, 34136 Trieste TS, Italy\label{aff12}
\and
ICSC - Centro Nazionale di Ricerca in High Performance Computing, Big Data e Quantum Computing, Via Magnanelli 2, Bologna, Italy\label{aff13}
\and
Institute of Physics, Laboratory for Galaxy Evolution, Ecole Polytechnique F\'ed\'erale de Lausanne, Observatoire de Sauverny, CH-1290 Versoix, Switzerland\label{aff14}
\and
Universit\'e C\^{o}te d'Azur, Observatoire de la C\^{o}te d'Azur, CNRS, Laboratoire Lagrange, Bd de l'Observatoire, CS 34229, 06304 Nice cedex 4, France\label{aff15}
\and
Department of Physics and Astronomy, University of the Western Cape, Bellville, Cape Town, 7535, South Africa\label{aff16}
\and
Tianjin Normal University, Binshuixidao 393, Tianjin 300387, China\label{aff17}
\and
Dipartimento di Fisica ``G. Occhialini", Universit\`a degli Studi di Milano Bicocca, Piazza della Scienza 3, 20126 Milano, Italy\label{aff18}
\and
INAF-Osservatorio Astronomico di Brera, Via Brera 28, 20122 Milano, Italy\label{aff19}
\and
Universit\'e Clermont Auvergne, CNRS/IN2P3, LPC, F-63000 Clermont-Ferrand, France\label{aff20}
\and
Universit\'e Claude Bernard Lyon 1, CNRS/IN2P3, IP2I Lyon, UMR 5822, Villeurbanne, F-69100, France\label{aff21}
\and
Leiden Observatory, Leiden University, Einsteinweg 55, 2333 CC Leiden, The Netherlands\label{aff22}
\and
Department of Physics and Astronomy, University of British Columbia, Vancouver, BC V6T 1Z1, Canada\label{aff23}
\and
MTA-CSFK Lend\"ulet Large-Scale Structure Research Group, Konkoly-Thege Mikl\'os \'ut 15-17, H-1121 Budapest, Hungary\label{aff24}
\and
Konkoly Observatory, HUN-REN CSFK, MTA Centre of Excellence, Budapest, Konkoly Thege Mikl\'os {\'u}t 15-17. H-1121, Hungary\label{aff25}
\and
Caltech/IPAC, 1200 E. California Blvd., Pasadena, CA 91125, USA\label{aff26}
\and
ESAC/ESA, Camino Bajo del Castillo, s/n., Urb. Villafranca del Castillo, 28692 Villanueva de la Ca\~nada, Madrid, Spain\label{aff27}
\and
School of Mathematics and Physics, University of Surrey, Guildford, Surrey, GU2 7XH, UK\label{aff28}
\and
INFN, Sezione di Trieste, Via Valerio 2, 34127 Trieste TS, Italy\label{aff29}
\and
Dipartimento di Fisica e Astronomia, Universit\`a di Bologna, Via Gobetti 93/2, 40129 Bologna, Italy\label{aff30}
\and
INFN-Sezione di Bologna, Viale Berti Pichat 6/2, 40127 Bologna, Italy\label{aff31}
\and
Dipartimento di Fisica, Universit\`a di Genova, Via Dodecaneso 33, 16146, Genova, Italy\label{aff32}
\and
INFN-Sezione di Genova, Via Dodecaneso 33, 16146, Genova, Italy\label{aff33}
\and
Department of Physics "E. Pancini", University Federico II, Via Cinthia 6, 80126, Napoli, Italy\label{aff34}
\and
INAF-Osservatorio Astronomico di Capodimonte, Via Moiariello 16, 80131 Napoli, Italy\label{aff35}
\and
Dipartimento di Fisica, Universit\`a degli Studi di Torino, Via P. Giuria 1, 10125 Torino, Italy\label{aff36}
\and
INFN-Sezione di Torino, Via P. Giuria 1, 10125 Torino, Italy\label{aff37}
\and
INAF-Osservatorio Astrofisico di Torino, Via Osservatorio 20, 10025 Pino Torinese (TO), Italy\label{aff38}
\and
INAF-IASF Milano, Via Alfonso Corti 12, 20133 Milano, Italy\label{aff39}
\and
INAF-Osservatorio Astronomico di Roma, Via Frascati 33, 00078 Monteporzio Catone, Italy\label{aff40}
\and
INFN-Sezione di Roma, Piazzale Aldo Moro, 2 - c/o Dipartimento di Fisica, Edificio G. Marconi, 00185 Roma, Italy\label{aff41}
\and
Centro de Investigaciones Energ\'eticas, Medioambientales y Tecnol\'ogicas (CIEMAT), Avenida Complutense 40, 28040 Madrid, Spain\label{aff42}
\and
Port d'Informaci\'{o} Cient\'{i}fica, Campus UAB, C. Albareda s/n, 08193 Bellaterra (Barcelona), Spain\label{aff43}
\and
Institute for Theoretical Particle Physics and Cosmology (TTK), RWTH Aachen University, 52056 Aachen, Germany\label{aff44}
\and
Deutsches Zentrum f\"ur Luft- und Raumfahrt e. V. (DLR), Linder H\"ohe, 51147 K\"oln, Germany\label{aff45}
\and
INFN section of Naples, Via Cinthia 6, 80126, Napoli, Italy\label{aff46}
\and
Institute for Astronomy, University of Hawaii, 2680 Woodlawn Drive, Honolulu, HI 96822, USA\label{aff47}
\and
Dipartimento di Fisica e Astronomia "Augusto Righi" - Alma Mater Studiorum Universit\`a di Bologna, Viale Berti Pichat 6/2, 40127 Bologna, Italy\label{aff48}
\and
Instituto de Astrof\'{\i}sica de Canarias, E-38205 La Laguna, Tenerife, Spain\label{aff49}
\and
Institute for Astronomy, University of Edinburgh, Royal Observatory, Blackford Hill, Edinburgh EH9 3HJ, UK\label{aff50}
\and
European Space Agency/ESRIN, Largo Galileo Galilei 1, 00044 Frascati, Roma, Italy\label{aff51}
\and
Institut de Ci\`{e}ncies del Cosmos (ICCUB), Universitat de Barcelona (IEEC-UB), Mart\'{i} i Franqu\`{e}s 1, 08028 Barcelona, Spain\label{aff52}
\and
Instituci\'o Catalana de Recerca i Estudis Avan\c{c}ats (ICREA), Passeig de Llu\'{\i}s Companys 23, 08010 Barcelona, Spain\label{aff53}
\and
Institut de Ciencies de l'Espai (IEEC-CSIC), Campus UAB, Carrer de Can Magrans, s/n Cerdanyola del Vall\'es, 08193 Barcelona, Spain\label{aff54}
\and
UCB Lyon 1, CNRS/IN2P3, IUF, IP2I Lyon, 4 rue Enrico Fermi, 69622 Villeurbanne, France\label{aff55}
\and
Mullard Space Science Laboratory, University College London, Holmbury St Mary, Dorking, Surrey RH5 6NT, UK\label{aff56}
\and
Departamento de F\'isica, Faculdade de Ci\^encias, Universidade de Lisboa, Edif\'icio C8, Campo Grande, PT1749-016 Lisboa, Portugal\label{aff57}
\and
Instituto de Astrof\'isica e Ci\^encias do Espa\c{c}o, Faculdade de Ci\^encias, Universidade de Lisboa, Campo Grande, 1749-016 Lisboa, Portugal\label{aff58}
\and
Department of Astronomy, University of Geneva, ch. d'Ecogia 16, 1290 Versoix, Switzerland\label{aff59}
\and
INFN-Padova, Via Marzolo 8, 35131 Padova, Italy\label{aff60}
\and
Universit\'e Paris-Saclay, Universit\'e Paris Cit\'e, CEA, CNRS, AIM, 91191, Gif-sur-Yvette, France\label{aff61}
\and
INFN-Bologna, Via Irnerio 46, 40126 Bologna, Italy\label{aff62}
\and
Institut d'Estudis Espacials de Catalunya (IEEC),  Edifici RDIT, Campus UPC, 08860 Castelldefels, Barcelona, Spain\label{aff63}
\and
Institute of Space Sciences (ICE, CSIC), Campus UAB, Carrer de Can Magrans, s/n, 08193 Barcelona, Spain\label{aff64}
\and
School of Physics, HH Wills Physics Laboratory, University of Bristol, Tyndall Avenue, Bristol, BS8 1TL, UK\label{aff65}
\and
University Observatory, LMU Faculty of Physics, Scheinerstr.~1, 81679 Munich, Germany\label{aff66}
\and
INAF-Osservatorio Astronomico di Padova, Via dell'Osservatorio 5, 35122 Padova, Italy\label{aff67}
\and
Universit\"ats-Sternwarte M\"unchen, Fakult\"at f\"ur Physik, Ludwig-Maximilians-Universit\"at M\"unchen, Scheinerstr.~1, 81679 M\"unchen, Germany\label{aff68}
\and
Institute of Theoretical Astrophysics, University of Oslo, P.O. Box 1029 Blindern, 0315 Oslo, Norway\label{aff69}
\and
Jet Propulsion Laboratory, California Institute of Technology, 4800 Oak Grove Drive, Pasadena, CA, 91109, USA\label{aff70}
\and
Felix Hormuth Engineering, Goethestr. 17, 69181 Leimen, Germany\label{aff71}
\and
Technical University of Denmark, Elektrovej 327, 2800 Kgs. Lyngby, Denmark\label{aff72}
\and
Cosmic Dawn Center (DAWN), Denmark\label{aff73}
\and
Max-Planck-Institut f\"ur Astronomie, K\"onigstuhl 17, 69117 Heidelberg, Germany\label{aff74}
\and
NASA Goddard Space Flight Center, Greenbelt, MD 20771, USA\label{aff75}
\and
Department of Physics and Astronomy, University College London, Gower Street, London WC1E 6BT, UK\label{aff76}
\and
Universit\'e de Gen\`eve, D\'epartement de Physique Th\'eorique and Centre for Astroparticle Physics, 24 quai Ernest-Ansermet, CH-1211 Gen\`eve 4, Switzerland\label{aff77}
\and
Department of Physics, P.O. Box 64, University of Helsinki, 00014 Helsinki, Finland\label{aff78}
\and
Helsinki Institute of Physics, Gustaf H{\"a}llstr{\"o}min katu 2, University of Helsinki, 00014 Helsinki, Finland\label{aff79}
\and
Laboratoire d'etude de l'Univers et des phenomenes eXtremes, Observatoire de Paris, Universit\'e PSL, Sorbonne Universit\'e, CNRS, 92190 Meudon, France\label{aff80}
\and
SKAO, Jodrell Bank, Lower Withington, Macclesfield SK11 9FT, UK\label{aff81}
\and
Centre de Calcul de l'IN2P3/CNRS, 21 avenue Pierre de Coubertin 69627 Villeurbanne Cedex, France\label{aff82}
\and
University of Applied Sciences and Arts of Northwestern Switzerland, School of Computer Science, 5210 Windisch, Switzerland\label{aff83}
\and
Universit\"at Bonn, Argelander-Institut f\"ur Astronomie, Auf dem H\"ugel 71, 53121 Bonn, Germany\label{aff84}
\and
Department of Physics, Institute for Computational Cosmology, Durham University, South Road, Durham, DH1 3LE, UK\label{aff85}
\and
Universit\'e Paris Cit\'e, CNRS, Astroparticule et Cosmologie, 75013 Paris, France\label{aff86}
\and
CNRS-UCB International Research Laboratory, Centre Pierre Bin\'etruy, IRL2007, CPB-IN2P3, Berkeley, USA\label{aff87}
\and
Institute of Physics, Laboratory of Astrophysics, Ecole Polytechnique F\'ed\'erale de Lausanne (EPFL), Observatoire de Sauverny, 1290 Versoix, Switzerland\label{aff88}
\and
Telespazio UK S.L. for European Space Agency (ESA), Camino bajo del Castillo, s/n, Urbanizacion Villafranca del Castillo, Villanueva de la Ca\~nada, 28692 Madrid, Spain\label{aff89}
\and
Institut de F\'{i}sica d'Altes Energies (IFAE), The Barcelona Institute of Science and Technology, Campus UAB, 08193 Bellaterra (Barcelona), Spain\label{aff90}
\and
European Space Agency/ESTEC, Keplerlaan 1, 2201 AZ Noordwijk, The Netherlands\label{aff91}
\and
DARK, Niels Bohr Institute, University of Copenhagen, Jagtvej 155, 2200 Copenhagen, Denmark\label{aff92}
\and
Space Science Data Center, Italian Space Agency, via del Politecnico snc, 00133 Roma, Italy\label{aff93}
\and
Centre National d'Etudes Spatiales -- Centre spatial de Toulouse, 18 avenue Edouard Belin, 31401 Toulouse Cedex 9, France\label{aff94}
\and
Institute of Space Science, Str. Atomistilor, nr. 409 M\u{a}gurele, Ilfov, 077125, Romania\label{aff95}
\and
Dipartimento di Fisica e Astronomia "G. Galilei", Universit\`a di Padova, Via Marzolo 8, 35131 Padova, Italy\label{aff96}
\and
Instituto de F\'isica Te\'orica UAM-CSIC, Campus de Cantoblanco, 28049 Madrid, Spain\label{aff97}
\and
Institut de Recherche en Astrophysique et Plan\'etologie (IRAP), Universit\'e de Toulouse, CNRS, UPS, CNES, 14 Av. Edouard Belin, 31400 Toulouse, France\label{aff98}
\and
Universit\'e St Joseph; Faculty of Sciences, Beirut, Lebanon\label{aff99}
\and
Departamento de F\'isica, FCFM, Universidad de Chile, Blanco Encalada 2008, Santiago, Chile\label{aff100}
\and
Universit\"at Innsbruck, Institut f\"ur Astro- und Teilchenphysik, Technikerstr. 25/8, 6020 Innsbruck, Austria\label{aff101}
\and
Department of Physics and Helsinki Institute of Physics, Gustaf H\"allstr\"omin katu 2, University of Helsinki, 00014 Helsinki, Finland\label{aff102}
\and
Instituto de Astrof\'isica e Ci\^encias do Espa\c{c}o, Faculdade de Ci\^encias, Universidade de Lisboa, Tapada da Ajuda, 1349-018 Lisboa, Portugal\label{aff103}
\and
Cosmic Dawn Center (DAWN)\label{aff104}
\and
Niels Bohr Institute, University of Copenhagen, Jagtvej 128, 2200 Copenhagen, Denmark\label{aff105}
\and
Universidad Polit\'ecnica de Cartagena, Departamento de Electr\'onica y Tecnolog\'ia de Computadoras,  Plaza del Hospital 1, 30202 Cartagena, Spain\label{aff106}
\and
Kapteyn Astronomical Institute, University of Groningen, PO Box 800, 9700 AV Groningen, The Netherlands\label{aff107}
\and
Dipartimento di Fisica e Scienze della Terra, Universit\`a degli Studi di Ferrara, Via Giuseppe Saragat 1, 44122 Ferrara, Italy\label{aff108}
\and
Istituto Nazionale di Fisica Nucleare, Sezione di Ferrara, Via Giuseppe Saragat 1, 44122 Ferrara, Italy\label{aff109}
\and
INAF, Istituto di Radioastronomia, Via Piero Gobetti 101, 40129 Bologna, Italy\label{aff110}
\and
Astronomical Observatory of the Autonomous Region of the Aosta Valley (OAVdA), Loc. Lignan 39, I-11020, Nus (Aosta Valley), Italy\label{aff111}
\and
Department of Physics, Oxford University, Keble Road, Oxford OX1 3RH, UK\label{aff112}
\and
Univ. Grenoble Alpes, CNRS, Grenoble INP, LPSC-IN2P3, 53, Avenue des Martyrs, 38000, Grenoble, France\label{aff113}
\and
Dipartimento di Fisica, Sapienza Universit\`a di Roma, Piazzale Aldo Moro 2, 00185 Roma, Italy\label{aff114}
\and
Aurora Technology for European Space Agency (ESA), Camino bajo del Castillo, s/n, Urbanizacion Villafranca del Castillo, Villanueva de la Ca\~nada, 28692 Madrid, Spain\label{aff115}
\and
Dipartimento di Fisica - Sezione di Astronomia, Universit\`a di Trieste, Via Tiepolo 11, 34131 Trieste, Italy\label{aff116}
\and
Department of Mathematics and Physics E. De Giorgi, University of Salento, Via per Arnesano, CP-I93, 73100, Lecce, Italy\label{aff117}
\and
INFN, Sezione di Lecce, Via per Arnesano, CP-193, 73100, Lecce, Italy\label{aff118}
\and
INAF-Sezione di Lecce, c/o Dipartimento Matematica e Fisica, Via per Arnesano, 73100, Lecce, Italy\label{aff119}
\and
Institut d'Astrophysique de Paris, 98bis Boulevard Arago, 75014, Paris, France\label{aff120}
\and
ICL, Junia, Universit\'e Catholique de Lille, LITL, 59000 Lille, France\label{aff121}
\and
CERCA/ISO, Department of Physics, Case Western Reserve University, 10900 Euclid Avenue, Cleveland, OH 44106, USA\label{aff122}
\and
Laboratoire Univers et Th\'eorie, Observatoire de Paris, Universit\'e PSL, Universit\'e Paris Cit\'e, CNRS, 92190 Meudon, France\label{aff123}
\and
Dipartimento di Fisica "Aldo Pontremoli", Universit\`a degli Studi di Milano, Via Celoria 16, 20133 Milano, Italy\label{aff124}
\and
INFN-Sezione di Milano, Via Celoria 16, 20133 Milano, Italy\label{aff125}
\and
Departamento de F{\'\i}sica Fundamental. Universidad de Salamanca. Plaza de la Merced s/n. 37008 Salamanca, Spain\label{aff126}
\and
Center for Data-Driven Discovery, Kavli IPMU (WPI), UTIAS, The University of Tokyo, Kashiwa, Chiba 277-8583, Japan\label{aff127}
\and
Jodrell Bank Centre for Astrophysics, Department of Physics and Astronomy, University of Manchester, Oxford Road, Manchester M13 9PL, UK\label{aff128}
\and
California Institute of Technology, 1200 E California Blvd, Pasadena, CA 91125, USA\label{aff129}
\and
Department of Physics \& Astronomy, University of California Irvine, Irvine CA 92697, USA\label{aff130}
\and
Departamento F\'isica Aplicada, Universidad Polit\'ecnica de Cartagena, Campus Muralla del Mar, 30202 Cartagena, Murcia, Spain\label{aff131}
\and
Instituto de F\'isica de Cantabria, Edificio Juan Jord\'a, Avenida de los Castros, 39005 Santander, Spain\label{aff132}
\and
Institute of Cosmology and Gravitation, University of Portsmouth, Portsmouth PO1 3FX, UK\label{aff133}
\and
Department of Computer Science, Aalto University, PO Box 15400, Espoo, FI-00 076, Finland\label{aff134}
\and
Universidad de La Laguna, Dpto. Astrof\'\i sica, E-38206 La Laguna, Tenerife, Spain\label{aff135}
\and
Ruhr University Bochum, Faculty of Physics and Astronomy, Astronomical Institute (AIRUB), German Centre for Cosmological Lensing (GCCL), 44780 Bochum, Germany\label{aff136}
\and
Department of Physics and Astronomy, Vesilinnantie 5, University of Turku, 20014 Turku, Finland\label{aff137}
\and
Finnish Centre for Astronomy with ESO (FINCA), Quantum, Vesilinnantie 5, University of Turku, 20014 Turku, Finland\label{aff138}
\and
Serco for European Space Agency (ESA), Camino bajo del Castillo, s/n, Urbanizacion Villafranca del Castillo, Villanueva de la Ca\~nada, 28692 Madrid, Spain\label{aff139}
\and
ARC Centre of Excellence for Dark Matter Particle Physics, Melbourne, Australia\label{aff140}
\and
Centre for Astrophysics \& Supercomputing, Swinburne University of Technology,  Hawthorn, Victoria 3122, Australia\label{aff141}
\and
Institut d'Astrophysique de Paris, UMR 7095, CNRS, and Sorbonne Universit\'e, 98 bis boulevard Arago, 75014 Paris, France\label{aff142}
\and
Departement of Theoretical Physics, University of Geneva, Switzerland\label{aff143}
\and
IRFU, CEA, Universit\'e Paris-Saclay 91191 Gif-sur-Yvette Cedex, France\label{aff144}
\and
INAF-Osservatorio Astrofisico di Arcetri, Largo E. Fermi 5, 50125, Firenze, Italy\label{aff145}
\and
Centro de Astrof\'{\i}sica da Universidade do Porto, Rua das Estrelas, 4150-762 Porto, Portugal\label{aff146}
\and
Instituto de Astrof\'isica e Ci\^encias do Espa\c{c}o, Universidade do Porto, CAUP, Rua das Estrelas, PT4150-762 Porto, Portugal\label{aff147}
\and
HE Space for European Space Agency (ESA), Camino bajo del Castillo, s/n, Urbanizacion Villafranca del Castillo, Villanueva de la Ca\~nada, 28692 Madrid, Spain\label{aff148}
\and
Department of Astrophysics, University of Zurich, Winterthurerstrasse 190, 8057 Zurich, Switzerland\label{aff149}
\and
INAF - Osservatorio Astronomico d'Abruzzo, Via Maggini, 64100, Teramo, Italy\label{aff150}
\and
Theoretical astrophysics, Department of Physics and Astronomy, Uppsala University, Box 516, 751 37 Uppsala, Sweden\label{aff151}
\and
Mathematical Institute, University of Leiden, Einsteinweg 55, 2333 CA Leiden, The Netherlands\label{aff152}
\and
School of Physics \& Astronomy, University of Southampton, Highfield Campus, Southampton SO17 1BJ, UK\label{aff153}
\and
Institute of Astronomy, University of Cambridge, Madingley Road, Cambridge CB3 0HA, UK\label{aff154}
\and
Center for Astrophysics and Cosmology, University of Nova Gorica, Nova Gorica, Slovenia\label{aff155}
\and
Institute for Particle Physics and Astrophysics, Dept. of Physics, ETH Zurich, Wolfgang-Pauli-Strasse 27, 8093 Zurich, Switzerland\label{aff156}
\and
Department of Astrophysical Sciences, Peyton Hall, Princeton University, Princeton, NJ 08544, USA\label{aff157}
\and
Space physics and astronomy research unit, University of Oulu, Pentti Kaiteran katu 1, FI-90014 Oulu, Finland\label{aff158}
\and
International Centre for Theoretical Physics (ICTP), Strada Costiera 11, 34151 Trieste, Italy\label{aff159}
\and
Center for Computational Astrophysics, Flatiron Institute, 162 5th Avenue, 10010, New York, NY, USA\label{aff160}}

\abstract{
The evolution of galaxies is profoundly influenced by the environment in which they reside. Cosmic voids, the most underdense regions of the Universe, serve as pristine laboratories for studying galaxy evolution 
in the relative absence or weaker influence of the complex physical processes that dominate denser environments. 
In this study, we investigate properties and merger histories of galaxies as a function of environment using the GAlaxy Evolution and Assembly (GAEA) mock-observation lightcone replicating the Euclid Deep Survey as foreseen at the epoch of the first data release. The \ha-selected galaxy sample spans the redshift range $0.4<z<1.8$, corresponding to the interval over which \ha\ is accessible to \Euclid\ slitless spectroscopy. We classify galaxies based on their void-centric distance and local density contrast, and we compare their stellar mass, specific star formation rate, bulge-to-total stellar mass ratio, and halo mass across different environments, while controlling for the dependence on stellar mass of the other intrinsic properties. We further analyse the merger histories of these galaxies to study their assembly evolution.
We find that galaxies located closer to void centres ($d_{\rm cc}\lesssim0.7 R_{\rm v}$) are, on average, less massive, more actively star-forming, and more disc-dominated than galaxies in denser regions, with their distributions statistically distinct at more than 4\textsigma\, confidence according to Kolmogorov--Smirnov tests. Merger histories indicate that void galaxies do not experience fewer mergers but that these occur later, leading to stellar mass assembly delayed by approximately 0.5--1 Gyr relative to galaxies in high-density regions. This supports a scenario in which the environment regulates the timing and nature of mergers rather than their overall frequency, producing a slower evolutionary path in low-density regions. We conclude by discussing the extent to which these trends are shaped by environmental parametrisation methods and observational selection effects. 
Our analysis provides a framework for interpreting forthcoming \Euclid data, and demonstrates that \Euclid\ will enable the identification of cosmic voids and the characterisation of their galaxy populations with unprecedented statistical power and redshift coverage, providing a quantitative forecast of \Euclid's capability to probe environmental effects on galaxy evolution.}


    \titlerunning{Probing galaxy evolution within cosmic voids in \Euclid-like simulations}
    \authorrunning{Euclid Collaboration: G. Papini et al.}
   
    \maketitle

\section{\label{sc:Intro}Introduction}

The large-scale structure of the Universe is shaped by the collapse of matter into a complex tapestry known as the `cosmic web', composed of nodes, filaments, walls, and voids that form in-between the other components. In the local Universe, voids occupy 70--80\% of the volume while containing only a small fraction of its mass \citep{sheth_hierarchy_2004, van_de_weygaert_cosmic_2011, pan_cosmic_2012, cautun_evolution_2014, libeskind_tracing_2018}. Due to their low-density environments, voids offer a unique laboratory for studying galaxy formation and evolution under minimal external influence. 

Observations and simulations agree that void galaxies occupy the low-mass and high star-forming end of the galaxy population, with morphologies biased towards disc-dominated systems. Numerous works report that galaxies in underdense regions are typically less massive and more actively star-forming than their counterparts in denser environments, from early analyses of void galaxies in the nearby Universe \citep{Grogin_Geller_Imaging_1999, Grogin_Geller_Imaging_2000, rojas_photometric_2004, Hoyle_luminosity_2005, rojas_spectroscopic_2005}, to more recent studies on wide and deep surveys that extended this picture to larger and higher-redshift samples \citep{ricciardelli_star_2014, liu_spectral_2015, davidzon_vimos_2016, curtis_properties_2024, rodriguez-medrano_evolutionary_2024}. This trend has been confirmed both locally \citep{liu_spectral_2015, ceccarelli_galaxy_2025, perez_galaxy_2025} and at intermediate redshifts \citep{peng_mass_2010, bolzonella_tracking_2010, davidzon_vimos_2016}, although its significance varies with the adopted definition of voids and the chosen method for reconstructing the environment that is generally optimised for the data set in analysis. Morphological studies further support this picture: galaxies in voids display lower bulge-to-total mass fractions and are more likely to retain extended discs, consistent with a reduced role of transformation mechanisms \citep{rojas_photometric_2004, ricciardelli_morphological_2017, rosas-guevara_revealing_2022, perez_galaxy_2025}. 
At the same time, halo mass distributions from cosmological simulations reveal that void galaxies preferentially occupy lower-mass dark matter haloes, and are more often centrals than satellites \citep{habouzit_properties_2020, rodriguez-medrano_evolutionary_2024}. Studies of the halo mass function also show that structure formation proceeds more slowly in voids than in the rest of the Universe \citep{verza_halo_2022}. Together, these trends point towards a scenario where the interplay between stellar mass growth, star formation activity, morphology, and central-satellite status is intimately linked to the large-scale density field.

Nonetheless, the precise role of the void environment in regulating galaxy properties remains unsettled. Some studies suggest that galaxies in voids may represent a special population, evolving primarily under the influence of internal processes and largely shielded from strong external interactions such as frequent encounters with other galaxies or the tidal influence of massive haloes \citep{kreckel_only_2011, beygu_void_2016}. However, comparisons at fixed stellar mass often show only subtle differences, suggesting that many of the apparent environmental signatures may instead be driven by the underlying dark matter distribution or by observational biases \citep{ricciardelli_star_2014, curtis_properties_2024}. Furthermore, the choice of environmental parametrisation strongly affects the strength of the inferred trends: void-centric distance is a widely used metric, but it implicitly assumes spherical symmetry, while local density estimators reveal substantial scatter that reflects the anisotropy and substructure of void interiors \citep{nadathur_beyond_2019, florez_void_2021}. \cite{Zaidouni_impact_2025} have also shown that different void definitions can lead to significantly different inferences on the environmental dependence of galaxy properties.

Additionally, recent studies based on cosmological simulations have questioned the classical picture of voids being so sparse that galaxies within them should undergo significantly fewer interactions, finding that void galaxies do not necessarily merge less often, but that their mergers tend to occur later in cosmic time and are more frequently highly unequal-mass interactions \citep{rosas-guevara_revealing_2022, rodriguez-medrano_evolutionary_2024}, with observational analyses at low redshift revealing a similar delayed evolutionary channel \citep{dominguez-gomez_galaxies_2023}.
The weaker tidal fields and lower relative velocities in voids allow satellites to survive longer before eventual coalescence, imprinting characteristic signatures on morphology and star formation. This has profound implications: if mergers in voids are systematically delayed, then void galaxies provide an observational window into evolutionary phases that, in denser environments, occurred earlier and are now obscured by subsequent transformations.

Moreover, recent work on the environmental histories of galaxies across the cosmic web reconstruct their trajectories back in time \citep{sarpa_tracing_2022}. They find that galaxies in voids have resided in these environments for most of cosmic time, whereas those in walls show a mix of long-term residency and gradual migration from voids. Filament and cluster galaxies follow more complex paths, with many originating in voids or walls before moving through filaments, and clusters accumulating most of their members only at $z<1$. Overall, these trends suggest that the present environment of a galaxy reflects both its local conditions and the cumulative impact of past migrations through the cosmic web.
In this context, constraining the merger histories of void galaxies becomes a key test of hierarchical growth. It links directly to fundamental questions about whether galaxies in underdense regions evolve along genuinely different pathways, or whether they simply follow the same evolutionary processes on shifted timescales. This is also important for cosmological studies using voids, as they provide a detailed characterisation of the galaxy populations that inhabit underdense regions. Understanding how galaxy properties and environmental histories influence void identification and measurements helps to quantify potential biases, improving the reliability of void-based cosmological constraints.

Effectively disentangling the influence of large-scale structure on galaxy evolution requires large spectroscopic surveys that are deep enough to detect faint galaxies even at high redshift, and allow substantial, contiguous regions of the Universe to be mapped.
The past decades have witnessed major progress thanks to wide-field spectroscopic surveys (e.g.: SDSS, \citealt{york_sloan_2000}; DEEP2, \citealt{davis_deep2_2001}; VVDS, \citealt{le_fevre_vimos_2005}; zCOSMOS, \citealt{lilly_zcosmos_2007}; BOSS, \citealt{dawson_baryon_2013}; VIPERS, \citealt{guzzo_vimos_2014, scodeggio_vimos_2018}; VUDS, \citealt{le_fevre_vimos_2015}; VANDELS, \citealt{mclure_vandels_2018, pentericci_vandels_2018}). While these surveys remain limited by trade-offs between depth, area, and redshift coverage, important works have been conducted during recent years on voids \citep[e.g.][]{hoyle_voids_2004, pan_cosmic_2012, sutter_public_2012, sutter_dark_2014, nadathur_beyond_2019, hamaus_precision_2020, contarini_cosmological_2023} and void galaxies \citep[e.g.][]{rojas_photometric_2004, micheletti_vimos_2014, tavasoli_galaxy_population_2015, beygu_void_2016, hawken_vimos_2017}, offering a proof of concept but also underscoring the need for larger, deeper samples. 

The \Euclid mission will take this step forward, combining depth, sky coverage, and redshift reach to map voids with unprecedented statistical power \citep{EuclidSkyOverview}. With the wide wavelength range of grism observations within its Deep Survey, \Euclid will span a redshift range that is not accessible by large spectrographs from the ground, thereby linking and surpassing our current knowledge from previous surveys at low and intermediate redshifts. This unique combination of depth, wavelength coverage, and wide area will provide the statistical power and sensitivity needed to probe galaxy formation, evolution, and the large-scale structure of the Universe across cosmic time.

Equally important as such state-of-the-art observations is their cross-validation with controlled predictions from cosmological simulations.
In this work, we study a \Euclid-like lightcone constructed from the semi-analytic model (SAM) GAEA (GAlaxy Evolution and Assembly; see \citealt{de_lucia_tracing_2024} and references therein) to evaluate the capability of the Euclid Deep Survey to recover the true distributions of galaxy properties across void environments. 

This study aims to (i) investigate how internal void structure and local environment impact galaxy properties, such as stellar masses, star formation activity, morphology, and dark matter haloes, (ii) evaluate whether the relative isolation of void galaxies translates into systematically different evolutionary pathways while assessing whether the assembly channels predicted by cosmological simulations can be captured in \Euclid-like data, (iii) assess the performance of an established void-finding algorithm under \Euclid-like deep observations. 
The results presented here provide theoretical groundwork for interpreting future observations from \Euclid, and contribute to the broader effort of understanding how cosmic environment influences galaxy evolution.
Building on previously identified qualitative trends, this work demonstrates for the first time that environmental studies of galaxy evolution in cosmic voids can be robustly performed with Euclid deep data.

This paper is structured as follows. In Sect.~\ref{sc:Simulations} we introduce the simulated data set and the `Euclidisation' procedure used to construct the lightcone. In Sect.~\ref{sc:Voids} we describe the void finder and our parametrisation of the underdense environment. Section~\ref{sc:galprops} examines the distribution of galaxy properties across void environments, while Sect.~\ref{sc:mergertrees} analyses the merger histories of void galaxies. We discuss possible biases in our methodology for environment parametrisation and observational selection effects in Sect.~\ref{sc:Discussion}. We summarise our findings and consider their implications for forthcoming \Euclid observations in Sect.~\ref{sc:Conclusions}.

\section{\label{sc:Simulations}Simulating \Euclid data}

Designed as a European Space Agency (ESA) mission to investigate the nature of dark energy and dark matter by accurately mapping the geometry and evolution of the Universe, \Euclid will deliver optical and near-infrared imaging, as well as slitless spectroscopy targeting \ha\ emission, across more than one-third of the extragalactic sky \citep{EuclidSkyOverview}. 
The telescope features two instruments: the visible instrument (VIS; \citealt{EuclidSkyVIS}), optimized for high-resolution wide-field imaging, and the Near Infrared Spectrometer and Photometer (NISP; \citealt{EuclidSkyNISP}), designed for photometry in three near-infrared (NIR) bands and for spectroscopic redshift measurements via emission line detection. Over its six-year nominal mission, \Euclid will carry out two complementary surveys probing billions of galaxies: the Wide Survey and the Deep Survey.

The Euclid Wide Survey (EWS) is the mission's primary observing program, covering 14~000\,deg$^2$ of the extragalactic sky in the optical and near-infrared and targeting \ha\ emission lines with a nominal line flux limit of $f_{\ha} \gtrsim  2 \times 10^{-16} \,\mathrm{erg\,s^{-1}\,cm^{-2}}$ (signal-to-noise ratio, ${\rm S/N}\! \gtrsim\! 3.5$). 
Complementing the EWS is the Euclid Deep Survey (EDS), designed to reach depths approximately two magnitudes fainter than the Wide Survey, with a nominal flux limit of $f_{\ha} \gtrsim 5 \times 10^{-17} \,\mathrm{erg\,s^{-1}\,cm^{-2}}$ (${\rm S/N}\! \gtrsim\! 3.5$). The EDS covers three independent fields, the Euclid Deep Fields (EDFs): North ($\approx\! 20$\,deg$^2$), South ($\approx\! 23$\,deg$^2$), and Fornax ($\approx\! 10$\,deg$^2$). In addition to their greater depth, the EDFs benefit from an extended wavelength range enabled by NISP's blue grism, broadening the accessible redshift range of the \ha\ emission line beyond that of the red grism alone (from $z\in[0.9,1.8]$ to $z\in[0.4,1.8]$)\footnote{At the time of our analysis, the upper redshift limit of $z=1.8$ was the available reference; subsequent updates extend this to $z=1.88$ \citep{EuclidSkyOverview}. Keeping the original boundary is a conservative choice and does not affect the conclusions.\label{note:zrange}}. For more details on the \Euclid\ mission see \cite{EuclidSkyOverview}.

While a preview of the typical, single-visit depth of the EWS can be found in the observations of the EDFs in the Euclid Quick Data Release (Q1; \citealp{Q1-TP001}), the forthcoming first major data release (DR1) will cover approximately 1900\,deg$^2$ of the EWS, along with multiple passes of the three EDFs with both red and blue grisms. In this way the EDFs will acquire an intermediate depth with respect to their nominal one.
\Euclid's design enables an unprecedented synergy between wide-area and deep-field observations with uniform depth and high resolution across the visible and near-infrared sky. In particular, the EDS will play a central role not only in calibration and systematics control but also as a foundational data set for legacy science. 

\subsection{\label{sc:GAEA}The GAEA semi-analytic model}

In this paper, we analyse a simulated data set prepared with the latest version of GAEA \citep{Hirschmann_2016, Xie_2017, fontanot_rise_2020, de_lucia_tracing_2024}. GAEA builds on the model first published in \cite{DeLucia_Blaizot_2007}. The latest rendition of the SAM presented by \cite{de_lucia_tracing_2024} combines the partitioning of cold gas into its molecular and neutral phases (\citealt{Xie_2017}), with an improved modelling of active galactic nuclei (AGN) activity in terms of cold gas accretion onto supermassive black holes (SMBH) and AGN-driven feedback (\citealt{fontanot_rise_2020}). 
The model is applied to substructure-based merger trees extracted from the Millennium cosmological simulation (\citealt{springel_simulations_2005}), a cube of $500\,\hMpc$ on each side that assumes Wilkinson Microwave Anisotropy Probe (WMAP1) concordance cosmology within a cold dark matter (\textLambda CDM) framework ($\OmLa\!=\!0.75$, $\Omm\!=\!0.25$, $\Omb\!=\!0.045$, $n_{\rm s}\!=\!1$, $\sigma_8\!=\!0.9$, and $H_0\!=\!73$\,\kmsMpc). 
From the output of the SAM, a lightcone was built (algorithm described in \citealt{Zoldan_2017}) with \HE-band AB limiting magnitude of 25 \citep{EP-Scharre} and an aperture area of $\approx21.8\,{\rm deg}^2$, consistent with the expected coverage of the EDFs North and South. 

Several galaxy properties are available from the model, among which the relevant for this work are stellar mass, star formation rate, mass and radius of the stellar bulge, virial mass $M_{200}^{\rm crit}$ of the dark matter halo of the system containing the galaxy, and galaxy merger trees (i.e. the full set of progenitors and descendants for each galaxy in the lightcone).  
We note that our predictions for the environmental dependence of these quantities are conditional on the specific feedback and environmental prescriptions implemented in GAEA. In particular, the model includes non instantaneous stripping of the cold and hot gaseous components associated with galaxies infalling onto larger systems \citep[see section~2 of][]{de_lucia_tracing_2024}, which contributes to reproducing observed trends in star-forming and passive populations as a function of environment (e.g. the star-forming and passive stellar mass functions in clusters shown in figure~7 of \citealt{de_lucia_tracing_2024}). The version of the model used in this analysis reproduces a wide range of galaxy observables over the redshift range relevant for \Euclid, and should therefore provide a reasonably accurate description of baryon cycling among the different components and of the resulting galaxy populations for our purposes.
Simulations do not generally provide ready-made catalogues of voids with well-defined properties such as location, size, or enclosed mass. This is partly due to the relatively recent use of voids in cosmological and evolutionary studies, and unlike haloes or clusters, void catalogues have only recently begun to be included in simulation releases.
As a result, a variety of approaches have been developed in recent years to reconstruct the large-scale environment and identify voids, many relying on different geometrical assumptions and sample selections. We return to this topic in Sect.~\ref{sc:Voids}, and discuss alternative void-finding methods in more detail in Appendix~\ref{apdx:VoidFinders}.

\subsection{\label{sc:Euclid-like}`Euclidisation' of the lightcone}

We have tailored the GAEA lightcone to reproduce the foreseen depth of the EDS at the epoch of DR1. 
To this end, we selected galaxies within the redshift range\footnotemark[1] covering \ha\ in the EDS, $0.4\!<\!z\!<\!1.8$, and we applied a series of modifications to the intrinsic \ha\ fluxes in the model in the following order.
First, we summed the three flux contributions available in GAEA: emission from star formation (SFR), AGN, and post-asymptotic giant branch stars (AGB), as these components cannot be disentangled in \Euclid spectra. The resulting intrinsic \ha\ flux distribution is consistent with the predictions of the empirical models 1 and 2 of \citet{pozzetti_modelling_2016}, which are built to reproduce the observed \ha\ luminosity functions, with differences within a factor of $\approx\!1.5$. These small discrepancies decrease with increasing redshift and become negligible around $z\approx1.7$.
Second, we summed the \ha\ line flux with the flux of the \Nii line doublet at $\lambda=6548,6584$\,\AA\,\citep{Acker_1989}, as the NISP spectral resolution does not allow the lines to be resolved and de-blended.
Third, we applied a uniform screen correction to account for dust extinction \citep{calzetti_dust_2000}, adopting a redshift-dependent nebular-to-continuum attenuation \citep{pannella_goods_herschel_2015, saito_synthetic_2020}, and finally we added an observational uncertainty to the simulated \ha\ flux consistent with the $3.5\sigma$ sensitivity limit expected for DR1 in the EDFs.
In the observational data set, the broad-band photometry will be used in the spectral energy distribution (SED) fitting procedure to derive galaxy properties such as stellar mass and star formation rate \citep{EP-Enia}, whereas in this simulated data set we rely on the intrinsic values directly provided by the model. It is important to note that properties derived from SED fitting can differ from the intrinsic ones due to observational uncertainties on photometry and redshift scatter. We quantify the impact of these uncertainties in Appendix~\ref{apdx:Observational_Uncertanties}.

For the depth and completeness we adopted a conservative approach in mimicking the expected depth of the EDFs at the epoch of DR1. We assumed an intermediate depth between the nominal depth of the EWS and the nominal depth of the EDS, $f_{\ha,\;\rm lim}^\mathrm{DR1} = 10^{-16} \, \mathrm{erg\;s^{-1}\;cm^{-2}}$, corresponding to approximately 10 telescope visits, as expected for DR1. The completeness is conservatively assumed to be 60\% at the maximum depth of DR1, but rising up to 80\% for brighter sources at the nominal flux limit of the EWS ($f_{\ha,\; \rm lim}^\mathrm{EWS} =2.0 \times 10^{-16} \, \mathrm{erg\;s^{-1}\;cm^{-2}}$). 
In the \Euclid slitless spectroscopy, data are expected to be contaminated by interlopers of two kinds:
galaxies at redshifts where \ha\ lies outside the wavelength coverage, and an emission line from a different transition is mistaken for \ha\ by the pipeline (line interlopers) and galaxies where a random spike of noise gets mistaken for \ha\ (noise interlopers). We mimicked a contamination of about 10\% of interloper galaxies, as an educated guess from forecasts available at the time of this analysis.\footnote{At the time this work was initiated, an interloper fraction of 10\% was adopted as a conservative forecast based on pre-launch estimates. Recent tests on the internal Euclid DR1 spectroscopic data (priv. comm.) now indicate a typical purity of $\sim\!95\%$ in the EDFN, implying an interloper fraction $\lesssim\!5\%$, with most residual interlopers identifiable because, when analysed at their (incorrect) spectroscopic redshift, their inferred physical properties are strongly inconsistent with those expected at that redshift.} We modelled interlopers as galaxies in the lightcone that are outside the \ha\ redshift range of DR1 to which a new random redshift was assigned within the boundaries set by \ha\ visibility. 
Interlopers were included in the tracer sample for the environment parametrisation to reflect the realistic occurrence in which a fraction of contaminants enters the construction of the large-scale density field. However, they were excluded from the analysis of galaxy physical properties, since assigning an incorrect redshift propagates into systematically biased estimates of quantities such as stellar mass and star formation rate. Their identification is non-trivial and, in real data, will rely on a combination of diagnostics, including spectroscopic quality flags \citep{Q1-TP007}, the inspection of spectroscopic redshift probability distribution functions to reject poorly constrained or multi-peaked solutions, and additional consistency checks on SED-derived quantities at $z_{\rm spec}$ (e.g. the presence of clear outliers in the stellar mass distribution). We therefore exclude interlopers here to mimic a realistic but conservative cleaning strategy that will be further tested and refined on \Euclid\ DR1 data.

An important consequence of our \ha\ flux-limited selection is that it causes a redshift dependence of the luminosity limit. At low redshift, the flux limit allows us to detect galaxies down to relatively low \ha\ luminosities ($L\!\approx\!10^{7}\,L_\odot$) and stellar masses ($\Mstellar\!\approx\!10^{8.5}\,M_\odot$), sampling a large fraction of the star-forming population and yielding a high density of tracers. However, at higher redshifts only galaxies with intrinsically brighter \ha\ emission, typically associated with larger stellar masses and higher star formation rates, remain above the flux threshold. This introduces a systematic evolution in the number density of tracers: the mean separation between galaxies increases with redshift, and the observed population becomes progressively biased towards massive, actively star-forming systems. These effects directly impact the measurement of environmental parameters, since quantities such as the local density contrast or 5\textsuperscript{th} nearest-neighbour distance are sensitive to the sampling density of the catalogue. In other words, part of the redshift evolution that we measure in environmental indicators is driven not by an intrinsic evolution of the cosmic web, but by the changing depth imposed by the flux limit. Therefore, in our analysis we normalise environmental parameters to their redshift-dependent mean values computed over the sample distribution, in order to disentangle genuine environmental trends from effects of the selection function (see Sect.~\ref{sc:envpars}).

Lastly, our focus on \ha-emitting galaxies introduces an inherent bias: we trace only the star-forming side of the cosmic web. Passive galaxies, which lack significant \ha\ emission, are not included in our analysis. For the purposes of this study, i.e. testing the ability of \Euclid to recover environmental trends at the lowest densities within the first data release, this limitation is not critical. Passive galaxies preferentially reside in dense environments, where processes such as multiple mergers, ram-pressure stripping, starvation, harassment, and tidal interactions efficiently remove gas, quenching star formation and producing red, early-type systems \citep{somerville_physical_2015}. Therefore, it is important to acknowledge this bias: in overdense regions, we are only seeing a subset of the complete galaxy population. We discuss these considerations further in Sects.~\ref{sc:galprops} and~\ref{sc:mergertrees}.

To evaluate how the previously listed observational constraints of our mock \Euclid\ DR1 sample affect void identification, we also created a reference catalogue, hereafter the Ground Truth (GT). The GT corresponds to the full simulated lightcone based on the model of \cite{de_lucia_tracing_2024}, to which we only applied a cut on stellar mass, $\Mstellar \geq 10^{9.2}\,M_\odot$. This limit was chosen to maximise the diversity of galaxy types, balancing the inclusion of the lowest-mass galaxies with the need to maintain high completeness, given the lightcone is cut at magnitude $\HE = 25$ by construction (see Sect.~\ref{sc:GAEA}). No observational effects were applied. It therefore provides an idealised representation of the intrinsic galaxy distribution, free from flux limits, sampling incompleteness, or interloper contamination. In Appendix~\ref{apdx:GroundTruth}, we compare the results obtained with the DR1-like lightcone to those from the GT. Throughout this work, the GT serves as a benchmark to validate the reliability of our measurements, and it is referenced at multiple stages of the analysis.

\section{\label{sc:Voids}The void environment}

Cosmic voids are broadly recognised as large regions of space corresponding to depressions in the cosmic density field, but there is no single operational definition. Instead, voids can be identified from different first principles, which naturally give rise to different classes of void finders, as described by \cite{lavaux_precision_2010}. Some algorithms search for spherical underdensities by growing or shrinking spheres at a fixed density threshold \citep{ceccarelli_clues_2013, paz_clues_2013, micheletti_vimos_2014, ruiz_clues_2015, paz_guess_2023}, while others reconstruct a discrete density field following the topology of the cosmic web, resulting in the identification of irregular void regions \citep{neyrinck_zobov_2008, sutter_vide_2015, Sousbie_2011_I, Sousbie_2011_II, nadathur_beyond_2019}. Most studies on simulations and spectroscopic redshift surveys rely on full 3D information, like all the works mentioned above, while others focusing on photometric-redshift catalogues operate on 2D redshift slices \citep{sanchez_cosmic_2017}. The choice of the method can significantly affect the detected void sizes, boundaries, and the definition of galaxy membership (see also \cite{Colberg_aspen-amsterdam_2008} for an early systematic comparison of void-finding algorithms). For our analysis we adopted the \Revolver implementation of \textsc{zobov} (fully described below in Sect.~\ref{sc:VoidFinder}), a parameter-free algorithm currently used within the Euclid Consortium for cosmological studies (\citealt{Radinovic23}, Radinovi\'c et al. in prep.). This choice followed from a comparison with other available algorithms, summarised in Appendix~\ref{apdx:VoidFinders}.

\subsection{\label{sc:VoidFinder}Void finder: \Revolver}

In this work, we employ the REal-space VOid Locations from surVEy Reconstruction toolkit, \Revolver \citep{nadathur_beyond_2019}. \Revolver is based on an enhanced version of the ZOnes Bordering On Voidness algorithm, \textsc{zobov} \citep{neyrinck_zobov_2008}, a parameter-free watershed algorithm \citep{platen_cosmic_2007} designed to identify cosmic voids through a topological analysis of the galaxy density field. It begins by constructing a Voronoi tessellation of the tracer distribution, where each galaxy is assigned a volume $V_j$ that comprises all the points closer to the given \textit{j}-th galaxy than to any other galaxy. The local number density of a galaxy is defined as the inverse of the volume of its associated Voronoi polyhedron, $\rho_j = 1/V_j$. Local density minima are identified as galaxies whose Voronoi volumes exceed those of their nearest neighbours. Around each minimum, \textsc{zobov} grows a region, or zone, by iteratively adding adjacent cells of increasing density (or decreasing volume), forming watershed basins that trace the basins in the density landscape. Adjacent zones are merged into larger voids if they are separated by sufficiently low-density ridges, based on a density contrast criterion. To ensure robustness, the algorithm introduces fake high-density buffer particles outside survey boundaries, preventing the formation of artificially large cells near masked or edge regions. The final void catalogue consists of irregularly shaped, non-overlapping regions defined purely by the topology of the density field (an illustration of a `Voronoi' void can be found in fig.~2 of \citealt{sutter_public_2012}).

For each void, \Revolver returns the coordinates of the centre and the total volume, without relying on arbitrary smoothing or shape assumptions. The void centre is defined as the centre of the largest sphere completely empty of galaxies that can be inscribed within the void. This corresponds to the circumcentre of the tetrahedron formed by the galaxy with the lowest local density and its three adjacent neighbours with the largest Voronoi cell volumes. This definition guarantees that the void centre corresponds to a true local density minimum, as it is defined from the immediate neighbourhood of the lowest-density galaxy \citep{nadathur_nature_2015,nadathur_testing_2016,nadathur_tracing_2017}. While this geometric definition makes the void centre sensitive to shot noise from the discrete tracer distribution, thereby introducing a level of contamination into our signal, \Revolver remains the best available approach to trace the immediate proximity of galaxies to local underdensities. Alternatives such as the volume-weighted barycentre \citep{sutter_vide_2015}, which depend on the global void structure, effectively smooth over the local density field. Emerging dynamical void finders applicable to data (e.g. BitVF; \citealp{sartori_back-in-time_2026}) are expected to strongly mitigate this shot-noise sensitivity, and their application to \Euclid-like samples is left for future work (Euclid Collaboration: Papini et al., in prep.).

\begin{figure}
    \centering
    \includegraphics[width=\linewidth]{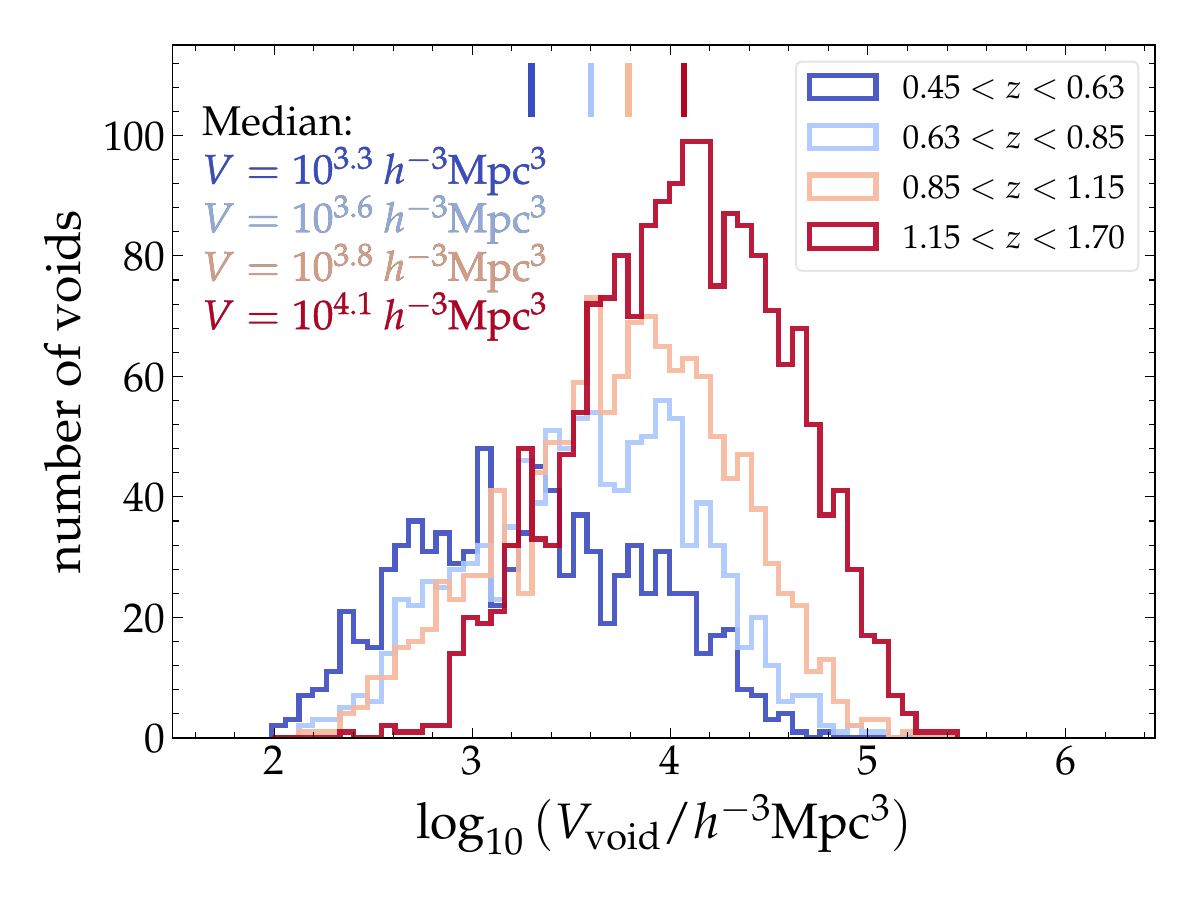}
    \caption{Volume distribution of voids recovered with \Revolver in comoving units. The four histograms represent redshift subsamples. Vertical segments indicate the median of each histogram; the value is also printed in the top left corner of the plot. }
    \label{fig:volvoidREVL}
\end{figure}

The void volume is computed as the sum of the Voronoi volumes of all galaxies within the void. In Fig.~\ref{fig:volvoidREVL}, we show the distribution of void volumes in four redshift bins. The increase of the median void volume with redshift observed in Fig.~\ref{fig:volvoidREVL} reflects the evolving density of tracers in our \ha\ flux-limited sample (see Sect.~\ref{sc:Euclid-like}), which leads to sparser sampling at higher redshifts and consequently to the identification of larger voids. 
Throughout this paper, as a proxy for the void size we use the effective radius, defined as the radius of a sphere with the same volume as the void: $R_{\rm v} = \left(3V_{\rm v} / 4\pi\right)^{1/3}$. 
The void finder returns a flag identifying voids that intersect the geometrical boundary of the lightcone. We exclude from our analysis these edge voids and the galaxies associated with them, as the density reconstruction is unreliable in these regions due to boundary effects. In addition, we also exclude any voids and galaxies at $z<0.45$ or $z>1.7$ to account for border effects on the line-of-sight axis. After verifying that, at all redshifts, the smallest identified voids have effective radii larger than the mean inter-particle separation, ensuring that they are statistically resolved, we chose not to impose any additional cut on the minimum void size, in order to preserve the full dynamic range of void scales in our analysis.
Lastly, we applied the same \Revolver\ procedure to the GT lightcone to ensure a consistent identification of voids and enable a direct comparison between the observationally biased and intrinsic galaxy distributions (see Appendix~\ref{apdx:VoidFinders}).

\subsection{\label{sc:envpars}Parametrising the underdense environment}
  
Having constructed the void catalogue with \Revolver, for the subsequent analysis we do not wish to rely solely on a binary in/out definition, since in \textsc{zobov}-based void finders the void boundaries correspond to overdense ridges. To distinguish galaxies in these edge regions from those deep in the void interiors, we require a more refined metric of environment.
Analogous to studies of galaxy clusters, where local environment parametrisations such as distance from the cluster centre or membership in substructures can be used to assign galaxy membership to clusters, we also aim to describe galaxies as a function of their specific position within their host voids.
Parametrising the environment, especially within underdense regions of the cosmic web, is neither straightforward nor uniquely defined. A wide range of environmental metrics have been developed, each offering a different perspective on how to interpret environmental effects. These methods differ in physical scale, underlying assumptions, and their sensitivity to observational biases, and can each bring different insights into the study of galaxy evolution.

In our analysis, we adopt void-centric distance as a first-order approximation of underdense environment \citep[see also][]{habouzit_properties_2020, wang_cosmology_2024}, as we aim to explore how galaxy properties change with proximity to the local density minimum within a void. As reported in Sect.~\ref{sc:VoidFinder}, the void centres identified by \Revolver have been shown to provide stable and accurate tracers of these minima \citep{nadathur_beyond_2019}. For each galaxy, we compute its distance to the centre of its associated void, $d_{\rm cc}$. To account for the wide range of void sizes, we normalise this distance by the effective radius of the void $R_{\rm v}$, which allows galaxies in voids of different sizes to be placed at the same relative scale. While this spherical approximation is a simplification, given that voids are known to have highly irregular shapes \citep{neyrinck_zobov_2008, sutter_vide_2015}, it offers a useful scale for comparing positions within the void volume. Indeed, in the inner regions of voids, well away from walls and filaments, the spherical approximation is generally considered valid (\citealt{icke_voids_1984, sheth_hierarchy_2004, hamaus_constraints_2016, curtis_properties_2024, perez_cavity_2024}). 

Nonetheless, the internal structure of voids is not necessarily featureless; previous studies have found that they may contain groups of galaxies, diffuse filaments, or low-density structures such as tendrils \citep{szomoru_hi_1996, elad_voids_1997, hoyle_voids_2004, sheth_hierarchy_2004, kreckel_void_2012, alpaslan_galaxy_2014}. For this reason, in Sect.~\ref{sc:Discussion} we also discuss the introduction of the local galaxy density as a control environmental parameter, both within and beyond void boundaries.

The local density contrast can provide a better characterisation of the galaxy distribution on smaller scales within voids. There are a variety of techniques to evaluate this quantity, both on fixed scales, such as by counting galaxies within a sphere of a fixed radius, and scale independent, for instance by tessellating the space or finding distances from neighbouring galaxies. In this work, we focus on methods that do not impose a scale in order to catch density fluctuations on small scales if they exist.

\begin{figure}
    \centering
    \includegraphics[width=1\linewidth]{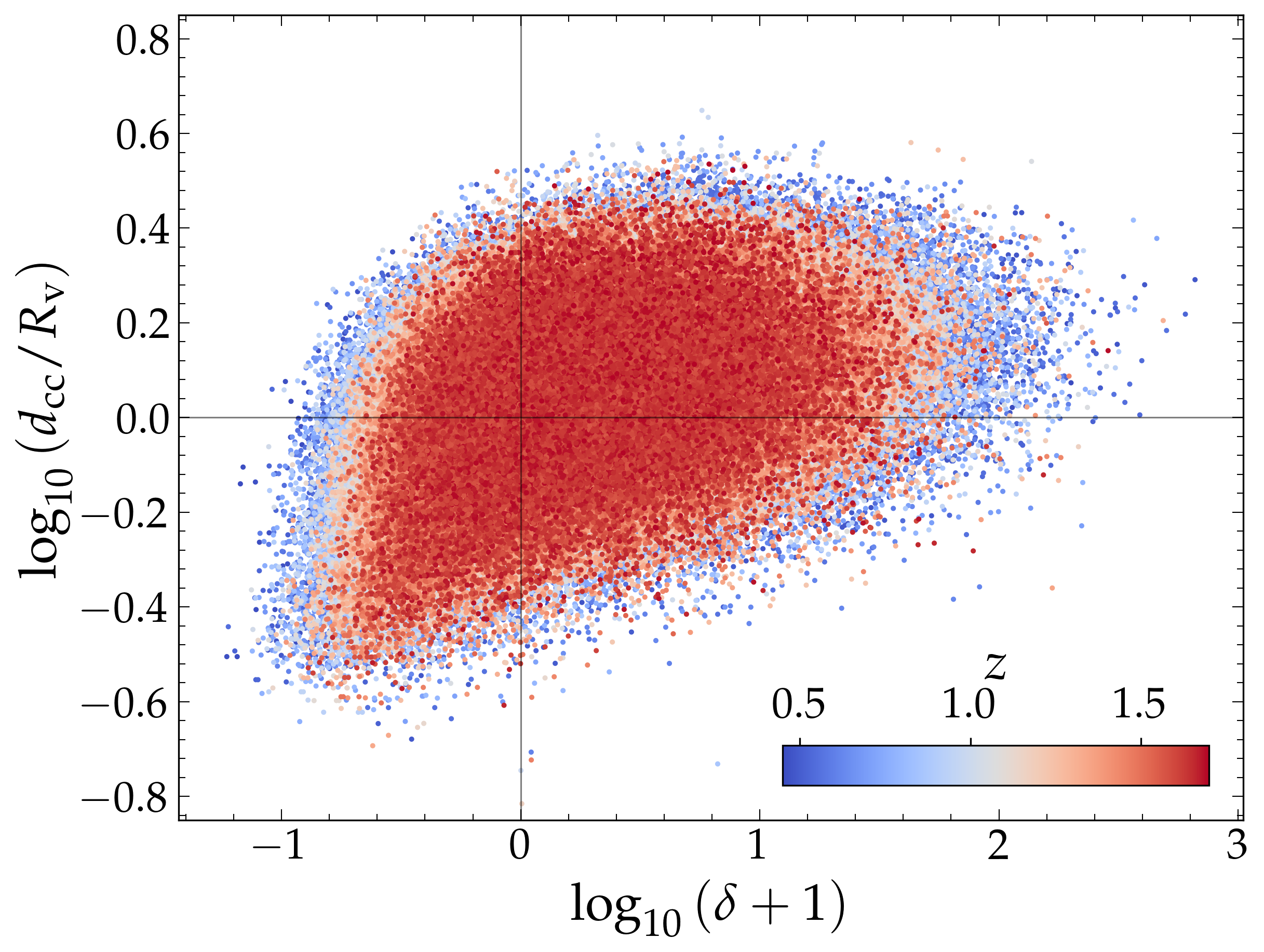}
    \caption{Correlation between the galaxy void-centric distance normalised to the void radius and its local density contrast, colour-coded by the redshift of the galaxy. Axes at (0,0) identify key standpoints in the parameter space: the local mean density ($\rho=\bar{\rho}$) and the spherical equivalent border of the void ($d_{\rm cc}=R_{\rm v}$). Points at low-redshift (blue) are plotted behind points at high-redshift (red).}
    \label{fig:dccRv_delta}
\end{figure}

Our void finder is based on 3D Voronoi tessellation, which assigns a Voronoi cell to each galaxy and computes density as the inverse of the cell volume (Sect.~\ref{sc:VoidFinder}). This method to parametrise the local density provides adaptive resolution and is robust against Poisson noise (\citealt{nadathur_beyond_2019}). 
To account for the redshift dependence introduced by our sample selection, we normalise the measured local density to the mean cosmic density at the galaxy redshift [$\rho\,(z)/\bar{\rho}\,(z)=1+\delta$].
As shown in Fig.~\ref{fig:dccRv_delta}, normalising the local density contrast by its redshift-dependent mean removes any systematic trend with redshift: galaxies at different redshifts occupy the same locus in the $[\log_{10}(\delta+1), \log_{10}(d_{\rm cc}/R_{\rm v})]$ plane. Both quantities are therefore dimensionless and effectively redshift-independent, so they can be compared across the entire sample.
The dispersion of the correlation becomes progressively larger both at higher densities and at larger distances from void centres, further indicating that void boundaries are not sharply defined and are sensitive to the sampling density of the tracer population. This reflects the intrinsic asymmetry and substructure within voids, where filamentary features and galaxy groups can exist at similar radial distances but in different local environments. These deviations from spherical symmetry and homogeneity suggest the need for multi-parameter environmental classifications, as we further discuss in Sect.~\ref{sc:localdensity}.

As an alternative estimator for local density, we also tested the distance to the 5\textsuperscript{th} nearest neighbour $d_\mathrm{5NN}$, an approach which is also scale-free and widely used in the literature \citep{bolzonella_tracking_2010, davidzon_vimos_2016}. This quantity was also normalised, in order to remove the redshift dependence, to the mean inter-particle separation (MIP) at the galaxy redshift ($d_\mathrm{5NN}/d_\mathrm{MIP}$). 
In agreement with recent studies \citep{haas_disentangling_2012, davidzon_vimos_2016, florez_void_2021}, we find a tight correlation between the two local density estimators, and we decide to rely on the local density contrast for our analysis.

\section{\label{sc:galprops}Galaxy properties in voids}

\begin{figure*}[htbp!]
\centering
\includegraphics[angle=0,width=1.0\hsize]{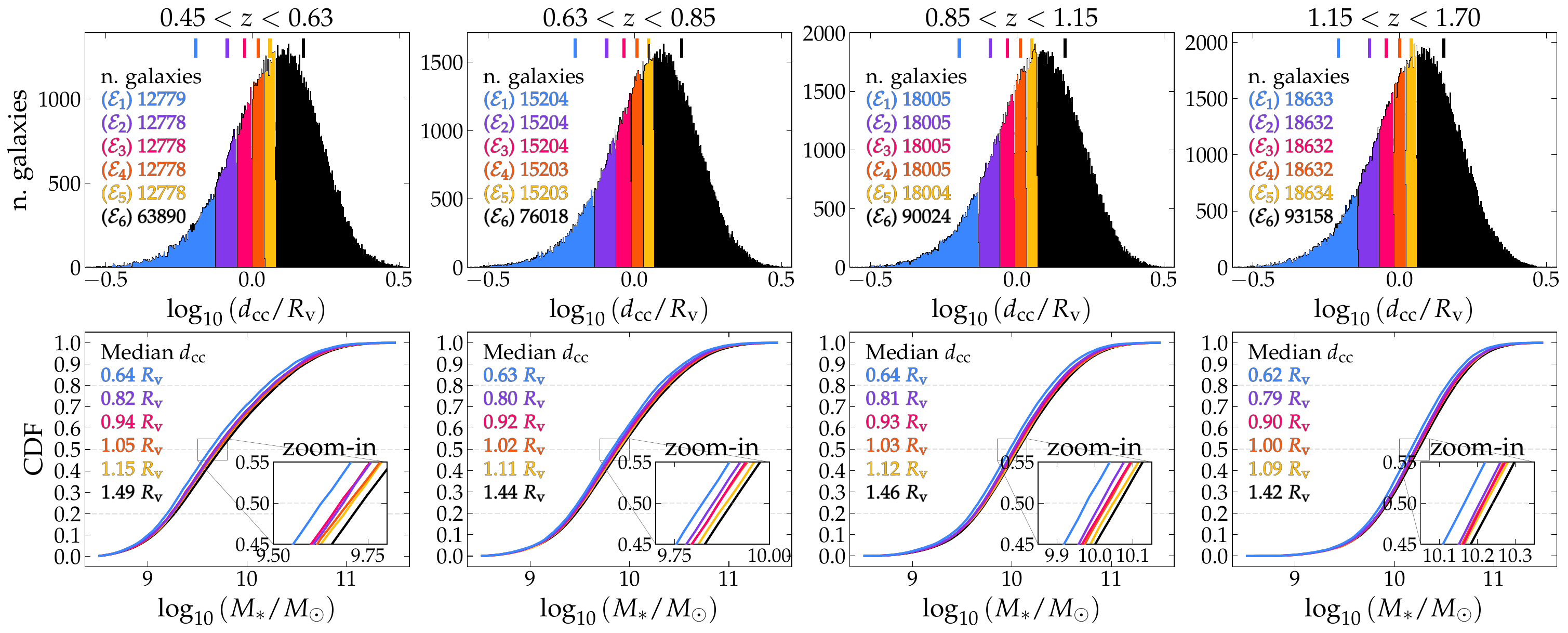}
\caption{Top row: Distributions of the normalised void-centric distance, colour-coded following the strategy described in Sect.~\ref{sc:envpars}. Coloured histograms represent galaxy populations progressively closer to the void centre, while the black histogram corresponds to the control sample at higher densities. Vertical tick marks at the top of each panel indicate the position of the median normalised distance in each bin. Bottom row: CDFs of stellar mass for the same environmental bins as defined in the top row. Median $d_{\mathrm{cc}}$ values in each bin are expressed in $R_{\rm v}$ units and indicated in the top left corner of each panel. Dashed grey lines are added at 20\%, 50\%, and 80\% of \Mstellar\ cumulative distributions as guidelines. The four columns of this figure stand for the 4 redshift bins described in the text.}
\label{fig:distCC_Mstar_CDF}
\end{figure*}

The interplay between galaxies and their surrounding environment plays a fundamental role in shaping their evolution. Using the parametrisation described in Sect.~\ref{sc:envpars}, we now compare the properties of galaxies across different environmental regimes in our \Euclid-like lightcone. The full sample is divided into four redshift bins, each spanning approximately 1.2 Gyr in lookback-time: $0.45\!<\!z\!<\!0.63$, $0.63\!<\!z\!<\!0.85$, $0.85\!<\!z\!<\!1.15$, and $1.15\!<\!z\!<\!1.70$. These intervals are chosen to uniformly cover the full \ha\ redshift range accessible in the EDFs. Within each redshift bin, we compute the median value of the normalised void-centric distance, and separately consider the distributions above and below the median. The lower half of the distribution, including galaxies located closer to the void centre, is further subdivided into five decile bins, each containing 10\% of the total sample, as illustrated in the top panel of Fig.~\ref{fig:distCC_Mstar_CDF}. The upper half of the distribution, consisting of galaxies at larger normalised distances, is utilised as a control sample representing regions of relatively higher local density. This percentile-based binning ensures uniform sample sizes across all environmental regimes and enables robust statistical comparisons. We compute cumulative distribution functions (CDFs) for key galaxy properties within each redshift bin to assess systematic trends as a function of void-centric environment (i.e. normalised void-centric distance $d_{\rm cc}/R_{\rm v}$).

Among the many physical quantities available in the mock lightcone, we focus on a representative set that both captures key aspects of galaxy evolution and can be robustly constrained with forthcoming \Euclid data from different techniques: (i) stellar mass and specific star formation rate (sSFR), which can be derived from the fitting of the SED, (ii) bulge-to-total stellar mass ratio, which serves as a proxy for galaxy morphology that will be \Euclid's show-piece thanks to the VIS high-resolution imaging (e.g. \citealp{Quilley25}), and (iii) mass of the host dark matter halo, a property not measured directly by \Euclid but potentially accessible through group and cluster identification, weak lensing, and galaxy-galaxy lensing and clustering. Together, these observables encompass galaxy facets such as star formation activity, structural evolution, and the link between galaxies and their dark matter haloes, offering a comprehensive framework to test environmental effects in underdense regions. In Appendix \ref{apdx:Observational_Uncertanties}, we assess how these trends respond to realistic observational uncertainties expected for the \Euclid\ DR1. The trends discussed below are broadly consistent with previous observational and theoretical works; a key outcome of this analysis is that they are robustly recovered within a \Euclid-like dataset, assuming that redshift interlopers have been effectively removed, demonstrating that \Euclid\ DR1 observations will already enable statistically meaningful studies of environmental effects on galaxy evolution.

\subsection{\label{sc:galprops_stellar_mass}Stellar mass, \Mstellar}

We begin by examining CDFs of stellar mass as a function of redshift and void-centric distance, as shown in the top panel of Fig.~\ref{fig:distCC_Mstar_CDF}. 
At all redshifts, galaxies located closer to the void centres tend to have lower stellar masses, with the median of the distribution shifting towards lower values as $d_{\rm cc}/R_{\rm v}$ decreases. This supports the view that galaxies in the most underdense regions are, on average, less massive than those in higher-density environments.

We quantitatively compared the stellar mass distributions in the different environments using a Kolmogorov--Smirnov (KS) test. For the sake of simplicity, we focused our attention on three representative environmental bins: (i) the innermost region, $d_{\rm cc}\leq0.6\,R_{\rm v}$, which probes the void interiors; (ii) the region just inside the void boundary, $d_{\rm cc}\approx0.90$--$0.95\,R_{\rm v}$, sensitive to the transition between underdense and overdense environment; and (iii) the bin above the median value of the normalised void-centric distance distribution, which traces galaxies in higher density regions.
As they are the first, third and sixth environmental bins of Fig.~\ref{fig:distCC_Mstar_CDF} respectively, we call them \Ei, \Eiii, and \Evi in the following, where $\mathcal{E}$ stands for \textit{environment}. We performed a KS test comparing the stellar mass distribution in \Ei with the distributions in \Eiii and \Evi, in each redshift bin (full results are provided in Appendix~\ref{apdx:KS_test_results}). For both comparisons, and at all redshifts, we find that the null hypothesis that the two distributions are drawn from the same parent distribution has a probability $p_{\rm KS} \ll 10^{-5}$ and can therefore be rejected. Throughout our analysis, we adopt a threshold of $p_{\rm KS}\!=\!10^{-5}$ as an indicator of statistical significance, corresponding approximately to a 4\textsigma\, confidence level. We also remark that, especially in the highest redshift bin, the fraction of relatively massive galaxies with $\log_{10}{(\Mstellar/M_\odot)} > 10.5$ increases continuously from $\approx\!20\%$ to $\approx\!30\%$ moving from \Ei to \Evi. This difference is less evident in the lower redshift bins.

\begin{figure}
    \centering
    \includegraphics[width=1.0\linewidth]{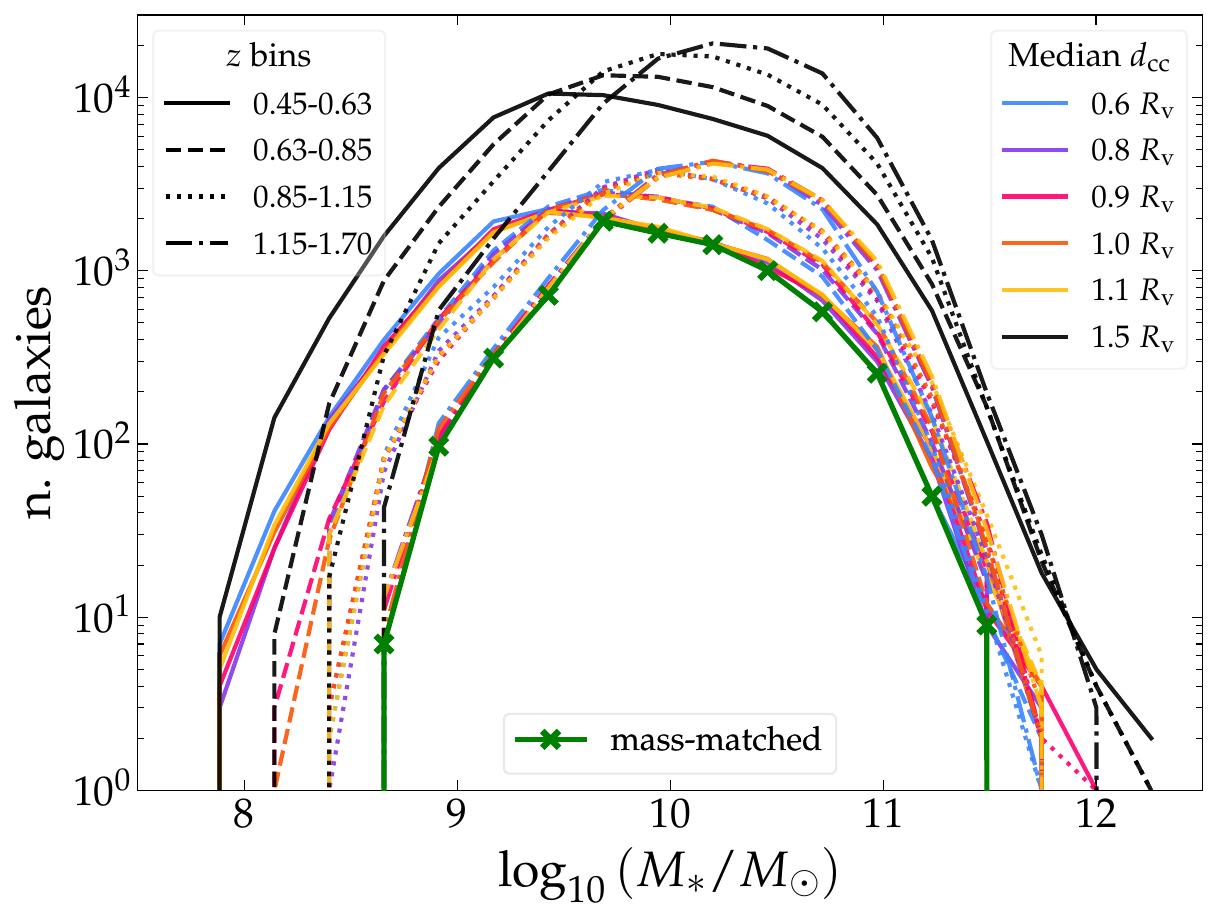}
    \caption{Distributions of stellar masses of all bins defined in Fig.~\ref{fig:distCC_Mstar_CDF}. Colours code for different environments, line styles identify redshift bins. Green line and crosses highlight the mass-matched distribution.}
    \label{fig:mass_match}
\end{figure}

\begin{figure*}[htbp!]
\centering
\includegraphics[angle=0,width=1.0\hsize]{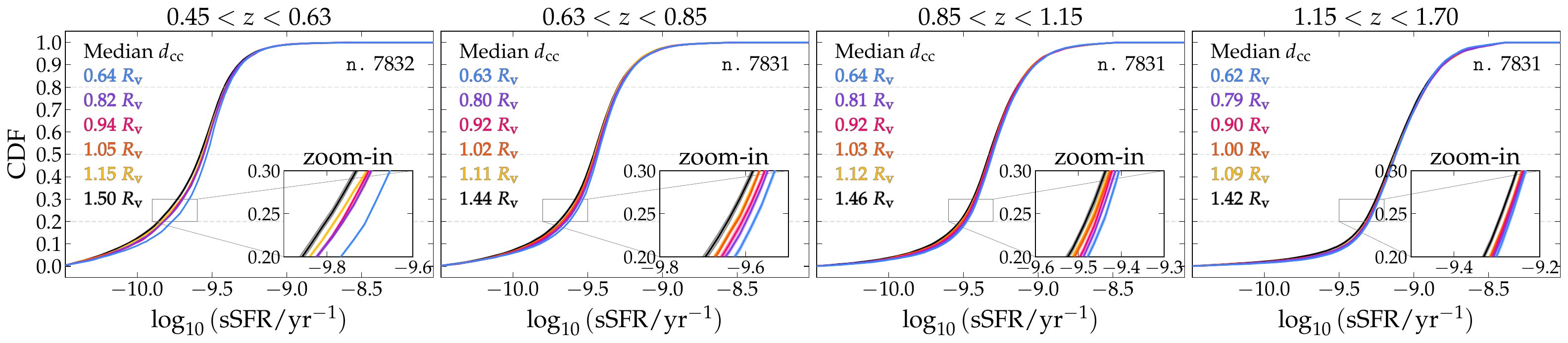}
\caption{sSFR distribution as a function of redshift and void-centric environment. Columns correspond to the four redshift bins described in the text. In each panel, we show CDFs of sSFR for the same environmental bins defined in Fig.~\ref{fig:distCC_Mstar_CDF} and described in Sect.~\ref{sc:galprops}. Stellar mass distributions have been matched across all environment-redshift bins, as outlined in Fig.~\ref{fig:mass_match}. The number of galaxies (\texttt{n.}) in each bin is printed in the top right corner of each panel. The CDFs were computed over 100 realisations of the mass-matching procedure; the shaded area represents their 1\textsigma\ dispersion.}
\label{fig:distCC_sSFR_CDF}
\end{figure*}

\begin{figure*}[htbp!]
\centering
\includegraphics[angle=0,width=1.0\hsize]{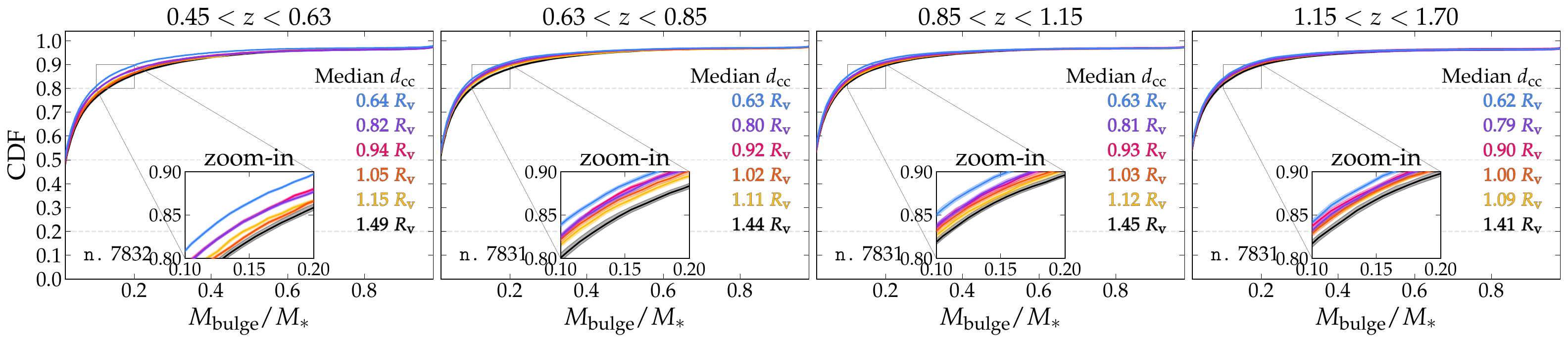}
\caption{As in Fig.~\ref{fig:distCC_sSFR_CDF} but for B/T stellar mass ratio. The number of galaxies (\texttt{n.}) in each bin is printed in the bottom left corner of each panel.}
\label{fig:distCC_MbulgetoMstar}
\end{figure*}

It is worth noting that, albeit the differences are statistically significant, their absolute amplitude remains small: for instance, the median stellar mass in \Ei differs from that in \Evi by only $\approx\!0.1$~dex across the four redshift intervals. This small difference is also expected given that our sample is \ha-selected and dominated by star-forming galaxies in all environments, so that strongly quenched, high-mass systems are largely absent from our sample.
Forecasts for the recovery of galaxy physical properties in preparation for real data suggest that such uncertainties can be sufficiently constrained to preserve the detectability of the underlying trends \citep{EP-Enia}. In Appendix~\ref{apdx:Observational_Uncertanties}, we explicitly test the impact of the expected DR1 uncertainties on stellar mass in the spectroscopic catalogue and find that the environmental trends persist, with galaxy populations in different environments remaining statistically distinct across all redshifts. Above all, this demonstrates that with \Euclid DR1 observations we will be able to resolve subtle, environment-driven differences in stellar mass distributions across void environments over a wide redshift range.

Many studies have demonstrated the strong dependence of galaxy properties on stellar mass, reflecting its fundamental role in galaxy evolution. This includes works on star formation and the main sequence \citep{brinchmann_physical_2004, noeske_star_2007, speagle_highly_2014, genzel_combined_2015, tacconi_phibss_2018}, on gas content and chemical enrichment \citep{daddi_multiwavelength_2007, elbaz_reversal_2007, scoville_evolution_2017, maiolino_re_2019, kewley_understanding_2019}, on structural and morphological trends \citep{franx_structure_2008, williams_detection_2009, peng_mass_2010}, and on the evolution of galaxy populations across environments and cosmic time \citep{davidzon_vimos_2016, cucciati_vimos_2017}.
In this work, we aim to investigate whether environmental effects, in addition to stellar mass, influence the evolution of galaxy properties, and how this influence changes with cosmic time. Since we have already shown in Fig.~\ref{fig:distCC_Mstar_CDF} that stellar mass depends on both environment and redshift, it is necessary to remove any stellar mass dependence from other galaxy properties in order to study the impact of environment as a function of redshift.

To isolate the role of environment in driving galaxy properties, we constructed mass-matched subsamples across all redshift and environmental bins. 
In Fig.~\ref{fig:mass_match} we show the stellar mass distributions of all 24 environment-redshift subsamples, computed in 0.2\,dex bins of $\log_{10}{(\Mstellar/M_\odot)}$. As anticipated in Sect.~\ref{sc:Euclid-like}, due to the \ha\ flux-limited nature of our total sample, the low-mass end is dominated by low-redshift galaxies, while the high-mass end is dominated by high-redshift ones. To construct mass-matched subsamples, we defined a common reference distribution by taking the lower envelope across all redshift and environment subsamples, i.e. the minimum galaxy count in each stellar mass bin. This lower envelope was used as the target for the mass-matching procedure, ensuring a uniform stellar mass distribution across environments and redshifts, and allowing unbiased comparisons of other galaxy properties. In practice, this is implemented by randomly down-sampling each subset's stellar-mass distribution so that every mass bin matches the target number set by the lower envelope.

\subsection{\label{sc:galprops_sSFR}sSFR}

We applied the mass-matching procedure to our sample to analyse the sSFR, shown in Fig.~\ref{fig:distCC_sSFR_CDF}.  As expected, the overall sSFR decreases towards lower redshifts, reflecting the general decline in star formation activity since $z\!\approx\!2$ \citep{madau_dickinson_cosmic_2014}.
The CDFs further show that, at all redshifts, galaxies located closer to void centres tend to exhibit higher sSFRs compared to those at larger void-centric distances. This indicates that galaxies in the most underdense environments have systematically higher sSFRs than their counterparts in denser regions at fixed stellar mass and redshift. 
Since the mass-matching procedure involves random down-sampling of more populated subsamples, we repeated the process 100 times by randomly extracting the same number of galaxies from the parent catalogue and computed the CDF for each realisation to estimate the dispersion. The dispersion of these iterations remains small in all bins, indicating that the environmental trends measured in these mock-observations are statistically robust and not driven by stochastic sampling.
These findings are consistent with previous results in the literature, which report enhanced star formation activity in low-density environments \citep{rojas_spectroscopic_2005, kreckel_void_2012, ricciardelli_star_2014, beygu_void_2016, curtis_properties_2024}. 

Furthermore, we observe that the separation between CDFs for different environments decreases with increasing redshift. This can be quantified again by a KS test  (full results are provided in Appendix~\ref{apdx:KS_test_results}). In the two lowest redshift bins, the sSFR distributions in \Ei and \Evi are different at a significance $p_{\rm KS} \ll 10^{-5}$ (computed as the median $p_{\rm KS}$ among the values derived from the 100 mass-matched realizations), and this holds also for the comparison between \Ei and \Eiii (see Sect.~\ref{sc:galprops_stellar_mass} for the definition of $\mathcal{E}$). 
At higher redshifts, the significance level remains $p_{\rm KS} \ll 10^{-5}$ between \Ei and \Evi, but for \Ei and \Eiii  it drops to $p_{\rm KS} \approx 10^{-4}$ and then $p_{\rm KS} \approx 0.1$ around $z\approx1$ and $z\approx1.5$ respectively. 
This shrinking separation between environments suggests a potential weakening of the trend with sSFR at earlier cosmic times. In future work, it would be interesting to determine whether the reversal of the sSFR-environment trend, previously found in both simulations \citep{chiang_galaxy_2017} and observations \citep{lemaux_vimos_2022} at $z>2$, also emerges as a prediction of this mock lightcone. This test is beyond the scope of the present study and cannot be performed with the DR1 \Euclid\ Deep H$\alpha$ spectroscopic sample, as H$\alpha$ is only traced up to $z\simeq1.8$, although \Euclid\ spectroscopy will in principle provide a limited number of high-redshift detections through other emission lines (e.g. H$\beta$ and [O\,\textsc{iii}] up to $z\sim2.9$; see fig.~1 of \citealt{EP-Quai}), and could be pursued in synergy with future dedicated high-redshift spectroscopic surveys (e.g. MOONRISE, \citealp{Maiolino_moonrise_2020}; PFS Galaxy Evolution Survey, \citealp{Greene_prime_2022}).
 
As for stellar mass, we tested the robustness of the sSFR trends against observationally motivated uncertainties that we expect in \Euclid\ DR1 data (Appendix~\ref{apdx:Observational_Uncertanties}). We find that the overall pattern of higher sSFRs in low-density environments remains visible, although the distributions are broadened and the separation between environments is reduced. Extreme low- and high-density environments remain statistically distinct ($p_{\rm KS} \ll 10^{-5}$), but differences between intermediate-density environments become minimal, especially at higher redshifts. 

\subsection{\label{sc:galprops_bulge_to_total}Bulge-to-total ratio, $M_{\rm bulge}/\Mstellar$}

We quantify galaxy morphology via the bulge-to-total (B/T) fraction, defined as the ratio of stellar mass in the bulge component to the total stellar mass of the galaxy as provided by the simulation, $M_{\rm bulge}/\Mstellar$. These B/T values represent the intrinsic structural decomposition predicted by the SAM; in contrast, forthcoming \Euclid\ data will deliver independent morphological constraints from high-resolution imaging and structural measurements, as demonstrated in recent preparatory and early-data studies \citep{Merlin-EP25, Bretonniere-EP26, Q1-SP047, Quilley25}.

The ratio between bulge mass and total stellar mass is found to also depend on environment, as shown in Fig.~\ref{fig:distCC_MbulgetoMstar}. The narrow spread among the 100 realisations observed at all redshifts indicates that the separation between CDFs is statistically robust and not driven by sampling variance. Due to our selection of \ha-emitting galaxies, the sample is dominated by disc-like systems: over 90\% of galaxies have B/T $< 0.2$ across all environments and redshifts.
Nonetheless, our parametrisation of environment via the normalised void-centric distance reveals a subtle but consistent trend: galaxies located closer to the void centre tend to have lower B/T values than those in higher-density regions, at all redshifts. 
The KS test performed between the distributions in \Ei and \Evi shows $p_{\rm KS}$ progressively increasing toward higher redshifts, with values $<10^{-5}$, $\approx\!10^{-5}$, $\approx\!10^{-4}$, $\approx\!10^{-3}$ from lowest to highest redshift bin, but still robustly rejecting the null hypothesis at least at $z<1.15$. We verified that the $p_{\rm KS}$ resulting from the comparison of \Ei and \Eiii is not small enough to reject the null hypothesis, at any redshift ($p_{\rm KS}>10^{-3}$).  Full results are provided in Appendix~\ref{apdx:KS_test_results}.

This trend aligns with previous studies reporting an excess of younger, disc-dominated morphologies in void environments \citep{rojas_photometric_2004, ricciardelli_morphological_2017, rosas-guevara_revealing_2022, perez_galaxy_2025}. One possible interpretation is that galaxies in voids experience fewer interactions or mergers due to their isolation, reducing the chances of morphological transformations through violent processes such as major mergers or tidal stripping. As a result, their discs are more likely to be preserved over cosmic time. This scenario will be further explored in Sect.~\ref{sc:mergertrees}.

\subsection{\label{sc:galprops_halo}Halo mass, $M_{\rm halo}$, and central/satellite distribution}

\begin{figure}[htbp!]
\centering
\includegraphics[angle=0,width=1.0\hsize]{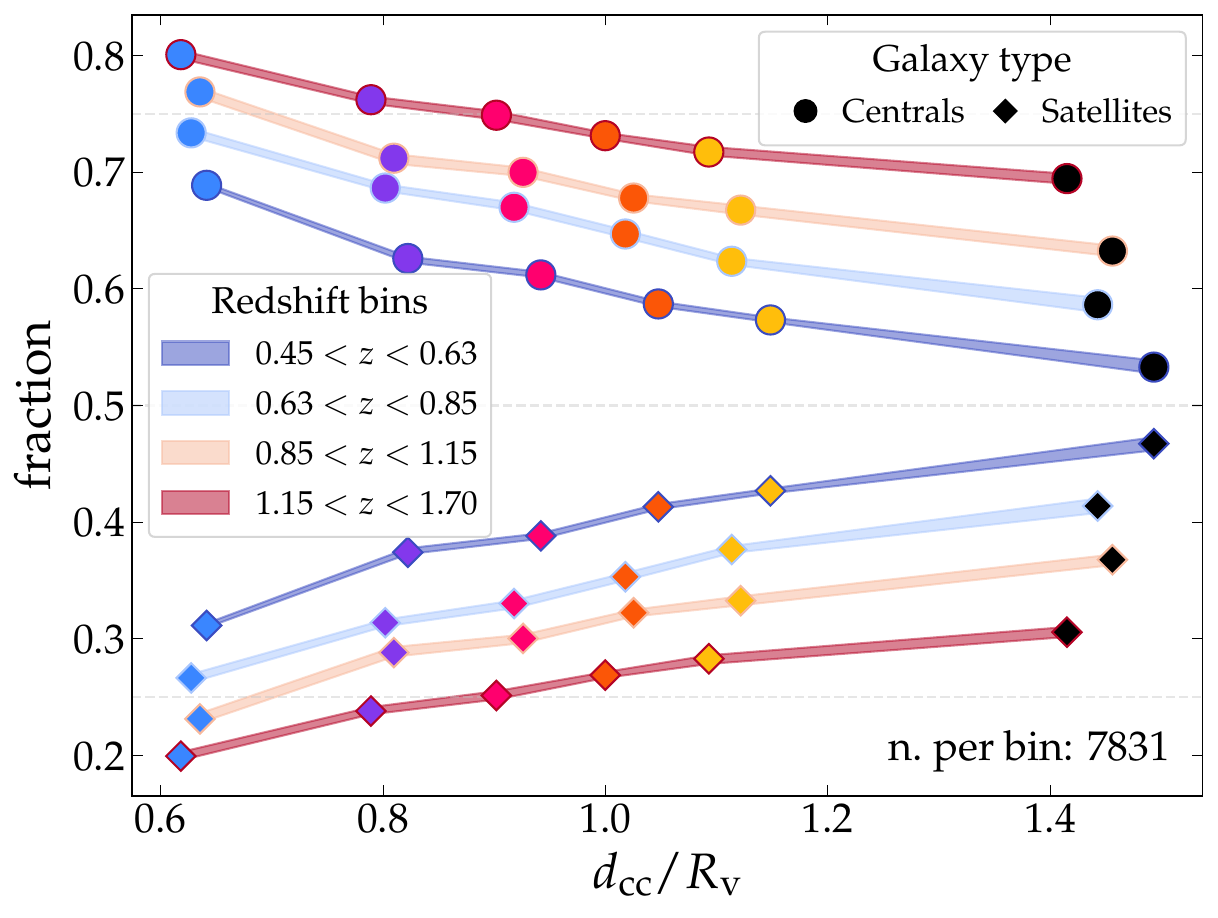}
\caption{Fraction of central and satellite galaxies as a function of redshift and environment. Marker colours represent bins of environment (as in Fig.~\ref{fig:distCC_Mstar_CDF}), defined according to the normalised void-centric distance as described in Sect.~\ref{sc:envpars}, while line colours indicate bins of redshift (as in Fig.~\ref{fig:dccRv_delta}). Circles represent centrals, while diamonds represent satellites. Stellar mass distributions are matched across all redshift and environment bins. Shaded areas represent uncertainties on fractions computed with the formula defined by \cite{gerke_deep2_2007}. Dashed grey lines are added at 25\%, 50\%, and 75\% as guidelines. The number of galaxies in each bin is reported in the bottom right corner of the figure.}
\label{fig:distCC_Type_fration}
\end{figure}

Lastly, we investigate how the dark matter halo mass, $M_{\rm halo}$, varies as a function of environment, redshift, and central/satellite galaxy type. In the GAEA model, galaxies are classified into centrals and satellites, and are associated to the virial mass of the overall dark matter halo where they reside. This corresponds to the halo of the galaxy for isolated  centrals, and to the halo of the group or cluster for galaxies in more complex systems. While halo masses can, in principle, be estimated observationally in dense environments -- for example through galaxy-galaxy lensing or velocity dispersion measurements in identified groups -- the available \Euclid\ DR1 data are not expected to yield reliable halo mass estimates for galaxies deep inside voids, where both lensing signals and group statistics become extremely limited. For this reason, in this section we use the model halo masses as a theoretical complement to interpret the environmental trends seen in the observable properties, and as a prediction for future constraints from lensing and group catalogues at later stages of the mission.

In Fig.~\ref{fig:distCC_Type_fration}, we show the fraction of centrals and satellites as a function of environment and redshift. Stellar mass distributions have been matched across all bins, as described in Sect.~\ref{sc:envpars}. We find a clear trend with both environment and redshift: the fraction of centrals decreases with increasing distance from void centre.
For instance, at the highest redshift bin around $z \approx 1.5$, centrals constitute $\approx\!80\%$ of galaxies within $0.6\,R_{\rm v}$ (\Ei) of void centres, decreasing to $\approx\!70\%$ at $1.4\,R_{\rm v}$ (\Evi). At lower redshifts around $z \approx 0.55$, the central fraction drops from 68\% in the innermost regions to 55\% at the largest distances from void centres. This is consistent with the expectation that galaxies in voids are more likely to be isolated, whereas in overdense regions such as groups and clusters, a single central is surrounded by multiple satellites \citep{liu_spectral_2015, habouzit_properties_2020, rodriguez-medrano_evolutionary_2024}.
We also find a clear redshift dependence: the fraction of centrals decreases towards lower redshifts as the satellite fraction in turn increases. As structures grow over cosmic time, more galaxies are incorporated into larger dark matter haloes, becoming satellites of more massive centrals. At the lowest redshift in our sample, centrals and satellites reach an approximately equal split (55\%--45\%) in regions of highest void-centric distance around $1.4\,R_{\rm v}$, while within $0.6\,R_{\rm v}$ near void centres the distribution remains closer to 32\% satellites and 68\% centrals.

\begin{figure}[htbp!]
\centering
\includegraphics[angle=0,width=1.0\hsize]{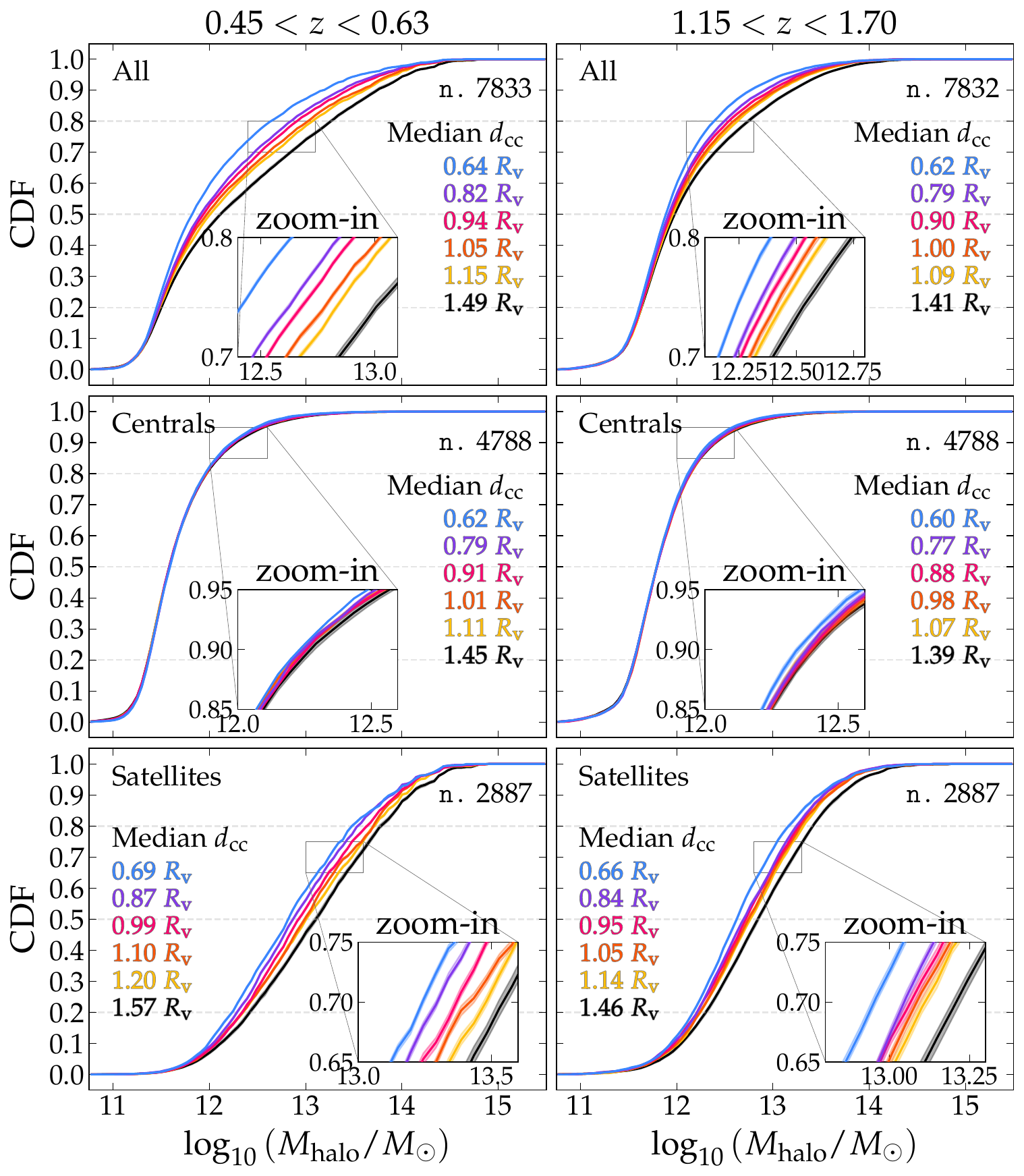}
\caption{As in Fig.~\ref{fig:distCC_sSFR_CDF} but for $M_{\rm halo}$. Only first and last bin of redshift are shown. The three rows refer to the total sample (top), and the two complementary subsamples of central (middle) and satellites (bottom).}
\label{fig:distCC_Mhalo}
\end{figure}

We further compute the CDFs of $M_{\rm halo}$ for the full galaxy population, as well as for centrals and satellites separately. The results are shown in Fig.~\ref{fig:distCC_Mhalo} for the first and last redshift bins, which bracket the range of evolutionary trends observed.
The parametrisation of environment follows that used in previous figures, with all subsamples mass-matched across redshift and environments. In Fig.~\ref{fig:distCC_Mhalo}, the shape of the total CDFs reflects the combined contribution of centrals and satellites. The sharp rise around $M_{\rm halo} \approx 10^{11.5}\,M_\odot$ is primarily due to centrals, whose halo mass is tightly correlated with stellar mass. In fact, after matching the stellar mass distribution, centrals display no significant dependence on environment.
In contrast, the satellite population exhibits a clear environmental trend. Satellites in void environments tend to reside in lower-mass haloes, while those in denser regions are more frequently associated with more massive haloes. The median of the satellite $M_{\rm halo}$ distribution shifts from below to above $10^{13}\,M_\odot$ as one moves from the centres of voids to the outer, denser regions. This variation drives the differences observed in the full $M_{\rm halo}$ CDFs across environments. 

More quantitatively, the KS tests performed between \Ei and \Eiii, and between \Ei and \Evi always return a median $p_{\rm KS} \ll 10^{-5}$ at all redshifts for satellite galaxies, while the same tests cannot reject the null hypothesis for central galaxies (full results are provided in Appendix~\ref{apdx:KS_test_results}). 
For satellite galaxies, we note in particular the differences in fractions of satellites hosted in haloes more massive than $10^{13}\,M_\odot$, i.e. the typical halo hosting galaxies with stellar mass similar to $M_\ast$ and above: the fraction varies from 25\% to 45\% moving from \Ei to \Evi in the highest redshift bin, and from 40\% to 55\% in the lowest redshift bin.

Our findings are consistent with expectations from large-scale structure formation and evolution: galaxy groups are uncommon within voids, so galaxies are more often found as isolated centrals. Moreover, the groups that do form in such underdense regions tend to occupy less massive haloes than their counterparts in denser environments. This explains why centrals show little environmental dependence once stellar mass is controlled for, while satellites clearly trace the underlying variation in host halo mass with environment. Overall, the environmental signal in halo mass is driven primarily by the satellite population, reflecting the rarity and lower mass of group haloes inside voids.

\section{\label{sc:mergertrees}Merger histories of galaxies in voids}

Understanding the evolution of galaxies within their environment requires not only a characterisation of their present-day properties but also a reconstruction of their history of interactions with one another. In particular, the low-density nature of cosmic voids has long suggested a quieter evolutionary path for galaxies residing within them, with lower merger frequencies with respect to denser regions of the Universe \citep{van_de_weygaert_cosmic_2011}. However, this picture is increasingly challenged by evidence from modern simulations which indicate that their merger rates may be comparable to those of galaxies in denser regions \citep{rosas-guevara_revealing_2022, dominguez-gomez_galaxies_2023, rodriguez-medrano_evolutionary_2024}, and that the key distinction instead lies in the timing and nature of these mergers.

Complementing our analysis of galaxy properties in cosmic voids, we examine the assembly history of stellar mass through both star formation and mergers. Although merger histories will not be directly measurable with \Euclid DR1, they represent a key element in interpreting the observable trends in stellar mass, sSFR, and morphology, and connecting \Euclid-like measurements to the underlying evolution of void galaxies. Since galaxy-galaxy interactions occur on sub-Mpc scales, we utilize the local density contrast as our primary environmental metric, as it is more sensitive to small-scale interactions than void-centric distance. Following the methodology described in Sect.~\ref{sc:envpars}, the sample in each redshift bin is divided into six subsamples, five bins each containing 10\% of the sample below the median, and one bin above the median to account for overdense regions. 

It is worth emphasising that our case study is based on an \ha-selected sample (see Sect.~\ref{sc:Euclid-like}). By construction, this selection favours star-forming galaxies and excludes quenched systems, which are more common in overdense regions and whose properties are often significantly affected by major mergers \citep{somerville_physical_2015}. As a result, our analysis of environmental trends discussed below pertain to the star-forming population only, and may underestimate the full contrast in merger histories across environments since the passive population, most sensitive to galaxy interactions, is missing. In Sect.~\ref{sc:mergers_passives} we will discuss this selection bias, by comparing our DR1-like sample with the GT lightcone, where the environment is reconstructed without observational biases and therefore contains a more complete diversity of galaxy star formation histories.

\begin{figure}[htbp!]
    \centering
    \includegraphics[width=1\linewidth]{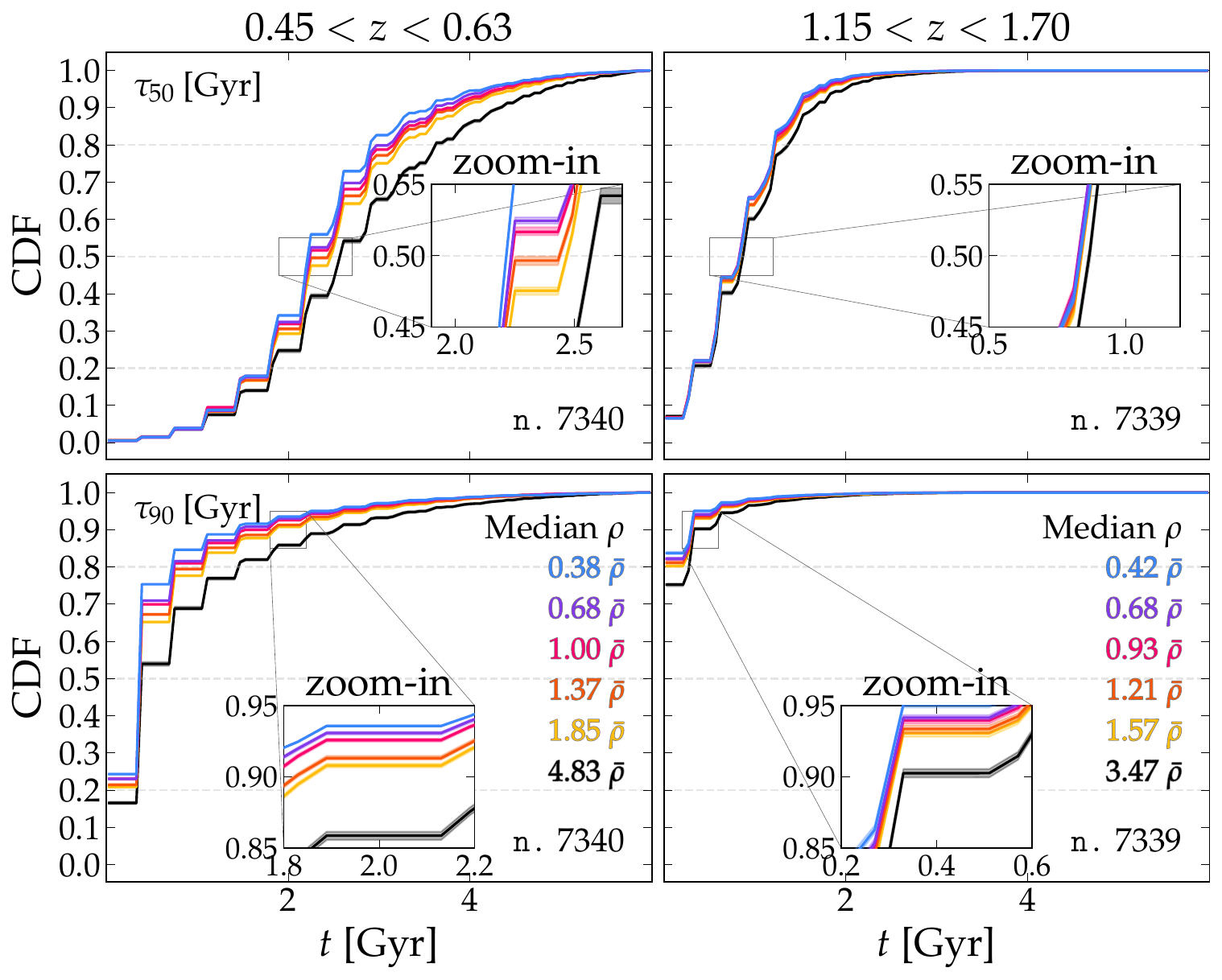}
    \caption{As in Fig.~\ref{fig:distCC_sSFR_CDF} but for the lookback-time at which the galaxy had a stellar mass equal to 50\% (top row) and 90\% (bottom row) of its observed value. Only the first and last bin of redshift are shown. The number of galaxies (\texttt{n.}) in each bin is printed in the bottom right corner of each panel.} 
    \label{fig:tau_50-90}
\end{figure}

We begin by quantifying the timescales over which galaxies assemble their stellar mass. By tracing the evolution of the main branch within each galaxy's merger tree, we track the stellar mass of the main progenitor across all snapshots, from its initial identification to the observed redshift of the galaxy in the lightcone. From this growth curve, we calculate the lookback time at which the main progenitor first reached a specific threshold of the total observed stellar mass recorded for that galaxy in the lightcone. Figure~\ref{fig:tau_50-90} shows the CDFs of $\tau_{50}$ and $\tau_{90}$, defined as the lookback-times when a galaxy first reached respectively 50\% and 90\% of its stellar mass, computed in mass-matched bins of local density contrast and redshifts. These quantities capture the combined effects of star formation and mergers on stellar mass growth. We display only the first and last redshift bins, which effectively bracket the evolutionary range of our sample. Galaxies in the lowest-density regions reach both $\tau_{50}$ and $\tau_{90}$ systematically later than their higher-density counterparts, by roughly 500 Myr to 1 Gyr on average, indicating that stellar mass assembly proceeds more slowly in underdense environments.

\begin{figure}[htbp!]
\centering
\includegraphics[angle=0,width=1.0\hsize]{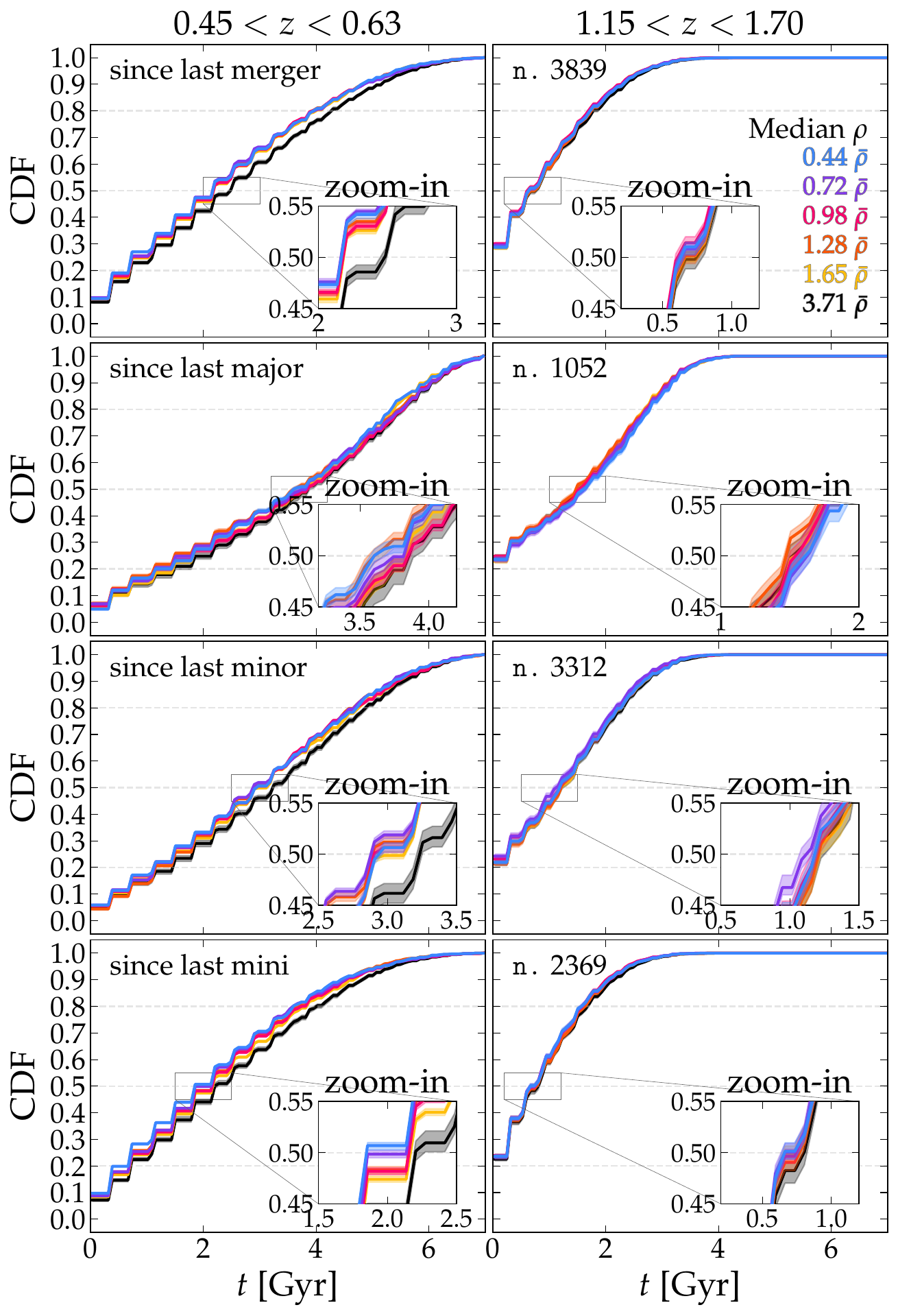}
\caption{As in Fig.~\ref{fig:distCC_sSFR_CDF}, but for the time since the last merger. Only the first and last bin of redshift are shown. In the top row all types of mergers are considered, while the other rows illustrate from top to bottom the time since the last major, minor, and mini merger. The number of galaxies (\texttt{n.}) in each bin is printed in the top left corner of each panel in the second column.} 
\label{fig:distCC_time_last_merger}
\end{figure}

The statistical significance of this trend is confirmed by KS tests: at low redshift, the distributions of both $\tau_{50}$ and $\tau_{90}$ differ markedly between \Ei--\Eiii and \Ei--\Evi ($p_{\rm KS}\ll10^{-5}$), while at high redshift the separation remains significant only between the lowest and highest density bins  (full results are provided in Appendix~\ref{apdx:KS_test_results}). This behaviour suggests that galaxies in underdense environments are on a delayed evolutionary path with respect to counterparts at higher densities, consistently with recent studies suggesting that void galaxies assemble their stellar mass later and more slowly than galaxies in denser regions \citep{dominguez-gomez_galaxies_2023}.

To further explore the assembly histories of galaxies, we quantify the type, frequency, and timing of individual merger events in their past merger trees. In this work, we define a merger as an interaction between two progenitors that satisfies two criteria: (i) a stellar mass ratio, $\mu=M_2/M_1$, greater than 1:100 ($\mu>0.01$) and (ii) a secondary galaxy with stellar mass $M_2 \geq 10^8\,M_\odot$, roughly corresponding to the resolution limit of the simulation. We further classify mergers into three categories based on the stellar mass ratio: major mergers ($\mu > 0.25$), minor mergers ($0.1 < \mu \leq 0.25$), and mini mergers ($0.01 < \mu \leq 0.1$). These categories are used throughout this section and follow commonly adopted conventions in the literature \citep{nipoti_evolution_2025}.

In this simulation, the frequency of mergers in a galaxy's past merger tree is found to show little dependence on the environment the galaxy resides in the lightcone. Furthermore, the fraction of galaxies that have never experienced a merger is statistically compatible between low-density and high-density regions within our uncertainties. We interpret this lack of environmental contrast as a potential consequence of our \ha-based, DR1-like sample selection (see Sect.~\ref{sc:Euclid-like}), a point we discuss thoroughly in Sect.~\ref{sc:mergers_passives} by comparing the DR1-like sample with the full GT lightcone.

However, a clear environmental signature emerges in the timing of these interactions. By selecting galaxies that have experienced at least one merger and calculating the lookback time of the most recent event, we find distinct trends across environments and redshifts.
Figure~\ref{fig:distCC_time_last_merger} shows the cumulative distribution of the look-back time to the most recent merger divided by type, as a function of density contrast, with stellar mass distributions matched across bins (see Sect.~\ref{sc:envpars}), for the first and last redshift bins. As expected, at high redshift the median of the distributions are at lower times, reflecting a younger Universe and higher merger frequency in general. 
At lower redshifts, the distributions become more distinct and environmental differences clearer. While major mergers show weak separation across different environments, minor and mini mergers occur later in underdense regions than in overdense ones. This is most pronounced for mini mergers: galaxies in low-density environments exhibit more recent interactions, with the distribution shifting towards earlier cosmic times as the local density increases.

These findings support a scenario in which delayed, low-mass interactions dominate the assembly of void galaxies. Fuelled by the accretion of low-mass satellites that do not disrupt morphology \citep{karademir_outer_2019}, star formation activity is not quenched but rather sustained \citep{kaviraj_importance_2014}.
This is quantitatively shown by the KS test: only for mini mergers, the distributions in \Ei and \Evi differ consistently ($p_{\rm KS}<10^{-5}$) up to $z\approx0.85$, while the null hypothesis cannot be rejected at $z>0.85$, and as well as for \Ei and \Eiii at any redshift (full results are provided in Appendix~\ref{apdx:KS_test_results}).

The picture emerging from this analysis suggests that void galaxies do not merge less frequently overall, but their mergers occur later and are predominantly minor ones. The evolutionary path of void galaxies is therefore not one of isolation, but of gradual transformation through delayed interactions over long timescales. The higher incidence of relatively recent minor mergers in voids implies that their observed properties retain stronger imprints of late-time interactions than those of galaxies in denser regions, explaining their enhanced star formation, disc-dominated morphologies, and prevalence of central galaxies. Ultimately, our findings indicate that the role of environment in galaxy evolution is not to determine whether galaxies merge, but to regulate when and how they do so. In voids, mergers happen later and are typically less violent, producing a slow evolutionary channel that complements the earlier, merger-driven assembly of galaxies in dense regions. 

\section{\label{sc:Discussion}Discussion}

The interpretation of galaxy properties in relation to their environment relies critically on how well that environment is reconstructed. Although our results show clear trends with void-centric distance, it is necessary to test their robustness against the limitations of this parametrisation. In parallel, observational biases such as flux limits, sampling incompleteness, and interlopers, can also affect the reconstruction, and to quantify their impact we compare the DR1-like lightcone with the GT catalogue (Appendix~\ref{apdx:GroundTruth}), which includes a complete tracer population and an ideal environmental reconstruction, enabling us to disentangle genuine physical trends from observational artefacts.

\subsection{\label{sc:localdensity}Parametrisation of environment}

Our analysis confirms that galaxy properties vary systematically with distance from the centre of their associated voids. However, the normalisation of this distance to the spherised radius of the void is known to be a simplification of the reality. On this matter, the local density contrast is a much more sensitive quantity to the subtle changes of the density field and has proven to be well suited in identifying galaxies that are truly isolated and far from dense regions of the cosmic web (see Sect.~\ref{sc:mergertrees}).
Motivated by this, we introduce a two-parameter framework which combines the normalised void-centric distance with the local density contrast to refine our environmental classification and trace both the local and global environment within underdense regions, similarly to previous theoretical studies based on simulated data \citep{habouzit_properties_2020, wang_cosmology_2024}. The normalised void-centric distance and the local density of a galaxy are correlated, especially within void-size spheres within the underdense regions ($d_{\rm cc}<R_{\rm v}$) and below the mean local density ($\rho<\bar{\rho}$), as evidenced in Fig.~\ref{fig:dccRv_delta}.

\begin{figure}[htbp!]
\centering
\includegraphics[angle=0,width=1.0\hsize]{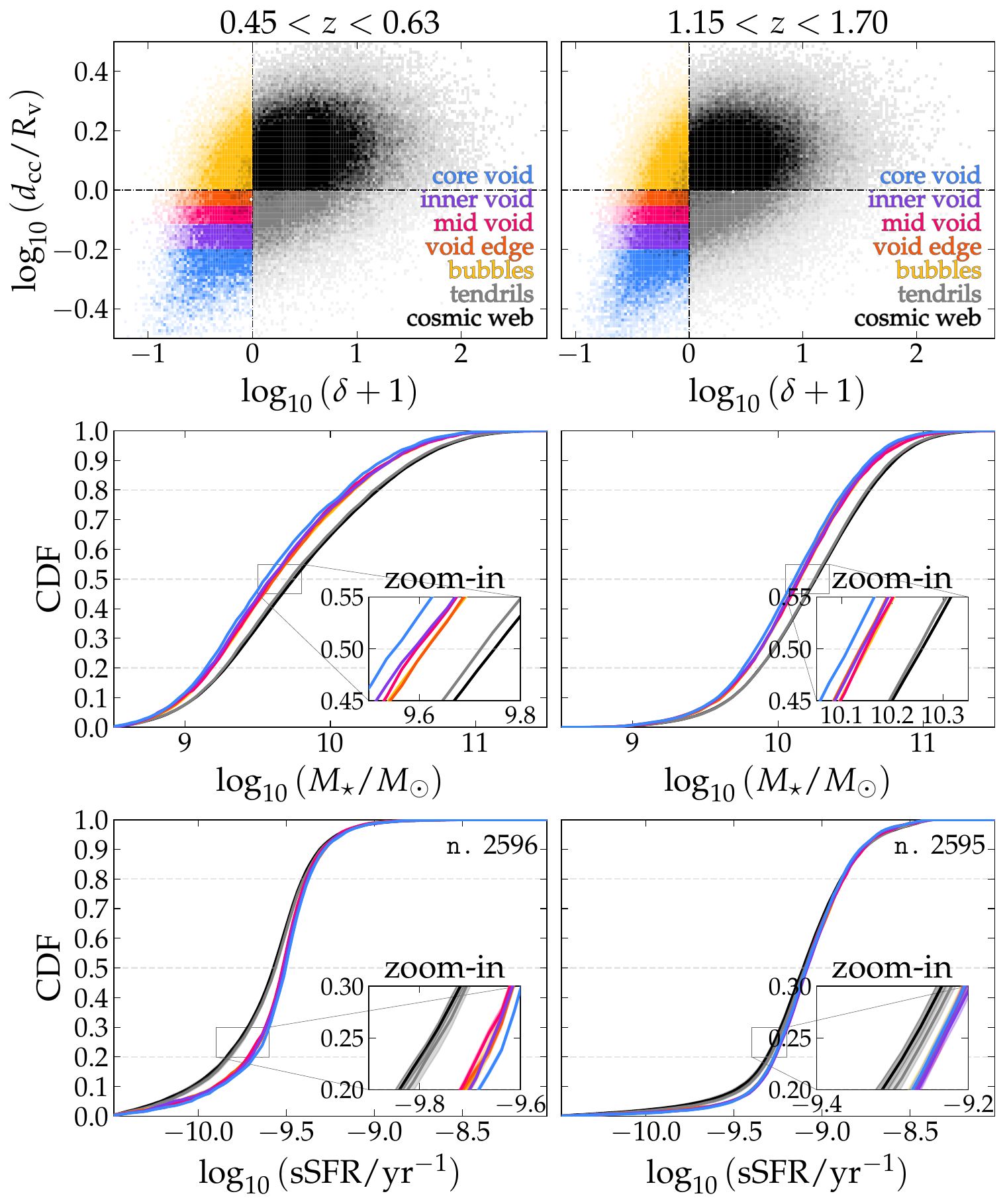}
\caption{Top row: Two-parameter environmental classification scheme based on normalised void-centric distance and local density contrast. The horizontal and vertical dashed lines denote the boundaries at $d_{\rm cc} = R_{\rm v}$ and $\rho = \bar{\rho}$, respectively. Regions of this plot are detailed in Sect.~\ref{sc:localdensity}. Middle and bottom rows: CDFs of stellar mass and sSFR, respectively, for the environmental classes defined above. Only first and last redshift bins are shown. The number of galaxies (\texttt{n.}) in each mass-matched bin is printed in the top right corner of bottom-row panels.} 
\label{fig:distCC_delta_mstar}
\end{figure}

By dividing this parameter space into four regions, we identify qualitatively distinct environments. Galaxies within $R_{\rm v}$ and with densities below the mean represent the true interior of voids. For this discussion, the inner void region is further subdivided into four equal-sized bins by normalised void-centric distance to track the transition from void core to edge. Within the same radius, however, some galaxies have local densities above the mean. These likely correspond to groups, `tendrils', or diffuse filaments that create very shallow overdensities. Conversely, some galaxies outside $R_{\rm v}$ lie in locally underdense regions, representing peripheral substructures, like low-density `bubbles' at the outskirts of the approximated sphere. This is a manifestation of the fact that voids are not spherical. Lastly, outside the radius and above the mean density are all the galaxies that belong to the denser regions of the cosmic web, such as nodes, filaments, and walls. The nomenclature introduced here is purely descriptive and serves to aid the visual interpretation of galaxy distributions within the density-distance plane (see Fig.~\ref{fig:distCC_delta_mstar}). It does not represent a formal or quantitative classification scheme, but rather a convenient framework to discuss environmental trends across different regions of the cosmic web.

This environmental classification is illustrated in Fig.~\ref{fig:distCC_delta_mstar}, where we use it to calculate cumulative distribution functions for stellar mass and sSFR across these subsamples, in the first and last redshift bin, which are representative of the full range. The CDFs of the sSFR are mass-matched across environments and redshift intervals, following the procedure adopted throughout the analysis (see Sect.~\ref{sc:galprops_stellar_mass}). The contrast between environments becomes even more pronounced under this refined parametrisation. Interestingly, galaxies located inside voids but with $\rho>\bar{\rho}$ (tendrils) exhibit properties akin to those in denser regions, while galaxies in underdense bubbles outside the void radius more closely resemble the inner void population, as evidenced by the increased difference of CDFs of underdense regions from the denser ones. 

This can be quantified by means of the KS test. Along the lines of the nomenclature defined in Sect.~\ref{sc:galprops_stellar_mass}, we performed the KS test comparing the inner void environment \Ei with bins at higher densities, in particular with newly introduced regions like bubbles (\Ev), tendrils (\Evi) and the overdense regions of the cosmic web (\Evii).
At all redshifts, the distributions of \Mstellar\ yield $p_{\rm KS} \ll 10^{-5}$ for all comparisons. 
For sSFR, distributions in \Ei and \Evii are different at a significance $p_{\rm KS} \ll 10^{-5}$ (computed as the median $p_{\rm KS}$ among the values derived from the 100 mass-matched realizations) up to $z\approx1.15$, and become $p_{\rm KS} \approx 10^{-5}$ at $z>1.15$. This is still enough to reject the null hypothesis.
The comparison between \Ei and \Evi yields $p_{\rm KS} \ll 10^{-5}$ up to $z\approx1.15$, and drops to $p_{\rm KS} \approx 10^{-3}$ at $z>1.15$. 
Lastly, we obtain $p_{\rm KS}>10^{-2}$ from the comparison of \Ei with \Ev at all redshifts.

\begin{figure}[htbp!]
    \centering
    \includegraphics[width=\linewidth]{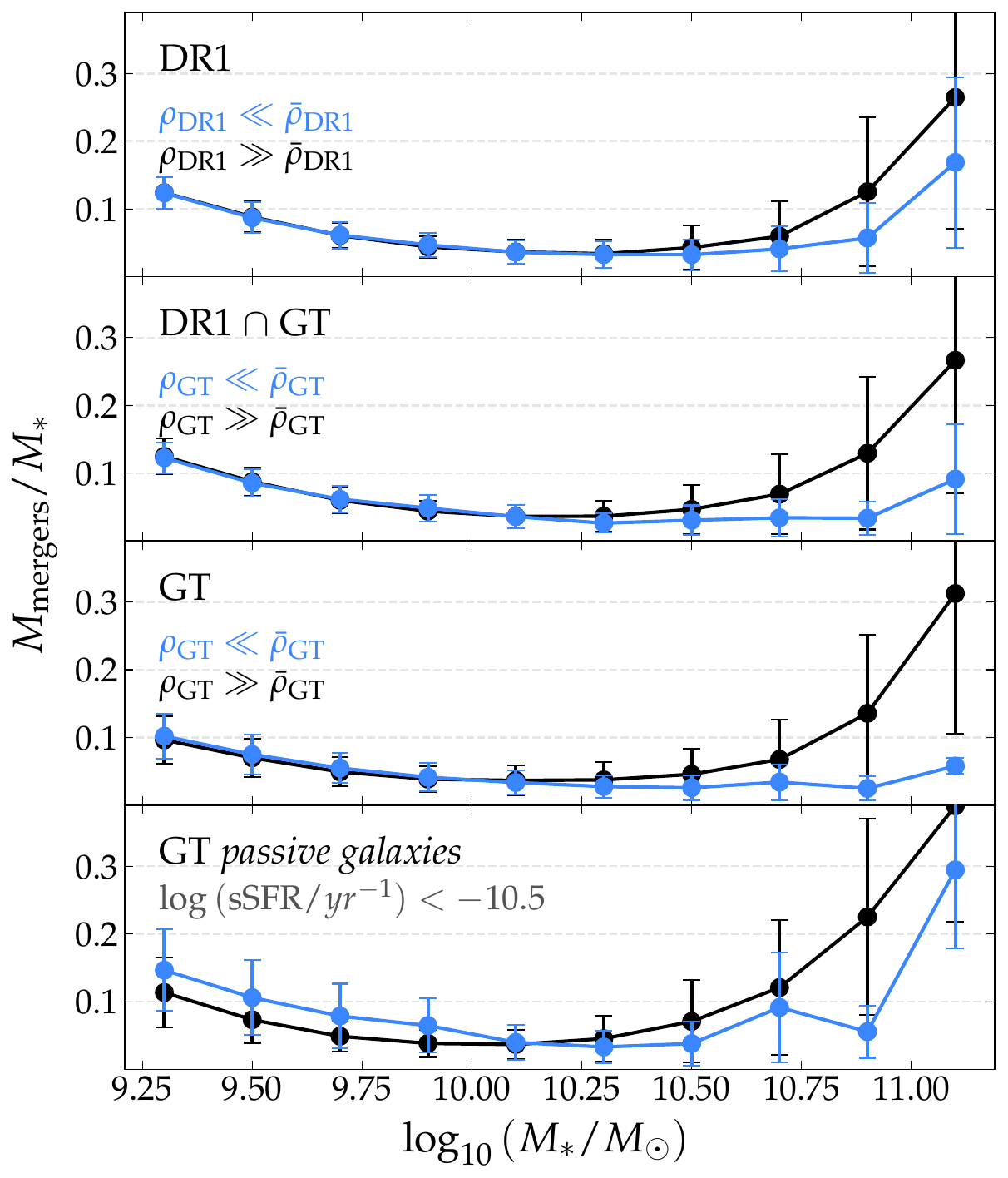}
    \caption{Median fraction of stellar mass accreted through mergers as a function of observed stellar mass in the bin of lowest redshift ($0.45<z\leq0.63$). Blue lines mark galaxies at the lowest local density, while black lines correspond to the highest densities (circa 5 times the mean density). Errors were computed as normalized median absolute deviations (NMAD). Panels were obtained from different samples and environmental reconstructions, from top to bottom: DR1-like sample with DR1 environment ($\rho_{\rm DR1}$), DR1-like sample with GT environment ($\rho_{\rm GT}$), GT sample with GT environment ($\rho_{\rm GT}$), and only passive galaxies within GT sample with GT environment ($\rho_{\rm GT}$). Dashed grey lines are added at 10\%, 20\%, and 30\% as guidelines.}
    \label{fig:mass_accreted_mergers}
\end{figure}

These findings suggest that multidimensional environmental metrics are useful to properly capture the complexity and diversity of low-density regions in the cosmic web. However, this approach comes at the cost of reduced statistical power: the number of galaxies per mass-matched subsample decreases to about 30\% of that obtained with a single-parameter metric. While the current statistics remain sufficient to demonstrate the potential of this methodology, they are still limited. Considering that the area of the GAEA lightcone corresponds only $\approx\!40\%$ of the total footprint covered by the EDFs, the forthcoming increase in sample size with future data releases will provide the statistical robustness required to fully exploit multi-parametric environmental analyses and effectively disentangle the more complex geometry in and around void edges. A key outcome of this test is the confirmation of previously reported trends, together with the demonstration that \Euclid\ DR1 will have the statistical power required to detect them.

\subsection{\label{sc:mergers_passives}The role of mergers in stellar mass growth}

Beyond stellar mass growth fuelled by star formation from accreted cold gas, galaxies can also build up their mass through mergers with other systems \citep{van_dokkum_growth_2010, kaviraj_coincidence_2011, lopez-sanjuan_dominant_2012, ferreras_constraints_2014}. The relative importance of mergers in building stellar mass depends strongly on the final stellar mass, with massive galaxies acquiring the bulk of their mass through mergers, whereas lower-mass galaxies predominantly grow via gas accretion \citep{cattaneo_how_2011}.
As demonstrated in Sect.~\ref{sc:mergertrees}, our analysis shows that galaxies in low-density environments experience mergers later on average, especially minor ones, leading to a slower and more extended assembly of stellar mass compared to galaxies in denser regions. 

However, in the DR1-like lightcone, we find that the frequency of mergers, and specifically the fraction of merger-free galaxies, shows negligible differences between environments. In other words, we find nearly as many galaxies without a merger in high-density regions as there are in void regions. We argue that this lack of environmental contrast is due to the bias introduced by the \ha-based, DR1-like selection function. To test this and better assess the role of mergers across the full galaxy population, we repeated our entire analysis using the GT catalogue, previously validated as an unbiased reference in Appendix~\ref{apdx:GroundTruth}. Applying \Revolver to the GT leads naturally to a different environmental reconstruction: we will use $\rho_{\rm GT}$ and $\rho_{\rm DR1}$ to distinguish them in this section. The GT dataset not only provides a more accurate reconstruction of the environment, since it includes all tracer particles, but also contains a broader range of galaxy types, including passive and weakly star-forming systems.

\begin{figure}[htbp!]
    \centering
    \includegraphics[angle=0,width=1.0\hsize]{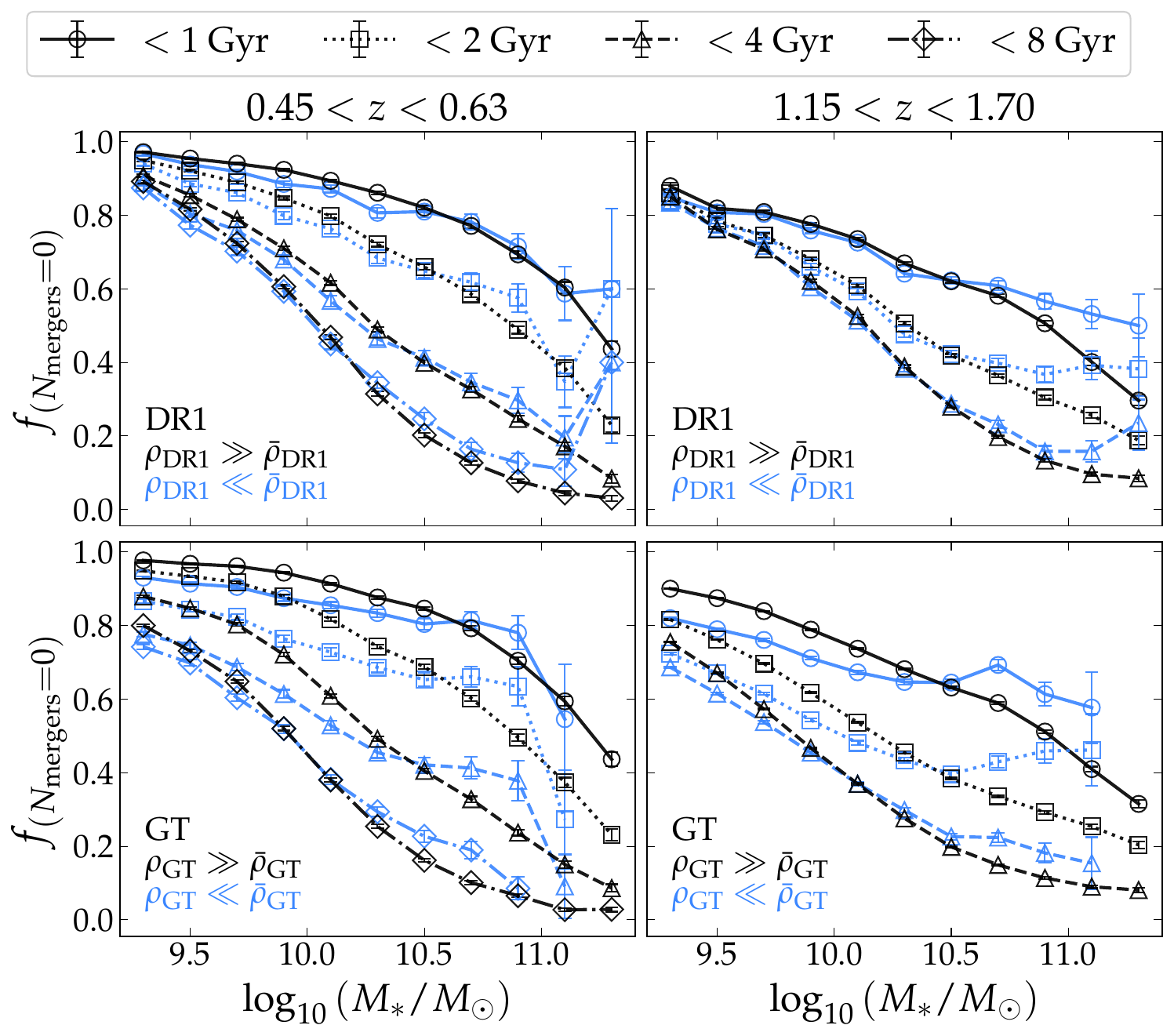}
    \caption{Fraction of merger-free galaxies with within fixed lookback-time intervals as a function of stellar mass. Columns identify first (left) and last (right) redshift bins. Colour scheme is the same as in Fig.~\ref{fig:mass_accreted_mergers}. Different line styles and markers indicate lookback intervals of 1, 2, 4, and 8\,Gyr (the latter is only calculable at the lowest redshift). Errors on fractions are obtained with the formula defined by \cite{gerke_deep2_2007}.}
    \label{fig:distCC_fmerg_eq_0_4tstep_DR1vsGT}
\end{figure}

To understand the role of mergers in the growth of galaxy stellar mass across environments, we first compute the fraction of stellar mass assembled through mergers as a function of total observed stellar mass in Fig.~\ref{fig:mass_accreted_mergers}. To isolate the impact of environment, we focus only on the two environments at the extremes of the local density contrast scale defined in Sect.~\ref{sc:mergertrees}: the lowest-density (\Ei) at $\rho \ll \bar{\rho}(z)$, and the highest-density regions (\Evi) at $\rho \gg \bar{\rho}(z)$. The figure shows four combinations of sample selection and environmental reconstruction (DR1-like vs. GT), also including a case restricted to passive galaxies only from the GT. We only display the representative result in the bin of lowest redshift. 

Regardless of sample selection, the divergence between environments emerges around $\log_{10}{(\Mstellar/M_\odot)}\!\approx\!10.5$. As expected, in both environments the impact of mergers is greater at high masses than at low masses. However, while the contribution of mergers to the total stellar mass remains below $\approx\!10\%$ for galaxies in low-density regions across all masses, it rises in denser environments, reaching $\approx\!25\%$ in the DR1-like sample and up to $\approx\!30\%$ in the GT. When considering only passive galaxies, this contribution increases further, reaching $\approx\!35\%$ at the highest stellar masses, clearly demonstrating that passives drive much of this trend when we consider the full GT sample. 

In Fig.~\ref{fig:distCC_fmerg_eq_0_4tstep_DR1vsGT}, we show the comparison of the fraction of merger-free galaxies, $f_{(N_{\rm mergers}=0)}$, within a fixed lookback-time interval $\Delta t$ as a function of stellar mass \Mstellar\ in steps of $\Delta\log_{10} {(\Mstellar/M_\odot)} = 0.2$, in bins of local density contrast and redshift for the two lightcones, DR1 and GT.  
In both lightcones and at all redshifts we observe a strong dependence on stellar mass: the merger-free fraction decreases towards higher masses, especially considering their merger history within the longest lookback interval ($<8$\,Gyr), in line with our previous result and hierarchical growth, where the assembly of massive systems is merger-driven \citep{blumenthal_formation_1984, cattaneo_how_2011, rosas-guevara_revealing_2022}.
With the inclusion of passive galaxies in the GT, an environmental inversion of the merger-free fraction trend at high stellar masses is visible across all lookback-time intervals: massive galaxies in low-density environments are significantly more likely to remain merger-free than their counterparts in dense regions. This confirms that in the regime where mergers dominate galaxy growth, the void environment indeed suppresses merger frequency. The reversal is also more evident if one considers the wider lookback-time intervals. 
This evidence demonstrates that the lack of environmental contrast in the DR1-like sample is a direct consequence of the \ha-based selection function, which targets star-forming systems, masking the underlying assembly signal typically preserved within the passive population.

Taken together, our findings paint a coherent picture. In this simulation, mergers generally play a modest role in stellar mass growth, becoming significant only for the most massive systems, particularly in high-density environments where passive galaxies dominate. For galaxies of the same final stellar mass, the lower merger contribution in voids ($\approx\!10\%$) implies that most of their stellar mass forms in situ through prolonged star formation. In these low-density environments, galaxies retain their cold gas more effectively due to the absence of ram-pressure stripping and tidal interactions \citep{kreckel_only_2011, beygu_void_2016, habouzit_properties_2020}, which allows a longer-lived conversion of gas into stars. This interpretation is consistent with early analyses of void galaxies within SDSS data, where their bluer colours and enhanced star formation are attributed to a more prolonged gas supply relative to wall galaxies, which are instead strangled and stripped of gas as they fall into clusters and groups \citep{rojas_photometric_2004, rojas_spectroscopic_2005}.

Furthermore, the specific dynamics of low-density regions facilitate a different mode of assembly. Satellites in voids experience weaker tidal fields and lower relative velocities, which extend their survival times and increase the probability of late-time mergers once a companion becomes bound. Conversely, the high velocity dispersions characteristic of dense environments can actually reduce merger efficiency \citep{ellison_galaxy_2010, deger_tidal_2018, benavides_accretion_2020, sureshkumar_galaxy_2024}. As shown in Sect.~\ref{sc:galprops_halo}, void galaxies are predominantly centrals, and the few satellites present typically inhabit lower-mass haloes (Fig.~\ref{fig:distCC_Mhalo}). These lower-mass hosts facilitate the coalescence of companions via shorter dynamical friction timescales, thus enhancing the rate of minor mergers for low-mass galaxies in voids. Consistent with recent literature, these predominantly minor, late-time mergers in low-density regions \citep{rosas-guevara_revealing_2022, rodriguez-medrano_evolutionary_2024} may help replenish gaseous reservoirs, effectively prolonging the star formation phase \citep{lin_where_2010}.

Environmental effects appear to regulate not the frequency of galaxy mergers but rather their timing and nature: in voids mergers, typically minor ones, occur later and produce a slower, prolonged evolutionary path that contrasts with the earlier, merger-driven growth characteristic of dense regions. This is consistent with findings by \cite{dominguez-gomez_galaxies_2023}, who identified a similar delayed evolutionary channel from the star formation histories of galaxies in local voids ($z\!<\!0.05$). By identifying this same environmental dependence at higher redshifts, our study suggests that the mechanisms shaping the delayed evolution of void galaxies were established early in cosmic history and have persisted across time.

\section{\label{sc:Conclusions}Conclusions}

In this work, we have used \Euclid-like mock lightcones based on the GAEA SAM to study the properties and merger histories of galaxies within cosmic voids, the underdense regions of the cosmic web. 
Most importantly, our framework demonstrates that the \Euclid\ Deep sample will enable the robust identification of voids and the characterisation of their galaxy populations across cosmic time. By mimicking the \Euclid\ DR1 observational biases, and by verifying the robustness of our environment parametrisation by means of an unbiased catalogue, we tested that the observational selection of \Euclid\ galaxies is not affecting this kind of analysis.
From a methodological perspective, we emphasise that while the quantitative strength of environmental signatures depends on the chosen parametrisation of void interiors, the qualitative trends remain robust. 
By parametrising environment through the normalised void-centric distance, we quantified how stellar mass, star formation activity, morphology, and halo mass vary across different regions of the large-scale structure, as well as the assembly history of void galaxies. Our main findings are as follows.

\begin{enumerate}[label=(\roman*)]
    \item Galaxies closer to void centres ($d_{\rm cc}/R_{\rm v} \lesssim 0.6$) are systematically less massive than those in denser environments, with the fraction of massive systems [$\log_{10}{(\Mstellar/M_\odot)}>10.5$] increasing from $\approx\!20\%$ in void interiors to $\approx\!30\%$ in dense regions. The difference is highly significant ($p_{\rm KS}\!\ll\!10^{-5}$) but small in amplitude, indicating that while mass segregation with environment is measurable, it is subtle.
    \item While matching their stellar mass distributions, galaxies in environments closer to void centres exhibit higher sSFRs than those in denser environments at all redshifts, showing enhanced star formation activity in voids. The environmental separation decreases with redshift, becoming statistically weak at $z\!\gtrsim\!1.15$, implying that environmental regulation of star formation strengthens only after cosmic structures mature. 
    \item As expected, most galaxies in our \ha-selected sample are disc-dominated, with bulge-to-total stellar mass ratios $<0.2$, with fractions of 80\% to 90\% of these galaxies from low to high redshift, independently of environment. Nonetheless, galaxies closer to void centres have lower B/T ratios than those in denser regions. The distinction weakens toward higher redshift but remains significant up to $z\!\approx\!1.15$, indicating that morphological transformations become more common in denser environments over time.
    \item The fraction of centrals increases towards lower redshifts and at smaller distances from void centres, with regions closest to void centres hosting between 70\% to 80\% centrals at low and high redshift, respectively. Among satellite galaxies that are found close to void centres, they are found to belong to less massive dark matter haloes than counterparts at higher densities. 
    \item Galaxies in low-density regions assemble their stellar mass more slowly, reaching 50\% and 90\% of their observed stellar mass about 0.5--1 Gyr later than their denser counterparts. This delay is statistically significant ($p_{\rm KS}\!\ll\!10^{-5}$) especially at low redshift, supporting a coherent picture of gradual, extended growth of stellar mass in underdense regions.
    \item The timing of galaxy interactions is environment-dependent. While major mergers occur early in every environment, minor and mini mergers occur more recently in low-density regions. This behaviour persists up to $z\!\approx\!1.15$ at high significance ($p_{\rm KS}<10^{-5}$), confirming that delayed minor interactions dominate void galaxies evolution.
    \item The contribution of mergers to stellar mass growth is strongly regulated by the environment, becoming a primary assembly channel only for massive systems [$\log_{10}(\Mstellar/M_\odot) \gtrsim 10.5$] in high-density regions. In voids, the stellar mass fraction accreted through mergers remains below $\approx\!10\%$ across all mass regimes, whereas it reaches $\approx\!30\%$ (and up to $35\%$ for passive systems) in dense environments. This is confirmed by the analysis of the GT lightcone, which reveals that massive void galaxies are significantly more likely to remain ``merger-free'' over long lookback-time intervals. Instead, these systems follow an evolutionary path dominated by sustained in situ star formation, permitted by the lack of external quenching processes in underdense regions.
\end{enumerate}

We stress that points (i)--(iii) involve quantities that will be directly measurable with \Euclid\ DR1 (stellar mass, sSFR, and morphology), whereas points (iv)--(vii) are primarily interpretative, complementing the overall picture by describing the physical evolution of void galaxies that underlies the properties that will actually be observed in the survey.
We conclude by highlighting that, while normalised void-centric distance provides a useful first-order description, incorporating the local density contrast mitigates some of the limitations of the spherical approximation. We argue that a multi-parametric approach is more robust, because while the two parameters are already good discerners of underdense environments, together they provide a more robust diagnostic, that is nonetheless limited by statistics. These considerations will be crucial for the analysis of the forthcoming \Euclid Deep data release, which will provide the opportunity to test these predictions on a statistically powerful data set combining depth, large area, and a wide redshift range within a optical-near-infrared imaging and spectroscopy survey.
Our framework demonstrates that the observational selection of \Euclid will not bias the reconstruction of the environment, indeed it will enable the robust identification of voids and the characterisation of their galaxy populations across cosmic time. 

\begin{acknowledgements}

\AckEC

GP, OC, and MB acknowledge the support and hospitality of the Institute for Fundamental Physics of the Universe - IFPU, that has been instrumental for to the development of this work, under the Focus Week Program ``Euclid Voids for Galaxy Evolution'' held on 25-29 November 2024. OC, GP, MM acknowledge support from the INAF mini-grant 2023 ``LION: Looking for the Imprint of Overdensity Networks''. MB acknowledges the support from INAF Minigrant 2023 ``ADIEU''. 
\end{acknowledgements}

\bibliography{bibliography} 

\begin{appendix}
    
    \section{Alternative methods of finding voids\label{apdx:VoidFinders}}
    
    Cosmic voids lack a universally agreed definition, beyond being large depressions of the density field. Different void-finding algorithms might recover voids with very different properties (sizes, shapes, galaxy membership...), based on the different working assumptions of the codes themselves. 
    Nonetheless, it is important to check that the results are qualitatively consistent across finders in order to confirm that observed trends in galaxy properties reflect genuine features of the underdense cosmic environment, rather than artefacts of a specific void-identification technique.
 
\begin{figure*}[htbp!]
        \centering
        \includegraphics[width=1\linewidth]{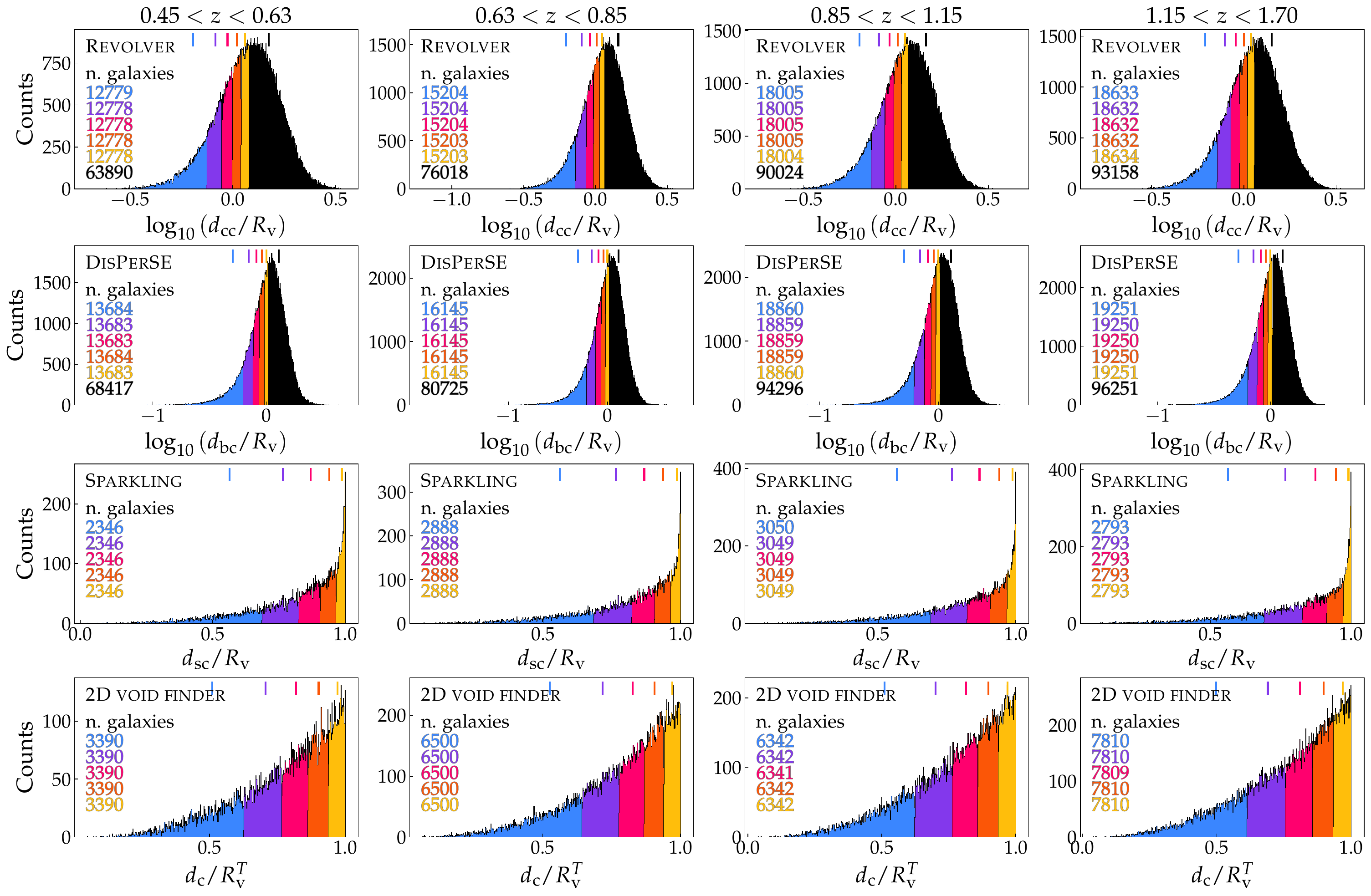 }
        \caption{Distribution of normalised void-centric distances, for galaxies identified with 4 different void-finding algorithms: from top to bottom row, \Revolver, \Disperse, \Sparkling, \twodvf. Parametrisation follows from what described in the main text (see Sect.~\ref{sc:envpars}): coloured histograms represent galaxy populations progressively closer to the void centre, while the black histogram corresponds to the control sample at $d_{\mathrm{cc}} / R_{\rm v} > 1$. The control sample is only available for algorithms that provide parametrisation beyond the spherical boundaries of the void. Vertical tick marks at the top of each panel indicate the position of the median normalised distance in each bin. Columns represent the four redshift bins utilised throughout the analysis. }
        \label{fig:voidfinders_voidcentricdistance_hist}
    \end{figure*}

    In this work we have adopted \Revolver, whose tessellation- and watershed-based approach provides a robust and reliable reconstruction of voids, faithfully tracing the irregularities of the density field (see Sect.~\ref{sc:VoidFinder}). However, \Revolver\ was not the only finder we tested. Since our aim was to identify the method best suited for environmental studies of galaxy evolution, we explored alternative finders with different dimensionality (2D vs. 3D) and geometric assumptions (spherical vs. non-spherical). Below we review our results with three alternative finders, and compare that the main statistical trends in void properties and galaxy populations are generally consistent. 
    In Fig.~\ref{fig:voidfinders_voidcentricdistance_hist}, we show the distributions of the void-centric distance normalised by void size for galaxies identified by the four different void-finding methods. Figures~\ref{fig:voidfinders_mass_CDF} and \ref{fig:voidfinders_ssfr_CDF} then show the corresponding CDFs of stellar mass and sSFR across these environmental reconstructions, allowing a direct comparison of the trends recovered by each algorithm.

    \subsection{\label{ssc:DisPerSE}\Disperse}
    
    \Disperse (Discrete Persistent Structures Extractor; \citealt{Sousbie_2011_I, Sousbie_2011_II}) is an open-source code designed for the automatic identification of topological structures, including voids, walls, filaments, and nodes. For a discrete input data set, it first computes the Delaunay tessellation of the input point set and then applies discrete Morse theory together with persistence analysis to identify statistically significant topological features, without requiring prior smoothing or fine-tuned parameters. The concept of persistence allows \Disperse to filter out features that are topologically less robust compared to the noise present in the input data set. This is controlled by the $N_{\sigma}$ parameter: increasing its value means that topologically more and more robust structures, with respect to the noise, will be retained.
    This makes it particularly well suited for mapping the full cosmic web. Indeed, it has already been tested on \Euclid-like mock catalogues for the reconstruction of the entire cosmic web (e.g. \citealt{EP-Kraljic, EP-Malavasi}), as well as on the first quick data release of the mission (\citealt{Q1-SP005, Q1-SP028}).
    For our purposes, it was informative to compare \Disperse\ with \Revolver, since both are tessellation-based. Yet, \Revolver\ was developed specifically for void detection, whereas in \Disperse\ voids emerge as a by-product of the global cosmic web reconstruction.

    After applying \Disperse\ in the configuration with persistence threshold $N_{\sigma}=3$ (standardly applied in this type of analysis) to our DR1-like lightcone, where galaxies are weighted by their stellar mass, we post-processed its void catalogue using the same analysis pipeline as for \Revolver (see Sect.~\ref{sc:envpars}). 3D structures identified as voids by \Disperse are made of collections of Delaunay tetrahedra. We computed the centre of each void as a barycentre, where each tetrahedron contributes to the barycentre with a weight equal to its volume. The void volume then is simply the sum of the volumes of all constituent tetrahedra. Galaxies are assigned to a void if they belong to the geometrical space composed of the ensemble of tetrahedra.
    Once galaxy membership is established, we computed the void-centric distance of each galaxy relative to the barycentre and normalised it by the spherical-equivalent radius of the void.
    
    \subsection{\label{ssc:Sparkling}\Sparkling}

    \Sparkling is a computational software designed to identify spherical cosmic voids in numerical $N$-body simulations and galaxy surveys \citep{ruiz_clues_2015, ruiz_structure_2019}. The method first estimates the density field via a Voronoi tessellation of the tracer distribution, which is only used to select underdense cells as initial candidates. Around each candidate, spheres are iteratively expanded until they reach a fixed integrated density contrast $\Delta$ set by the user, defining in this way a well-defined void radius. To refine void centres, the algorithm tests random displacements, retaining those centroids that yield the largest expansion. Finally, overlapping spheres are removed, keeping the largest candidates to build a clean and non-overlapping void catalogue. The algorithm has been widely employed to study void profiles, both in context of cosmology and galaxy evolution \citep{ceccarelli_sparkling_2016, lambas_sparkling_2016, lares_sparkling_2017, correa_redshift-space_2021, correa_redshift-space_2022, curtis_properties_2024, rodriguez-medrano_local_2023, rodriguez-medrano_traces_2025}. As \Sparkling operates on an integrated density contrast basis, it ensures that both the void interiors and their surroundings represent a large-scale underdense environment. For this reason, it was interesting to compare the interior of these spherical voids with what we found for tessellation based finders.
    
    For our analysis, we applied \Sparkling with a fixed integrated density contrast $\Delta = -0.7$. Galaxies were associated to voids if they lay geometrically inside the spherical volume of a detected void. As in Sect.~\ref{sc:envpars}, we parametrised the environment using void-centric distance. However, in this case the distribution was divided into five equal-sized bins that all fall within the void radius, by construction of the algorithm's spherical geometry. Unlike in tessellation-based finders, this approach does not naturally provide a control sample at distances beyond the void edge.
    
    \subsection{\label{ssc:2Dvoidfinder}\twodvf}
    
    \cite{sanchez_cosmic_2017} introduced a void-finding algorithm specifically tailored for photometric surveys. The method projects galaxies into two-dimensional redshift slices and identifies voids within the smoothed 2D density field of each slice. By setting the slice thickness to at least twice the photometric redshift scatter (e.g. $\approx100\,\hMpc$), the algorithm mitigates line-of-sight position errors in photometric data. Tests with simulations have shown that this approach can recover void statistics consistent with spectroscopic samples, particularly for larger voids.
    It is interesting to include such a finder in our comparison, since unlike tessellation-based algorithms that rely on precise 3D information, this method requires only photometric redshifts and could result suitable for work on \Euclid photometric catalogues, as well as surveys with complex geometries, internal gaps, or limited declination coverage, which can be challenging for contiguous volume-based 3D finders.

    For our analysis, we applied \twodvf\ to the DR1-like lightcone. Due to its 2D nature, the algorithm measures a transverse radius for each void, while the radius along the line of sight is fixed at $50\,\hMpc$ (half the slice thickness). Given the density of our galaxy sample, the recovered voids are relatively small, with no transverse radii above $50\,\hMpc$. We therefore approximated the transverse radius as a spherical radius, associating galaxies to voids within spheres of that size, centred at the redshift of the slice. 
    
    \subsection{\label{ssc:comparison}Comparing finders}

     \begin{figure*}[htbp!]
        \centering
        \includegraphics[width=0.99\linewidth]{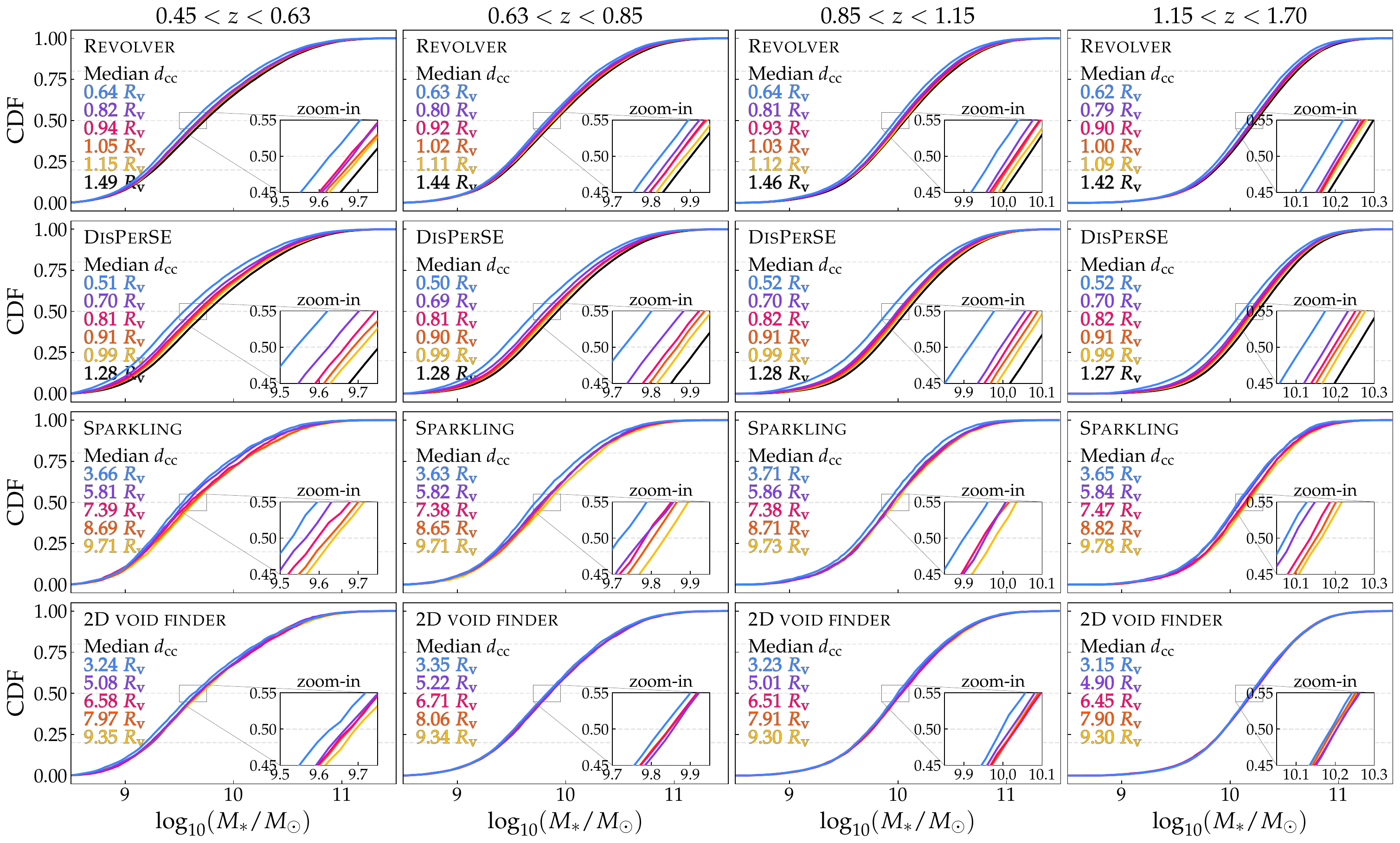}
        \caption{CDFs of stellar mass for the same environmental bins as defined in Fig.~\ref{fig:voidfinders_voidcentricdistance_hist}. Rows identify the finders, columns identify the redshift bin. Median $d_{\mathrm{cc}}$ values in each bin are expressed in $R_{\rm v}$ units and indicated in the legend.}
        \label{fig:voidfinders_mass_CDF}
    \end{figure*}
    
    \begin{figure*}[htbp!]
        \centering
        \includegraphics[width=0.99\linewidth]{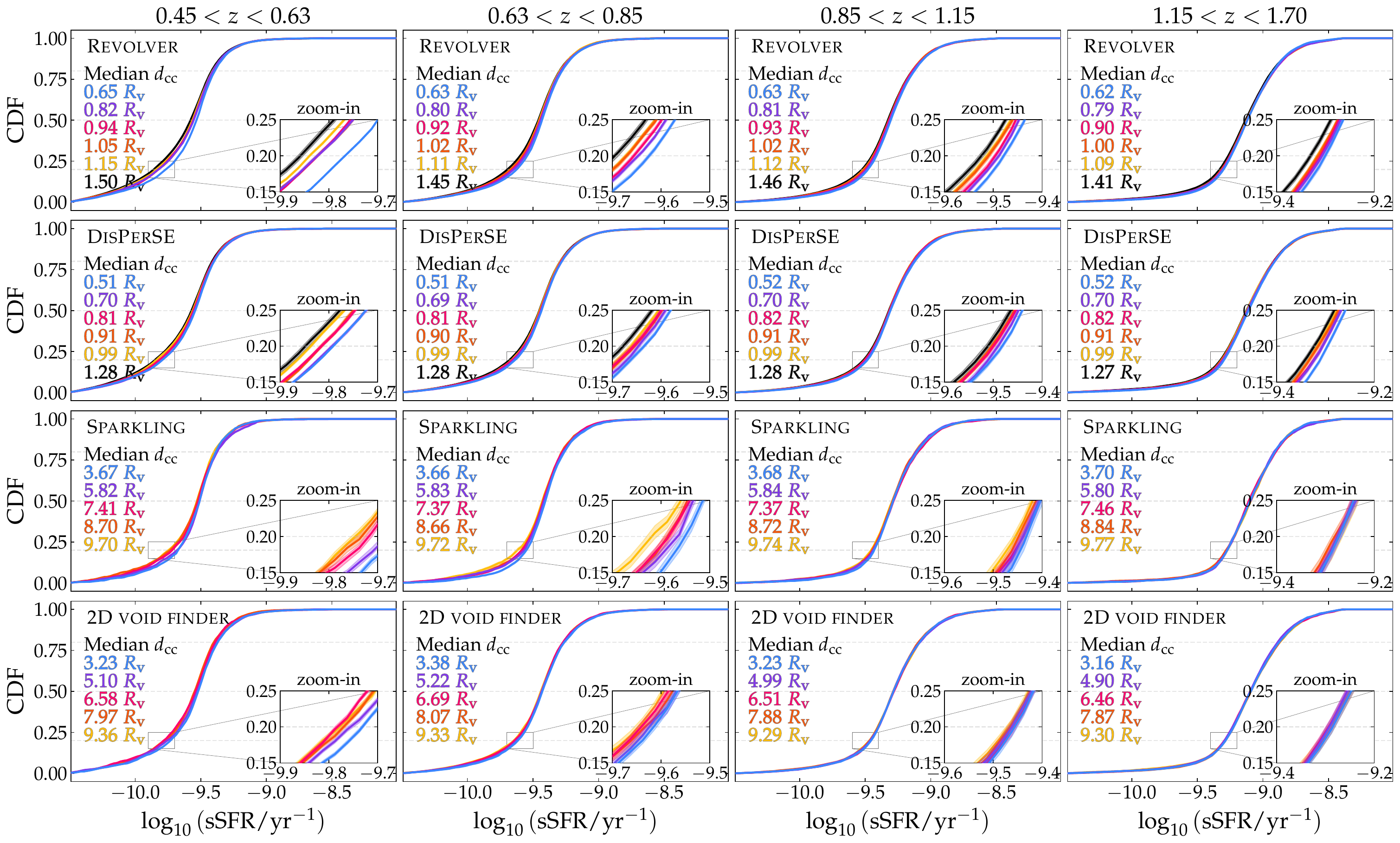}
        \caption{CDFs of sSFR for the same environmental bins as defined in Fig.~\ref{fig:voidfinders_voidcentricdistance_hist}. Rows identify the finders, columns identify the redshift bin. Stellar mass distributions have been matched across all environment-redshift bins, as outlined in Fig.~\ref{fig:mass_match}. The CDFs were computed over 100 realisations of the mass-matching procedure; the shaded area represents their 1\textsigma\ dispersion.}
        \label{fig:voidfinders_ssfr_CDF}
    \end{figure*}

    After showing in Fig.~\ref{fig:voidfinders_voidcentricdistance_hist} the distributions of our chosen parameter for environmental classification, the normalised void-centric distance, Figs.~\ref{fig:voidfinders_mass_CDF} and \ref{fig:voidfinders_ssfr_CDF} present the cumulative distribution functions (CDFs) of stellar mass and the mass-matched sSFR for all void finders tested, compared with our reference reconstruction from \Revolver.  
    Overall, the same qualitative trends with environment and redshift are recovered across all methods: galaxies located closer to void centres tend to be less massive and more actively star-forming.  
    In the case of \Disperse, the stellar-mass segregation with environment appears slightly more pronounced than in \Revolver. This difference arises because the configuration used for the topological reconstruction was mass-weighted, enhancing the contrast between dense and underdense regions. However, after mass-matching the samples, the environmental trend in sSFR remains clearly visible, confirming that the correlation seen with \Revolver is physically meaningful rather than an artefact of the tessellation scheme.
    The spherical algorithm \Sparkling\ also yields results in excellent agreement with the tessellation-based finders, reproducing the same trends in both stellar mass and sSFR, providing an independent cross-check of our conclusions. This consistency indicates that the observed dependence of galaxy properties on void-centric distance is not tied to the geometric assumptions of a particular finder.
    For the \twodvf, the signal is weaker, as expected given that projection effects and limited line-of-sight resolution dilute environmental contrasts. Nonetheless, the fact that even this photometric, two-dimensional method recovers the same qualitative trend is highly encouraging.  
    Taken together, the agreement among tessellation-based (\Revolver, \Disperse), spherical (\Sparkling), and 2D photometric (\twodvf) approaches demonstrates that the observed dependence of stellar mass and star formation activity on environment is a robust and genuine imprint of the underdense large-scale structure, rather than an artefact of any specific void-identification technique.

    \newpage
    \section{\label{apdx:GroundTruth}Testing reconstruction on a lightcone without observational biases}

    In this work, we have studied the reconstruction of the void environment in the EDFs at the epoch of the first full public data release (DR1), identifying clear trends in galaxy properties as a function of their distance from the centre of the associated void.
    However, the sample used for this analysis does present limitations in terms of completeness and purity. First, the selection is based on the assumption that galaxy spectra have an \ha\ emission line and the sample is flux-limited (see Sect.~\ref{sc:Euclid-like}). This introduces a redshift-dependent bias: galaxies with fainter \ha\ luminosity are only detected at low redshift, while brighter galaxies become progressively over-represented at higher redshift.
    Additionally, the completeness reaches the nominal value of 60\% at the flux limit of DR1, which means that the completeness of the sample is low and dependent on the flux sensitivity of the survey.
    Lastly, the presence of interlopers also degrades the purity of the sample, by artificially `filling in' underdensities in the galaxy distribution and potentially masking or weakening the signal of cosmic voids.
    To assess the robustness of our void identification under these observational constraints, we introduce a reference catalogue, referred to as the `Ground Truth' (GT). This consists of the full lightcone based on the model presented in \cite{de_lucia_tracing_2024}, to which we only applied a cut on stellar mass, $\Mstellar \geq 10^{9.2}\,M_\odot$. No observational effects were applied. This catalogue provides an idealised view of the underlying galaxy distribution. We apply the code \Revolver to the GT, obtain a catalogue of fiducial voids with their properties, and we estimate $d_{\rm cc}/R_{\rm v}$ for all the galaxies.
    
\begin{figure}[htbp!]
    \centering
    \includegraphics[angle=0,width=\hsize]{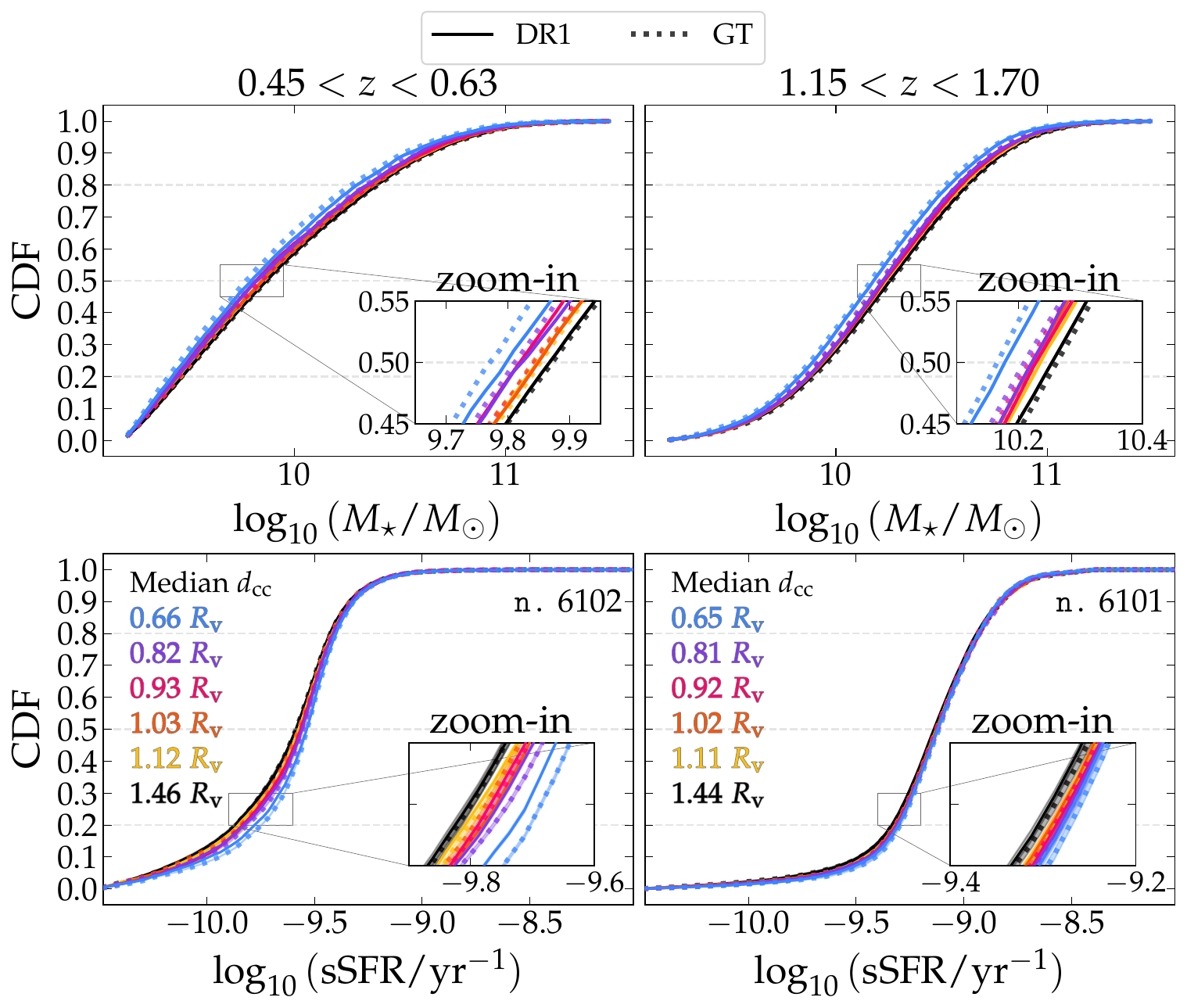}
    \caption{Cumulative distribution functions of \Mstellar\ (top) and sSFR (bottom) in DR1-like and GT catalogues. Columns identify first (left) and last (right) redshift bins of the lightcone. In every panel, solid lines identify DR1 while dotted lines stand for GT. The environmental binning strategy follows the same procedure described in Sect.~\ref{sc:envpars}. The galaxy sample is restricted to objects present in the DR1 catalogue, ensuring that the same set of galaxies is analysed under both environmental reconstructions. The number of galaxies (\texttt{n.}) in each mass-matched bin is printed in the top right corner of bottom-row panels.} 
    \label{fig:distCC_mstar_ssfr_DR1vsGT}
    \end{figure}

    In Fig.~\ref{fig:distCC_mstar_ssfr_DR1vsGT}, we compare CDFs of two key galaxy properties -- stellar mass and sSFR -- using environmental bins defined in the DR1 lightcone and the GT. The environmental binning strategy follows the same procedure described in Sect.~\ref{sc:envpars}. The galaxy sample used in the comparison plot is restricted to objects present in the DR1 catalogue, ensuring that the same set of galaxies is analysed under both environmental reconstructions. In this way, we explicitly test the robustness of the environment reconstruction. For each galaxy, the void-centric distance is computed relative to the centre of its associated void, as identified independently in the DR1 and GT catalogues.
    We find that the trends observed in the GT are qualitatively recovered in DR1. In the GT, the separation between environmental bins is more pronounced, suggesting that a pure and complete sampling leads to a more precise reconstruction of the density field. Nonetheless, the persistence of environmental trends in the DR1 lightcone confirms that, despite the presence of interlopers, flux limitations, and incomplete sampling, our methodology remains effective at identifying the underlying structure of cosmic voids.
    
    \section{\label{apdx:Observational_Uncertanties}Estimated effect of observational uncertainties}
    
\begin{figure*}[htbp!]
        \centering
        \includegraphics[width=1\linewidth]{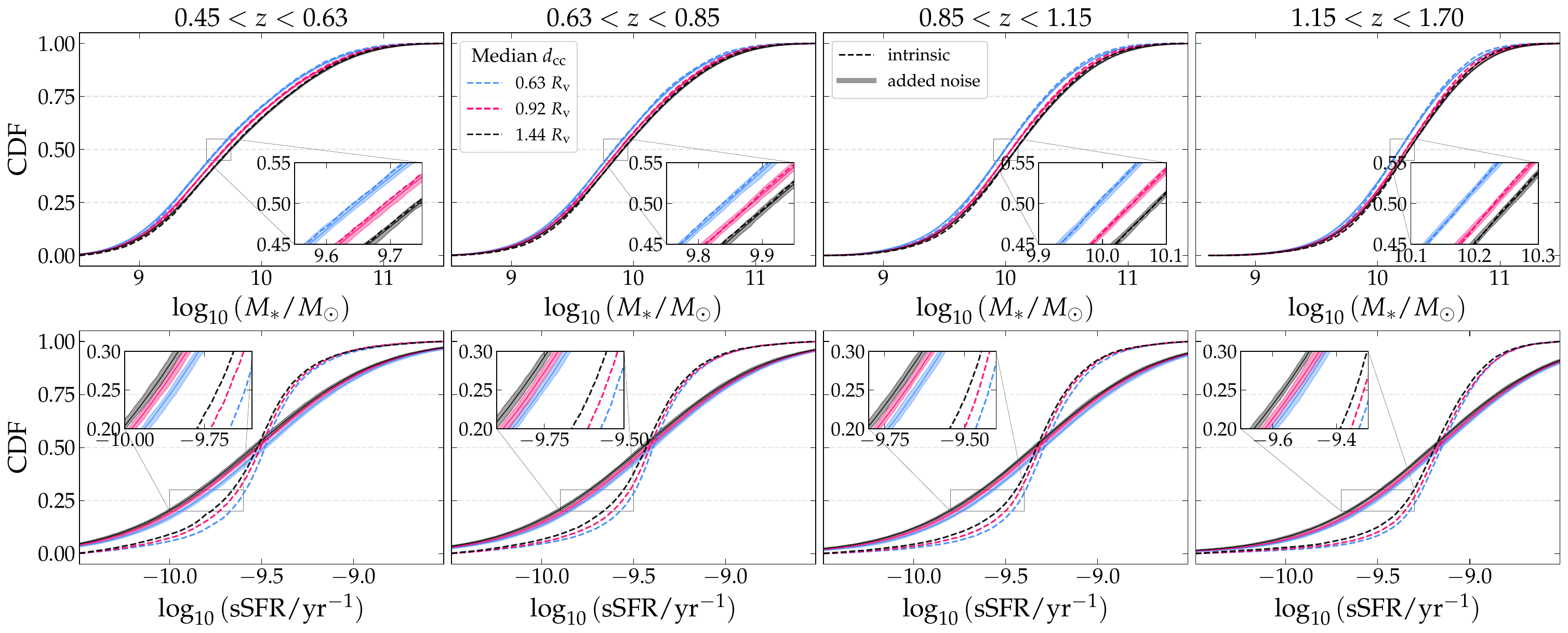}
        \caption{Impact of observational uncertainties on the stellar mass and sSFR distributions in different environments. Columns correspond to the four redshift bins analysed in the main text. Rows show the CDFs of \Mstellar\ (top) and sSFR (bottom) for the three environments used in the KS tests: \Ei, \Eiii, and \Evi. In each panel, we display the intrinsic CDFs (dashed lines) together with the 3\textsigma\ dispersion of 100 realisations obtained after adding observational uncertainties, 0.15\,dex for stellar mass and 0.45\,dex for sSFR (shaded areas).}
        \label{fig:obsvuncertanties}
        \label{LastPage}        
    \end{figure*}

    We tested the robustness of the environmental trends presented in Sect.~\ref{sc:galprops} by introducing observationally motivated uncertainties to mimic those expected in DR1 data, specifically on physical galaxy properties such as stellar mass and star formation rate. 
    We adopted uncertainty estimates consistent with those anticipated for DR1, based on SED-fitting tests performed on 4402 DESI galaxies at their spectroscopic redshift (\citealt{EP-Enia}, private communication). Following these results, we applied a Gaussian scatter of 0.15\,dex to stellar masses and 0.45\,dex to sSFR.
    Figure~\ref{fig:obsvuncertanties} summarises these tests, showing both the intrinsic CDFs and those obtained after adding uncertainties (for 10 representative realisations per property), across all redshift bins.
    
    For stellar mass, we added 0.15 dex of Gaussian scatter and recomputed the stellar-mass CDFs for all environments, alongside KS tests between the density regimes discussed in the main text (\Ei--\Eiii and \Ei--\Evi). In all redshift bins, the distributions remain clearly distinct, with $p_{\rm KS}\!\ll\!10^{-5}$, indicating that the mass-environment segregation predicted by the simulation is robust to the level of noise expected in DR1.
    
    For sSFR, we applied the larger uncertainty of 0.45\,dex. As anticipated, the distributions get smoothed out and broadened (visible in the decreased steepness of each CDF), and once mass-matching is combined with this level of noise, the separation between environments decreases, particularly at higher redshift, where the intrinsic trends are already weaker. Within the low-density regime (\Ei--\Eiii), differences are no longer statistically separable (with $p_{\rm KS}\approx10^{-3}$), but low-density and high-density regions remain distinguishable at all redshifts, with $p_{\rm KS}\!\ll\!10^{-5}$ for \Ei versus \Evi.
    
    Overall, these tests show that the qualitative environmental trends discussed in the main text are robust to realistic observational uncertainties. Even under conservative assumptions, the contrast between void interiors ($d_{\rm cc} < 0.7 R_{\rm v}$) and overdense regions persists, suggesting that the key signatures of environmental regulation identified in this work should remain detectable in forthcoming DR1 data.
    
    \newpage
    \section{\label{apdx:KS_test_results}KS-test results}
    
    In this section, we provide the complete results of the Kolmogorov--Smirnov (KS) tests associated with the analysis presented in Sects.~\ref{sc:galprops} and~\ref{sc:mergertrees}. Table~\ref{tab:galprops} summarises the results for galaxy properties in Sect.~\ref{sc:galprops} parametrised by the normalised void-centric distance ($d_{\rm cc}/R_{\rm v}$), while Table~\ref{tab:mergerprops} is focused on the environmental dependence of mass assembly and merger histories based on the local density contrast $(1+\delta)$ (see  Sect.~\ref{sc:mergertrees}). 

    \begin{table}[htbp!]
    \centering
    \caption{KS-test p-values for galaxy properties. The environments (\textcolor{voids}{\Ei}, \textcolor{walls}{\Eiii}, \Evi) are defined with respect to $d_{\rm cc}/R_{\rm v}$. Cell colour coding indicates significance levels: green for $\leq 10^{-5}$, yellow for $10^{-5} < p \leq 10^{-3}$, orange for $10^{-3} < p \leq 0.05$, and red for $> 0.05$. All reported values are median p-values derived from 100 random realisations of the stellar mass-matching procedure. The exception is \Mstellar, for which the p-value refers to the single test performed on the intrinsic distribution.}
    \label{tab:galprops}
    {
    \setlength{\tabcolsep}{3pt}
    \begin{tabular}{lcccc}
        \hline
        \noalign{\vskip 1pt}
         & $0.45-0.63$ & $0.63-0.85$ & $0.85-1.15$ & $1.15-1.70$ \\
        \noalign{\vskip 1pt}
        \hline
        \noalign{\vskip 1pt}
        \multicolumn{5}{l}{Property: \Mstellar} \\
        \textcolor{voids}{\Ei}-\textcolor{walls}{\Eiii} & \cellcolor{green!30}$6.7\times10^{-8}$ & \cellcolor{green!30}$8.5\times10^{-9}$ & \cellcolor{green!30}$5.4\times10^{-18}$ & \cellcolor{green!30}$2.1\times10^{-21}$ \\
        \textcolor{voids}{\Ei}-\Evi & \cellcolor{green!30}$5.1\times10^{-46}$ & \cellcolor{green!30}$1.7\times10^{-39}$ & \cellcolor{green!30}$3.8\times10^{-69}$ & \cellcolor{green!30}$4.8\times10^{-68}$ \\
        \noalign{\vskip 1pt}
        \hline
        \noalign{\vskip 1pt}
        \multicolumn{5}{l}{Property: sSFR} \\
        \textcolor{voids}{\Ei}-\textcolor{walls}{\Eiii} & \cellcolor{green!30}$9.8\times10^{-12}$ & \cellcolor{green!30}$9.0\times10^{-6}$ & \cellcolor{yellow!40}$5.2\times10^{-5}$ & \cellcolor{red!30}$2.0\times10^{-1}$ \\
        \textcolor{voids}{\Ei}-\Evi & \cellcolor{green!30}$1.6\times10^{-35}$ & \cellcolor{green!30}$7.2\times10^{-26}$ & \cellcolor{green!30}$4.4\times10^{-15}$ & \cellcolor{green!30}$6.8\times10^{-8}$ \\
        \noalign{\vskip 1pt}
        \hline
        \noalign{\vskip 1pt}
        \multicolumn{5}{l}{Property: B/T} \\
        \textcolor{voids}{\Ei}-\textcolor{walls}{\Eiii} & \cellcolor{orange!40}$2.2\times10^{-3}$ & \cellcolor{red!30}$2.6\times10^{-1}$ & \cellcolor{orange!40}$1.4\times10^{-2}$ & \cellcolor{red!30}$6.0\times10^{-1}$ \\
        \textcolor{voids}{\Ei}-\Evi & \cellcolor{green!30}$6.3\times10^{-9}$ & \cellcolor{green!30}$6.8\times10^{-6}$ & \cellcolor{yellow!40}$1.8\times10^{-4}$ & \cellcolor{yellow!40}$9.1\times10^{-4}$ \\
        \noalign{\vskip 1pt}
        \hline
        \noalign{\vskip 1pt}
        \multicolumn{5}{l}{Property: $M_{\rm halo}$ (all galaxies)} \\
        \textcolor{voids}{\Ei}-\textcolor{walls}{\Eiii} & \cellcolor{green!30}$1.5\times10^{-21}$ & \cellcolor{green!30}$8.7\times10^{-16}$ & \cellcolor{green!30}$4.1\times10^{-17}$ & \cellcolor{green!30}$6.5\times10^{-10}$ \\
        \textcolor{voids}{\Ei}-\Evi & \cellcolor{green!30}$2.0\times10^{-86}$ & \cellcolor{green!30}$7.3\times10^{-76}$ & \cellcolor{green!30}$2.2\times10^{-63}$ & \cellcolor{green!30}$3.0\times10^{-39}$ \\
        \noalign{\vskip 1pt}
        \hline
        \noalign{\vskip 1pt}
        \multicolumn{5}{l}{Property: $M_{\rm halo}$ (centrals)} \\
        \textcolor{voids}{\Ei}-\textcolor{walls}{\Eiii} & \cellcolor{red!30}$6.0\times10^{-1}$ & \cellcolor{red!30}$6.0\times10^{-1}$ & \cellcolor{red!30}$7.8\times10^{-1}$ & \cellcolor{red!30}$3.3\times10^{-1}$ \\
        \textcolor{voids}{\Ei}-\Evi & \cellcolor{red!30}$2.7\times10^{-1}$ & \cellcolor{red!30}$2.6\times10^{-1}$ & \cellcolor{red!30}$4.1\times10^{-1}$ & \cellcolor{red!30}$2.7\times10^{-1}$ \\
        \noalign{\vskip 1pt}
        \hline
        \noalign{\vskip 1pt}
        \multicolumn{5}{l}{Property: $M_{\rm halo}$ (satellites)} \\
        \textcolor{voids}{\Ei}-\textcolor{walls}{\Eiii} & \cellcolor{green!30}$5.7\times10^{-6}$ & \cellcolor{green!30}$2.4\times10^{-7}$ & \cellcolor{green!30}$4.2\times10^{-9}$ & \cellcolor{green!30}$1.1\times10^{-6}$ \\
        \textcolor{voids}{\Ei}-\Evi & \cellcolor{green!30}$5.2\times10^{-28}$ & \cellcolor{green!30}$1.3\times10^{-24}$ & \cellcolor{green!30}$1.1\times10^{-27}$ & \cellcolor{green!30}$6.0\times10^{-28}$ \\
        \hline
        \noalign{\vskip 1pt}
    \end{tabular}
    }
    \end{table}

    These tables provide a quantitative assessment of the statistical distinctness of galaxy populations across different environments in terms of the specific galaxy properties studied throughout our analysis. The implications of these statistical signals, or the lack thereof, are discussed in the context of galaxy evolution within the corresponding sections of the main text. Collectively, these results provide a quantitative reference for the level of environmental differentiation that is foreseen to be detectable under the observational constraints and selection functions anticipated for \Euclid\ DR1.

    \begin{table}[htbp!]
    \centering
    \caption{KS-test p-values for mass assembly and merger histories. The environments (\textcolor{voids}{\Ei}, \textcolor{walls}{\Eiii}, \Evi) are defined with respect to $1+\delta$. Cell colour coding indicates significance levels: green for $\leq 10^{-5}$, yellow for $10^{-5} < p \leq 10^{-3}$, orange for $10^{-3} < p \leq 0.05$, and red for $> 0.05$. All reported values are median p-values derived from 100 random realisations of the stellar mass-matching procedure.}
    \label{tab:mergerprops}
    {
    \setlength{\tabcolsep}{3pt}
    \begin{tabular}{lcccc}
        \hline
        \noalign{\vskip 1pt}
         & $0.45-0.63$ & $0.63-0.85$ & $0.85-1.15$ & $1.15-1.70$ \\
        \noalign{\vskip 1pt}
        \hline
        \noalign{\vskip 1pt}
        \multicolumn{5}{l}{Property: $\tau_{50}$} \\
        \textcolor{voids}{\Ei}-\textcolor{walls}{\Eiii} & \cellcolor{green!30}$1.2\times10^{-9}$ & \cellcolor{yellow!40}$5.2\times10^{-5}$ & \cellcolor{orange!40}$2.6\times10^{-2}$ & \cellcolor{red!30}$1.8\times10^{-1}$ \\
        \textcolor{voids}{\Ei}-\Evi & \cellcolor{green!30}$2.1\times10^{-129}$ & \cellcolor{green!30}$1.3\times10^{-83}$ & \cellcolor{green!30}$2.5\times10^{-36}$ & \cellcolor{green!30}$3.2\times10^{-18}$ \\
        \noalign{\vskip 1pt}
        \hline
        \noalign{\vskip 1pt}
        \multicolumn{5}{l}{Property: $\tau_{90}$} \\
        \textcolor{voids}{\Ei}-\textcolor{walls}{\Eiii} & \cellcolor{green!30}$9.3\times10^{-13}$ & \cellcolor{yellow!40}$9.7\times10^{-4}$ & \cellcolor{orange!40}$8.1\times10^{-2}$ & \cellcolor{red!30}$3.3\times10^{-1}$ \\
        \textcolor{voids}{\Ei}-\Evi & \cellcolor{green!30}$1.6\times10^{-161}$ & \cellcolor{green!30}$1.3\times10^{-69}$ & \cellcolor{green!30}$2.3\times10^{-45}$ & \cellcolor{green!30}$1.0\times10^{-25}$ \\
        \noalign{\vskip 1pt}
        \hline
        \noalign{\vskip 1pt}
        \multicolumn{5}{l}{Property: $t_{\rm last\,merger}$} \\
        \textcolor{voids}{\Ei}-\textcolor{walls}{\Eiii} & \cellcolor{red!30}$4.7\times10^{-1}$ & \cellcolor{red!30}$2.6\times10^{-1}$ & \cellcolor{red!30}$8.2\times10^{-1}$ & \cellcolor{red!30}$5.1\times10^{-1}$ \\
        \textcolor{voids}{\Ei}-\Evi & \cellcolor{green!30}$1.2\times10^{-6}$ & \cellcolor{green!30}$2.8\times10^{-6}$ & \cellcolor{orange!40}$5.5\times10^{-3}$ & \cellcolor{red!30}$1.7\times10^{-1}$ \\
        \noalign{\vskip 1pt}
        \hline
        \noalign{\vskip 1pt}
        \multicolumn{5}{l}{Property: $t_{\rm last\,major}$} \\
        \textcolor{voids}{\Ei}-\textcolor{walls}{\Eiii} & \cellcolor{red!30}$1.7\times10^{-1}$ & \cellcolor{red!30}$5.2\times10^{-1}$ & \cellcolor{red!30}$6.1\times10^{-1}$ & \cellcolor{red!30}$5.2\times10^{-1}$ \\
        \textcolor{voids}{\Ei}-\Evi & \cellcolor{orange!40}$8.4\times10^{-2}$ & \cellcolor{red!30}$4.2\times10^{-1}$ & \cellcolor{red!30}$6.1\times10^{-1}$ & \cellcolor{red!30}$5.2\times10^{-1}$ \\
        \noalign{\vskip 1pt}
        \hline
        \noalign{\vskip 1pt}
        \multicolumn{5}{l}{Property: $t_{\rm last\,minor}$} \\
        \textcolor{voids}{\Ei}-\textcolor{walls}{\Eiii} & \cellcolor{red!30}$9.9\times10^{-1}$ & \cellcolor{red!30}$9.6\times10^{-1}$ & \cellcolor{red!30}$8.3\times10^{-1}$ & \cellcolor{red!30}$9.3\times10^{-1}$ \\
        \textcolor{voids}{\Ei}-\Evi & \cellcolor{orange!40}$7.2\times10^{-2}$ & \cellcolor{orange!40}$3.3\times10^{-2}$ & \cellcolor{orange!40}$6.8\times10^{-2}$ & \cellcolor{red!30}$3.7\times10^{-1}$ \\
        \noalign{\vskip 1pt}
        \hline
        \noalign{\vskip 1pt}
        \multicolumn{5}{l}{Property: $t_{\rm last\,mini}$} \\
        \textcolor{voids}{\Ei}-\textcolor{walls}{\Eiii} & \cellcolor{red!30}$5.9\times10^{-2}$ & \cellcolor{red!30}$1.2\times10^{-1}$ & \cellcolor{red!30}$2.3\times10^{-1}$ & \cellcolor{red!30}$7.1\times10^{-1}$ \\
        \textcolor{voids}{\Ei}-\Evi & \cellcolor{green!30}$1.9\times10^{-6}$ & \cellcolor{green!30}$3.8\times10^{-6}$ & \cellcolor{orange!40}$3.0\times10^{-3}$ & \cellcolor{orange!40}$3.9\times10^{-2}$ \\
        \hline
        \noalign{\vskip 1pt}
    \end{tabular}
    }
    \end{table}

\end{appendix}

\end{document}